\documentclass[letterpaper]{article}

\newcommand{\chapter}[2][]{}

\usepackage[margin=1.5in]{geometry}

\usepackage{fix-cm}
\usepackage[english]{babel}
\usepackage{amssymb}
\usepackage{amsmath}
\usepackage{amsthm}

\newtheorem{theorem}{Theorem}

\usepackage{stmaryrd}
\usepackage{graphicx}
\usepackage[dvipsnames]{xcolor}
\usepackage{colortbl}

\usepackage{caption}
\usepackage{subcaption}
\usepackage{makeidx}
\usepackage{multicol}
\usepackage{todonotes}
\usepackage{wrapfig}
\usepackage{inputenc}
\usepackage{xspace}
\usepackage{stmaryrd}
\usepackage[LGR,OT1]{fontenc}
\usepackage{mathptmx}
\usepackage{graphicx}
\usepackage{listings}

\usepackage{mathrsfs}
\usepackage{stmaryrd}
\usepackage{bussproofs}
  
\usepackage{authblk}
\usepackage{multirow}

\usepackage{hyperref}
\hypersetup{
    colorlinks=true,
    linkcolor=blue,
    filecolor=magenta,      
    urlcolor=cyan,
    citecolor=ForestGreen
}

\usepackage{fancyhdr}
\date{}
\include{preamble}

\title{Formal methods for quantum algorithms: A Survey}

\author[4]{Christophe Chareton}
\author[4]{Sébastien Bardin}
\author[2,4]{Dongho Lee}
\author[2]{Benoit Valiron}
  \author[3]{Renaud Vilmart}
  \author[1,5]{Zhaowei Xu}

\affil[1]{Université Paris-Saclay,  CNRS, ENS Paris-Saclay, Laboratoire Méthodes Formelles, 91190, Gif-sur-Yvette, France }
\affil[2]{Université Paris-Saclay, Inria, CentraleSupélec, CNRS, ENS Paris-Saclay, Laboratoire Méthodes Formelles, 91190, Gif-sur-Yvette, France }
\affil[3]{Université Paris-Saclay, Inria, CNRS, ENS Paris-Saclay, Laboratoire Méthodes Formelles, 91190, Gif-sur-Yvette, France }
\affil[4]{Université Paris-Saclay, CEA, LIST, Palaiseau, France}
\affil[5]{University of Tartu, Estonia}

\begin{document}

\maketitle


\begin{abstract}
    While recent progress in quantum hardware open the door
for significant speedup in cryptography as well as additional key areas (biology, chemistry, optimization, machine learning, etc), quantum algorithms are still
hard to implement right, and the validation of such quantum programs
is a challenge. Moreover, importing the testing and debugging practices
at use in classical programming is extremely difficult in the quantum case, due to the destructive
aspect of quantum measurement.
As an alternative strategy,  formal methods     are  prone to play a decisive role in
the emerging field of quantum software. 
Recent works initiate solutions for problems occurring at every stage of the development process: high-level program design, implementation, compilation, etc.
We review the induced challenges for an efficient use of formal methods in quantum computing and the current most promising research directions.
\end{abstract}
\setcounter{tocdepth}{5}

\newpage

\tableofcontents

\usetikzlibrary{shapes}
 \definecolor{mygrey}{rgb}{0.9, 0.95, 0.95}
 \definecolor{yell}{RGB}{246,187,0}
\definecolor{yell}{RGB}{246,187,0}
\definecolor{blu}{RGB}{0,20,100}
\definecolor{re}{RGB}{148,6,39}
\definecolor{gree}{RGB}{0,100,0}

\newcommand{\meassymb}{\mathrel{\mbox{\text{${\frown}\hspace{-1.5ex}{/}\hspace{1ex}$}}}}
\newcommand{\ps}{\ensuremath{\texttt{pps}}}
\newcommand{\polyshor}{\ensuremath{\texttt{Shor-poly}}}
\newcommand{\Ca}{\ensuremath{\texttt{C\_angle}}}
\newcommand{\Cw}{\ensuremath{\texttt{C\_width}}}
\newcommand{\Cr}{\ensuremath{\texttt{C\_range}}}
\newcommand{\Ck}{\ensuremath{\texttt{C\_ket}}}
\newcommand{\isabelle}{\ensuremath{\textsc{Isabelle/HOL}}\xspace}

\newcommand{\prange}{\ensuremath{\texttt{pps\_range}}}
\newcommand{\pwidth}{\ensuremath{\texttt{pps\_width}}}
\newcommand{\prangeab}{\ensuremath{\texttt{p\_r}}}
\newcommand{\pwidthab}{\ensuremath{\texttt{p\_w}}}
\newcommand{\pangleab}{\ensuremath{\texttt{p\_a}}}
\newcommand{\pketab}{\ensuremath{\texttt{p\_k}}}
\newcommand{\pangle}{\ensuremath{\texttt{pps\_angle}}}
\newcommand{\pket}{\ensuremath{\texttt{pps\_ket}}}
\newcommand{\flathangle}{\ensuremath{\texttt{flat\_h\_angle}}}
\newcommand{\flathket}{\ensuremath{\texttt{flat\_h\_ket}}}
\newcommand{\flatangle}{\ensuremath{\texttt{flat\_angle}}}
\newcommand{\flatket}{\ensuremath{\texttt{flat\_ket}}}
\newcommand{\iogate}{\ensuremath{\texttt{circuit\_io}}}
\newcommand{\ketlength}{\ensuremath{\texttt{ket\_length}}}
\newcommand{\bvtoket}{\ensuremath{\texttt{bv\_to\_ket}}}
\newcommand{\contfrac}{\ensuremath{\texttt{cont\_frac}}}
\newcommand{\coprime}{\ensuremath{\texttt{co\_prime}}}
\newcommand{\bvlength}{\ensuremath{\texttt{bv\_length}}}
\newcommand{\applyps}{\ensuremath{\texttt{pps\_apply}}}
\newcommand{\csize}{\ensuremath{\texttt{csize}}}
\newcommand{\noeg}{\ensuremath{\texttt{size}}}
\newcommand{\noegc}{\ensuremath{\texttt{size}}_c}
\newcommand{\getket}{\ensuremath{\texttt{ket\_get}}}
\newcommand{\getbv}{\ensuremath{\texttt{bv\_get}}}
\newcommand{\makebv}{\ensuremath{\texttt{make\_bv}}}
\newcommand{\makeket}{\ensuremath{\texttt{make\_ket}}}
\newcommand{\noegs}{\ensuremath{\texttt{noGatesSet}}}
\newcommand{\noo}{\ensuremath{\texttt{noOperations}}}
\newcommand{\noinit}{\ensuremath{\texttt{noInit}}}
\newcommand{\nomeas}{\ensuremath{\texttt{noMeas}}}
\newcommand{\qsize}{\ensuremath{\texttt{qsize}}}
\newcommand{\cmem}{\ensuremath{\texttt{cmem}}}
\newcommand{\ppsequiv}{\ensuremath{\texttt{ppsequiv}}}
\newcommand{\return}{\ensuremath{\texttt{return}}}
\newcommand{\ereturn}{\ensuremath{\texttt{exec\_return}}}
\newcommand{\qmem}{\ensuremath{\texttt{qmem}}}
\newcommand{\ppssequence}{\ensuremath{\texttt{pps\_seq}}}
\newcommand{\ppsctl}{\ensuremath{\texttt{pps\_ctl}}}
\newcommand{\ppsanc}{\ensuremath{\texttt{pps\_ancilla}}}
\newcommand{\ppspar}{\ensuremath{\texttt{pps\_par}}}
\newcommand{\ppshad}{\ensuremath{\texttt{pps\_had}}}
\newcommand{\ppsrz}{\ensuremath{\texttt{pps\_rz}}}
\newcommand{\ppsphase}{\ensuremath{\texttt{pps\_phase}}}
\newcommand{\ppsswap}{\ensuremath{\texttt{pps\_swap}}}
\newcommand{\ppsid}{\ensuremath{\texttt{pps\_id}}}
\newcommand{\ppscnot}{\ensuremath{\texttt{pps\_cnot}}}
\newcommand{\et}{\ensuremath{\text{ET}}}
\newcommand{\extree}{\ensuremath{\texttt{exec\_tree}}}
\newcommand{\tolist}{\ensuremath{\texttt{to\_list}}}
\newcommand{\proba}{\ensuremath{\texttt{proba}}\xspace}
\newcommand{\probmeas}{\ensuremath{\texttt{proba\_measure}}\xspace}
\newcommand{\ancillas}{\ensuremath{\texttt{ancillas}}\xspace}
\newcommand{\ancilla}{\ensuremath{\texttt{ancilla}}\xspace}
\newcommand{\nogates}{\ensuremath{\texttt{size}}\xspace}
\newcommand{\contnogates}{\ensuremath{\ctlconst}}
\newcommand{\probmeasp}{\ensuremath{\texttt{proba\_measure\_p}}\xspace}
\newcommand{\probmeaspart}{\ensuremath{\texttt{proba\_measure}}\xspace}
\newcommand{\probmeaspartp}{\ensuremath{\texttt{proba\_measure\_p}}\xspace}
\newcommand{\implements}{\ensuremath{\texttt{implements}}\xspace}
\newcommand{\balanced}{\ensuremath{\texttt{balanced}}\xspace}
\newcommand{\constant}{\ensuremath{\texttt{constant}}\xspace}
\newcommand{\circuittrace}{\ensuremath{\texttt{circuit\_trace}}\xspace}
\newcommand{\circtopps}{\ensuremath{\texttt{circ\_to\_pps}}\xspace}
\newcommand{\gate}{\ensuremath{\texttt{gate}}\xspace}
\newcommand{\gencircuit}{\ensuremath{\texttt{gen\_circuit}}\xspace}
\newcommand{\gc}{\ensuremath{\texttt{GC}}\xspace}
\newcommand{\measuretrace}{\ensuremath{\texttt{meas\_trace}}\xspace}
\newcommand{\probatrace}{\ensuremath{\texttt{proba\_trace}}\xspace}
\newcommand{\evalue}{\ensuremath{\texttt{value}}\xspace}
\newcommand{\bvtoint}{\ensuremath{\texttt{bv\_to\_int}}\xspace}
\newcommand{\sep}{\ensuremath{\ \vert \ }}
\newcommand{\scal}{\ensuremath{\ \cdot \ }}
\newcommand{\zero}{\ensuremath{\ \mathbf{0} \ }}
\newcommand{\un}{\ensuremath{\ \mathbf{1} \ }}
\newcommand{\cnot}{\ensuremath{\texttt{cnot}}\xspace}
\newcommand{\had}{\ensuremath{\texttt{had}}}
\newcommand{\If}{\ensuremath{\mathbf{if}}\ }
\newcommand{\Then}{\ensuremath{\mathbf{then}}\ }
\newcommand{\Else}{\ensuremath{\mathbf{else}}\ }
\newcommand{\rx}{\ensuremath{\mathit{Rx}}}
\newcommand{\true}{\ensuremath{\mathit{true}}}
\newcommand{\BV}{\ensuremath{\text{bv}}}
\newcommand{\results}{\ensuremath{\text{results}}}
\newcommand{\lmr}{\ensuremath{\text{lmr}}}
\newcommand{\CR}{\ensuremath{\mathit{CR}}}
\newcommand{\rtoc}{\ensuremath{\text{r\_to\_c}}}
\newcommand{\ctoa}{\ensuremath{\text{c\_to\_a}}}
\newcommand{\itoa}{\ensuremath{\text{i\_to\_a}}}
\newcommand{\atoc}{\ensuremath{\text{a\_to\_c}}}
\newcommand{\cascade}{\ensuremath{\text{S}}\xspace }
\newcommand{\modulus}[1]{\ensuremath{ |#1|}}
\newcommand{\qftline}{\ensuremath{\text{QFT\_line}}\xspace }
\newcommand{\qftcoeffs}{\ensuremath{\text{QFT\_coeffs} }\xspace }
\newcommand{\qftlines}{\ensuremath{\text{QFT\_lines}}\xspace }
\newcommand{\Bv}{\ensuremath{\text{BV}}}
\newcommand{\inv}[1]{\ensuremath{-_a{#1}}}
\newcommand{\invind}[1]{\ensuremath{\textit{inv}^{#1}}}
\renewcommand{\dfrac}{\ensuremath{\text{angle}}\xspace }
\newcommand{\add}[2]{\ensuremath{#1 +_a #2}}
\newcommand{\Rcorrect}[2]{\ensuremath{\texttt{ppscircequiv}\Big(#1,  #2\Big)}}
\newcommand{\rcorrect}[2]{\ensuremath{(#2 \triangleright  #1)}}
\newcommand{\ccorrect}[2]{\ensuremath{(#2 \triangleright^c #1)}}
\newcommand{\invd}[2]{\ensuremath{\inv{#1}^{#2}}}
\newcommand{\dsum}[2]{\ensuremath{\stackrel{\tiny{a}}{\sum}\stackrel{}{_{#1}^{#2}}}}
\newcommand{\vdyadic}{\ensuremath{\mathit{Dyad}}}
\newcommand{\defin}{\ensuremath{\mathit{def}}}
\newcommand{\tofset}[1]{\ensuremath{\llbracket0, #1 \llbracket}\xspace }
\newcommand{\tofsett}[2]{\ensuremath{\llbracket #1, #2 \llbracket}\xspace }
\newcommand{\tofsetr}[1]{\ensuremath{\llbracket0, #1 \llbracket}\xspace }
\newcommand{\rows}[1]{\ensuremath{r( #1)}\xspace }
\newcommand{\col}[1]{\ensuremath{c( #1)}\xspace }
\newcommand{\bitrev}[1]{\ensuremath{\overleftrightarrow{#1}}\xspace }
\newcommand{\bvl}{\ensuremath{\mathit{l}}\xspace }
\newcommand{\place}{\ensuremath{\texttt{place}}\xspace }
\newcommand{\dcos}{\ensuremath{\mathit{d\_cos}}\xspace }
\newcommand{\ope}[1]{\ensuremath{O_{#1}\xspace }}
\newcommand{\dsin}{\ensuremath{\mathit{d\_sin}}\xspace }
\newcommand{\diagonal}{diagonal\ }
\newcommand{\ketl}{\ensuremath{\mathit{ketl}}}
\newcommand{\kettoint}{\ensuremath{\mathit{int}}}
\newcommand{\bvx}{\ensuremath{\mathit{bvx}}\xspace }
\newcommand{\bvy}{\ensuremath{\mathit{bvy}}\xspace }
\newcommand{\bvi}{\ensuremath{\mathit{bvi}}\xspace }
\newcommand{\bvk}{\ensuremath{\mathit{bvk}}\xspace }
\newcommand{\bin}{\ensuremath{\texttt{binary}}\xspace }
\newcommand{\rz}{\ensuremath{\texttt{rz}}}
\newcommand{\val}{\ensuremath{\textit{val}}}
\newcommand{\unitrt}[1]{\ensuremath{\mathit{unityRt}_{#1}}}
\newcommand{\avalue}[1]{\ensuremath{e^{2\pi i #1 }}}
\newcommand{\advalue}[2]{\avalue{\frac{#1}{2^{#2}}}}
\newcommand{\zerod}{\ensuremath{0_a}}
\newcommand{\rzp}{\ensuremath{\textit{R}}}
\newcommand{\intoexp}[2]{\ensuremath{\frac{2 \pi  #1}{2^{#2}}}}
\newcommand{\intoexpi}[2]{\ensuremath{\frac{2 \pi i #1}{2^{#2}}}}
\newcommand{\intounitm}[2]{\ensuremath{\omega_n^{-k}}}
\newcommand{\intocos}[2]{\ensuremath{\textit{cos} (\intoexp{#1}{#2})}}
\newcommand{\intosin}[2]{\ensuremath{\textit{sin} (\intoexp{#1}{#2})}}

\newcommand{\hadid}[2]{\ensuremath{\text{create\_superposition}_{#2}^{#1}}}
\newcommand{\operas}[2]{\ensuremath{(#1 \mapsto_O #2)}}
\newcommand{\operasi}[3]{\ensuremath{(#1 \mapsto_{O,#3} #2)}}
\newcommand{\superid}[2]{\ensuremath{\text{rewrite\_superposition}_{#2}^{#1}}}
\newcommand{\checkm}{\ensuremath{\includegraphics[width = 0.02\textwidth]{check.png}}}
\newcommand{\xmark}{\ensuremath{\includegraphics[width = 0.015\textwidth]{x.png}}}
\newcommand{\size}[1]{\ensuremath{\texttt{width}(#1)}\xspace }
\newcommand{\width}{\ensuremath{\texttt{width}}\xspace }
\newcommand{\ctlconst}{\ensuremath{\texttt{ctl-size-const}}\xspace }

\newcommand{\depth}{\ensuremath{\text{size}}\xspace }
\newcommand{\depthpre}{\ensuremath{\textit{depth\_pre}}\xspace }
\renewcommand{\d}{\ensuremath{\textit{d}}\xspace }
\newcommand{\pathsemantics}{\ensuremath{\mathit{Path}}\xspace }
\newcommand{\matsemantics}{\ensuremath{\mathit{Mat}}\xspace }
\newcommand{\interl}{\vspace{1.5mm}}
\newcommand{\pathfull}{\ensuremath{\textit{path\_sum}}}
\newcommand{\requires}[1]{$\qquad$\textcolor{blu}{requires\{\textcolor{black}{#1}\}}}
\newcommand{\ensures}[1]{$\qquad$\textcolor{re}{ensures\{\textcolor{black}{#1}\}}}
\newcommand{\ensurestf}[1]{$\qquad$\textcolor{re}{ensures\{\textcolor{black}{#1}}}
\newcommand{\ensurests}[1]{$\qquad\qquad$ #1\textcolor{re}{\}}}

\newcommand{\invariant}[1]{$\qquad$\textcolor{gree}{invariant\{\textcolor{black}{#1}\}}}
\newcommand{\variant}[1]{$\qquad$\textcolor{gree}{variant\{\textcolor{black}{#1}\}}}
\newcommand{\Matrix}{\ensuremath{\text{matrix}}\xspace }
\newcommand{\matrixtype}{\ensuremath{\texttt{matrix}} }
\newcommand{\kettype}{\ensuremath{\texttt{ket}} }
\newcommand{\Rows}{\ensuremath{\text{rows}}\xspace }
\newcommand{\Columns}{\ensuremath{\text{columns}}\xspace }
\newcommand{\Get}{\ensuremath{\text{get}}\xspace }

\newcommand{\csequ}{\ensuremath{\textit{Sequ}}}
\newcommand{\cpar}{\ensuremath{\texttt{circ\_par}}}

\renewcommand{\mod}{\ensuremath{\texttt{mod}}}
\renewcommand{\div}{\ensuremath{\texttt{div}}\xspace}
\newcommand{\modu}[2]{\ensuremath{#1[#2..]}}
\newcommand{\adjoint}[1]{\ensuremath{\mathit{adjoint}(#1)}}
\newcommand{\daggered}[1]{\ensuremath{#1^{\dagger}}}
\newcommand{\divi}[2]{\ensuremath{#1[..#2]}}
\newcommand{\vproj}{\ensuremath{\texttt{proj}}}
\newcommand{\Cnot}{\ensuremath{\textit{CNOT}}}
\newcommand{\norme}{\ensuremath{\textit{norme}}}
\newcommand{\norm}[1]{\ensuremath{| #1|}}
\newcommand{\N}{\ensuremath{\textit{N}}}
\newcommand{\Cnnot}{\ensuremath{\textit{CNNOT}}}
\newcommand{\ccnot}[1]{\ensuremath{\textit{CNNOT}^{#1}}}
\newcommand{\cnnot}{\ensuremath{\textit{cnnot}}}
\newcommand{\Id}{\ensuremath{\textit{Id}}}
\newcommand{\id}{\ensuremath{\textit{id}}}
\newcommand{\R}{\ensuremath{\textit{R}}\ }
\newcommand{\Rd}{\ensuremath{\textit{R}^d}\ }
\newcommand{\Rdi}[1]{\ensuremath{\textit{R}^{d_{#1}}}\ }
\renewcommand{\Re}{\ensuremath{\textit{R}^e}\ }

\newcommand{\shorof}{\ensuremath{\texttt{Shor-OF}}\xspace}
\newcommand{\qbrickCORE}{\ensuremath{\textsc{Qbricks}}}
\newcommand{\qbrick}{\qbrickCORE\xspace}
\newcommand{\pselt}{\ensuremath{\textit{AbsPS}}\xspace}
\newcommand{\qbricks}{\qbrickCORE\xspace}
\newcommand{\sqir}{\ensuremath{\textsc{Sqir}}\xspace}
\newcommand{\oqasm}{\ensuremath{\textsc{OpenQASM}}\xspace}
\newcommand{\qiskit}{\ensuremath{\textsc{Qiskit}}\xspace}
\newcommand{\silq}{\ensuremath{\textsc{Silq}}\xspace}
\newcommand{\voqc}{\ensuremath{\textsc{VOQC}}\xspace}
\newcommand{\coq}{\ensuremath{\text{Coq}}\xspace}
\newcommand{\quipper}{\ensuremath{\text{Quipper}}\xspace}
\newcommand{\protoquipper}{\ensuremath{\text{ProtoQuipper}}\xspace}
\newcommand{\protoquippers}{\ensuremath{\text{ProtoQuippers}}\xspace}
\newcommand{\qwire}{\ensuremath{\textsc{Qwire}}\xspace}
\newcommand{\qbrickDSL}{\qbrickCORE-\textsc{DSL}\xspace}
\newcommand{\qbrickOPE}{\qbrickCORE-\textsc{OPE}\xspace}
\newcommand{\qbrickSPEC}{\qbrickCORE-\textsc{Spec}\xspace}
\newcommand{\hqhl}{HQHL}
\newcommand{\hqhllong}{Hybrid Quantum Hoare Logic} 
\newcommand{\superp}{\ensuremath{\textit{Sup}}\xspace}
\newcommand{\Rde}{\ensuremath{\textit{R}^{d\times e}}\ }
\newcommand{\RP}{\ensuremath{\textit{R}^{\Pi d_j}}\ }
\newcommand{\qubit}{\ensuremath{\textit{Qubit}}\ }
\newcommand{\alice}{\ensuremath{\textit{alice}}\ }
\newcommand{\bob}{\ensuremath{\textit{bob}}\ }
\newcommand{\tif}{\ensuremath{\textit{if}}\ }
\newcommand{\tthen}{\ensuremath{\textit{then}}\ }
\newcommand{\telse}{\ensuremath{\textit{else}}\ }
\newcommand{\kp}{\ensuremath{\textit{kP}}\xspace }
\newcommand{\pp}{\ensuremath{\textit{pP}}\xspace }
\newcommand{\circuit}{\ensuremath{\text{circuit}}\ }
\newcommand{\mycirc}{\circ}
\renewcommand{\circuit}{\ensuremath{\text{circ}}\ }
\newcommand{\fcirc}{\ensuremath{\text{fcirc}}\ }
\newcommand{\func}{\ensuremath{\textit{f}}\ }
\newcommand{\make}{\ensuremath{\text{make}}\xspace }
\newcommand{\precirc}{\ensuremath{\text{preCircuit}}\xspace }
\newcommand{\spec}{\ensuremath{\textit{spec}}\xspace }
\newcommand{\PO}{\ensuremath{\textit{PO}}\xspace }
\newcommand{\precircs}{\ensuremath{\text{preCircuits}}\xspace }
\newcommand{\phasepart}{\ensuremath{\textit{phasePart}}\ }
\newcommand{\diagphase}{\ensuremath{\text{d\_ph}}\ }
\newcommand{\diagrz}{\ensuremath{\text{d\_rz}}\ }
\newcommand{\diagsequ}{\ensuremath{\text{d\_sequence}}\ }
\newcommand{\diagsequit}{\ensuremath{\text{d\_sequ\_iter}}\ }
\newcommand{\diagpar}{\ensuremath{\text{d\_parallel}}\ }
\newcommand{\qbit}{\ensuremath{\textit{Qubit}}\ }
\newcommand{\For}{\ensuremath{\textit{for}}\ }
\newcommand{\Int}{\ensuremath{\text{int}}\ }
\newcommand{\complex}{\ensuremath{\text{complex}}\ }
\newcommand{\Do}{\ensuremath{\textit{do}}\ }
\newcommand{\apply}{\ensuremath{\texttt{apply}}\ }
\newcommand{\measure}{\ensuremath{\texttt{measure}}\ }
\newcommand{\countpop}{\ensuremath{\textit{popcount}}}
\newcommand{\popcount}{\ensuremath{\textit{popcount}}}
\newcommand{\semantics}{\ensuremath{\textit{sem}}}
\newcommand{\protoquaml}{Proto-Quaml\ }
\newcommand{\liquid}{\text{Liqu}\textit{i}$|\rangle$}
\newcommand{\projectq}{\textsc{ProjectQ}}
\newcommand{\HH}
           {\ensuremath{\scalebox{.8}{\begin{tikzpicture}
    \clip (-.3,-.24) rectangle (1.1,.24);
    \node[draw,rectangle](1) at (0,0){H};
    \node[draw,rectangle](2) at (.8,0){H};
    \draw (1)to(2);
  \end{tikzpicture}}\ }}
\newcommand{\AND}{\ensuremath{\ \&\ }}
\newcommand{\opnot}{\ensuremath{\textit{not}\ }}
\newcommand{\cons}{\ensuremath{\textit{cons}}\ }

\newcommand{\cont}{\ensuremath{\texttt{control}}\ }
\newcommand{\even}{\ensuremath{\textit{even}}\ }
\newcommand{\odd}{\ensuremath{\textit{odd}}\ }
\newcommand{\ct}{\ensuremath{\textit{ct}}\ }
\newcommand{\qft}{\ensuremath{\text{QFT}}\ }
\newcommand{\eqdef}{\ensuremath{=_{\textit{def}}}\ }

\newcommand{\clone}{\textbf{clone}\xspace }
\newcommand{\export}{\textbf{export}\xspace }
\newcommand{\with}{\textbf{with}\xspace }
\newcommand{\type}{\textbf{type}\xspace }
\newcommand{\function}{\textbf{function}\xspace }
\newcommand{\predicate}{\textbf{predicate}\xspace }
\newcommand{\axiomb}{\textbf{axiom}\xspace }
\newcommand{\valb}{\textbf{val}\xspace }
\newcommand{\rec}{\textbf{rec}\xspace }
\newcommand{\ghost}{\texttt{ghost}\xspace }
\newcommand{\letb}{\textbf{let}\xspace }
\newcommand{\lemmab}{\textbf{lemma}\xspace }

\newcommand{\bit}{\ensuremath{\textit{bit}}\ }
\newcommand{\meas}{\ensuremath{\textit{Meas}}\ }
\newcommand{\Meas}{\ensuremath{\textit{Meas}}\ }
\newcommand{\List}[1]{\ensuremath{\textit{list}^{#1}}\ }
\newcommand{\tens}[1]{\ensuremath{\textit{Tens}(#1)}\ }
\newcommand{\Tens}[1]{\ensuremath{(\textit{Tens}(#1))}}
\newcommand{\base}[2]{\ensuremath{\overline{#2}^{#1}}}
\newcommand{\basedix}[1]{\base{10}{#1} }
\newcommand{\basedeux}[1]{\base{2}{#1} }

\newcommand{\vect}[1]{\ensuremath{\textit{vect}_{#1}}\ }

\newcommand{\init}{\ensuremath{\texttt{init}}\ }
\newcommand{\tlet}{\ensuremath{\textit{let}}\ }
\newcommand{\tin}{\ensuremath{\textit{in}}\ }
\newcommand{\get}{\ensuremath{\texttt{get}}\ }

\newcommand{\dec}{\ensuremath{\textit{dec}} }
\newcommand{\lrangle}[1]{\ensuremath{\langle  #1 \rangle}}
\newcommand{\parcomp}{\ensuremath{\ /\hspace{-2pt}/ \ }}
\newcommand{\bigparcomp}{\ensuremath{\ \big/\hspace{-3pt}\big/ \ } }
\newcommand{\bigparcompa}[1]{\ensuremath{\ \big/\hspace{-3pt}\big/_{ #1 } \ } }
  \newcommand\tab[1][1cm]{\hspace*{#1}}

\newcounter{magicrownumbers}
\newcommand\rownumber{
  \refstepcounter{magicrownumbers}\arabic{magicrownumbers}}
\newcommand{\ket}[2]{\ensuremath{\vert #1 \rangle}_{#2}}
\newcommand{\bra}[2]{\ensuremath{\ \langle #1 \vert}_{#2}}
\newcommand{\braket}[2]{\ensuremath{\ \langle #1 \vert{#2}\rangle}}
\newcommand{\ketbra}[2]{\ket{#1}{}\!\!\!\!\bra{#2}{}}
\newcommand{\rotation}[3]{\ensuremath{\mathit{Rot}_{#1,#2}(#3)}}
\newcommand{\reflexion}[1]{\ensuremath{\mathit{Refl}_{#1}}}
\newcommand{\card}[1]{\ensuremath{\mid #1 \mid}}
\newcommand{\hid}[3]{\pgfmathsetmacro{\total}{pow(2,#1)}\pgfmathsetmacro{\figscale}{#2}\pgfmathsetmacro{\nodescale}{#3}\input{fig/hid}}
\newcommand{\cnotmac}[3]{\pgfmathsetmacro{\total}{pow(2,#1)}\pgfmathsetmacro{\figscale}{#2}\pgfmathsetmacro{\nodescale}{#3}\input{fig/cnot}}
\newcommand{\hidcnot}[3]{\pgfmathsetmacro{\total}{pow(2,#1)}\pgfmathsetmacro{\figscale}{#2}\pgfmathsetmacro{\nodescale}{#3}\input{fig/hidcnot}}
\newcommand{\entrees}[3]{\pgfmathsetmacro{\total}{pow(2,#1)-1}\pgfmathsetmacro{\figscale}{#2}\pgfmathsetmacro{\nodescale}{#3}\input{fig/pos}}
\newcommand{\sorties}[3]{\pgfmathsetmacro{\total}{pow(2,#1)-1}\pgfmathsetmacro{\figscale}{#2}\pgfmathsetmacro{\nodescale}{#3}\input{fig/resint}}
\newcommand{\sortiesun}[3]{\pgfmathsetmacro{\total}{pow(2,#1)-1}\pgfmathsetmacro{\figscale}{#2}\pgfmathsetmacro{\nodescale}{#3}\input{fig/resintun}}
\newcommand{\sortieszero}[3]{\pgfmathsetmacro{\total}{pow(2,#1)-1}\pgfmathsetmacro{\figscale}{#2}\pgfmathsetmacro{\nodescale}{#3}\input{fig/resintzero}}

\newcommand{\projzero}[3]{\pgfmathsetmacro{\total}{pow(2,#1)-1}\pgfmathsetmacro{\figscale}{#2}\pgfmathsetmacro{\nodescale}{#3}\input{fig/projzero}}
\newcommand{\projun}[3]{\pgfmathsetmacro{\total}{pow(2,#1)-1}\pgfmathsetmacro{\figscale}{#2}\pgfmathsetmacro{\nodescale}{#3}\input{fig/projun}}

\newcommand{\mathid}[3]{\ensuremath{\begin{pmatrix} \pgfmathsetmacro{\total}{pow(2,#1)}\pgfmathsetmacro{\figscale}{#2}\pgfmathsetmacro{\nodescale}{#3}\input{fig/hid}\end{pmatrix}}}
\newcommand{\shift}[2]{\ensuremath{\overleftarrow{#1}^{#2}}\xspace }
\newcommand{\mathidcnot}[3]{\ensuremath{\begin{pmatrix} \pgfmathsetmacro{\total}{pow(2,#1)}\pgfmathsetmacro{\figscale}{#2}\pgfmathsetmacro{\nodescale}{#3}\input{fig/hidcnot}\end{pmatrix}}}
\newcommand{\matentrees}[3]{\ensuremath{\begin{matrix} \pgfmathsetmacro{\total}{pow(2,#1)-1}\pgfmathsetmacro{\figscale}{#2}\pgfmathsetmacro{\nodescale}{#3}\input{fig/pos}\end{matrix}}}
\newcommand{\matsorties}[3]{\ensuremath{\begin{matrix} \pgfmathsetmacro{\total}{pow(2,#1)-1}\pgfmathsetmacro{\figscale}{#2}\pgfmathsetmacro{\nodescale}{#3}\input{fig/resint}\end{matrix}}}
\newcommand{\matsortiesun}[3]{\ensuremath{\begin{matrix} \pgfmathsetmacro{\total}{pow(2,#1)-1}\pgfmathsetmacro{\figscale}{#2}\pgfmathsetmacro{\nodescale}{#3}\input{fig/resintun}\end{matrix}}}
\newcommand{\matsortieszero}[3]{\ensuremath{\begin{matrix} \pgfmathsetmacro{\total}{pow(2,#1)-1}\pgfmathsetmacro{\figscale}{#2}\pgfmathsetmacro{\nodescale}{#3}\input{fig/resintzero}\end{matrix}}}
\newcommand{\matprojzero}[3]{\ensuremath{\begin{pmatrix} \pgfmathsetmacro{\total}{pow(2,#1)-1}\pgfmathsetmacro{\figscale}{#2}\pgfmathsetmacro{\nodescale}{#3}\input{fig/projzero}\end{pmatrix}}}
\newcommand{\matprojun}[3]{\ensuremath{\begin{pmatrix} \pgfmathsetmacro{\total}{pow(2,#1)-1}\pgfmathsetmacro{\figscale}{#2}\pgfmathsetmacro{\nodescale}{#3}\input{fig/projun}\end{pmatrix}}}

 \newcommand{\pre}{\ensuremath{\math{pre}}\xspace }
 \newcommand{\isaket}{\ensuremath{\text{isAKet}}\xspace }
 \newcommand{\isaketbasis}{\ensuremath{\text{is\_a\_ket\_basis\_elt}}\xspace }
 \newcommand{\isaketl}{\ensuremath{\text{isAKetL}}\xspace }

 \newcommand{\real}{\ensuremath{\texttt{Re}}\xspace }
 \newcommand{\isreal}{\ensuremath{\text{Real}}\xspace }
 \newcommand{\phaseinv}{\ensuremath{\mathit{Inv}}\xspace }
 \newcommand{\imag}{\ensuremath{\texttt{Im}}\xspace }
\newcommand{\intocosd}[2]{\ensuremath{\real(\intounit{#1}{#2})}}
\newcommand{\intosind}[2]{\ensuremath{\imag(\intounit{#1}{#2})}}

 \newcommand{\correct}{\ensuremath{\text{sequenceCorrect}}\xspace }
 \newcommand{\firstc}{\ensuremath{\mathit{first}_c}\xspace }
 \newcommand{\matsem}{\ensuremath{\mathit{matSem}}\xspace }
 \newcommand{\topre}{\ensuremath{\text{Pre}}\xspace }
 \newcommand{\toqc}{\ensuremath{\text{Qc}}\xspace }
 \newcommand{\eigen}{\ensuremath{\texttt{Eigen}}\xspace }
 \newcommand{\eigenseq}{\ensuremath{\text{eigen\_seq}}\xspace }
 \newcommand{\eigensquare}{\ensuremath{\text{eigen\_square}}\xspace } 
 \newcommand{\eigencomp}{\ensuremath{\text{eigen\_comp}}\xspace }
 \newcommand{\eigenpow}{\ensuremath{\text{eigen\_pow\_2}}\xspace }
 \newcommand{\conteigen}{\ensuremath{\text{control\_eigen}}\xspace }
 \newcommand{\conteigend}{\ensuremath{\text{control\_eigen\_dyad}}\xspace }
 \newcommand{\conteigenscal}{\ensuremath{\text{control\_eigen\_scalar}}\xspace }
 \newcommand{\conteigenseq}{\ensuremath{\text{control\_eigen\_seq\_iter}}\xspace }
 \newcommand{\flatphase}{\ensuremath{\mathit{flatPhase}}\xspace }
 \newcommand{\flatk}{\ensuremath{\mathit{flat}}\xspace }
 \newcommand{\diag}{\ensuremath{\texttt{diag}}\xspace }
 \renewcommand{\flat}{\ensuremath{\texttt{flat}}\xspace }
 \newcommand{\result}{\ensuremath{\texttt{result}}\xspace }
\newcommand{\sem}[3]{\ensuremath{(#1: #2 \mapsto #3)}}

 \newcommand{\idc}{\ensuremath{\texttt{Id}}\xspace }
 \newcommand{\ido}{\ensuremath{I}\xspace }
 \newcommand{\dz}{\ensuremath{0_d}\xspace }

\newcommand{\applyqft}{\ensuremath{\text{apply\_inverse\_QFT}}\xspace }
\newcommand{\initial}{\ensuremath{\mathit{initial\_state}}\xspace }
\newcommand{\diagseq}{\ensuremath{\mathit{diagSeqIter}}\xspace }
\newcommand{\applybox}{\ensuremath{\mathit{apply\_black\_box}}\xspace }
\newcommand{\create}{\ensuremath{\mathit{create\_superposition}}\xspace }
\newcommand{\cascadec}{\ensuremath{\mathit{cascCEO}}\xspace }
\newcommand{\bcascadec}{\ensuremath{\mathit{boundedCascCEO}}\xspace }
\newcommand{\ec}{\ensuremath{\mathit{cEO}}\xspace }
\newcommand{\idrevq}{\ensuremath{\mathit{idRevQFT}}\xspace }
\newcommand{\idsup}{\ensuremath{\mathit{idSup}}\xspace }
\newcommand{\eigenexp}{\ensuremath{\mathit{eigenExp}}\xspace }
\newcommand{\bound}{\ensuremath{\text{bound}}\xspace }
\newcommand{\fun}{\ensuremath{\texttt{fun}}\xspace }
\newcommand{\supfunc}{\ensuremath{\text{superposition}}\xspace }
\newcommand{\ift}{\ensuremath{\text{inverse\_Fourier\_transform}}\xspace }
\newcommand{\ie}{\ensuremath{\textit{i.e.}}\xspace }
\newcommand{\wrt}{\ensuremath{\textit{wrt}}\xspace }
\newcommand{\bb}{\ensuremath{\text{apply\_black\_box}}\xspace }
\newcommand{\abb}{\ensuremath{\textit{apply\_black\_box}}\xspace }
\newcommand{\initialisation}{\ensuremath{\textit{initialisation}}\xspace }
\newcommand{\qftrew}{\ensuremath{\text{place\_QFT}}\xspace }
\newcommand{\revqftrew}{\ensuremath{\text{rev\_place\_Qft}}\xspace }
\newcommand{\phasepre}{\ensuremath{\text{phase\_estim\_pre}}\xspace }
\newcommand{\phasestim}{\ensuremath{\text{phase\_estim}}\xspace }
\newcommand{\ft}{\ensuremath{\textit{ft}}\xspace }

 \newcommand{\sequence}{\ensuremath{\texttt{sequence}}\xspace }
 \renewcommand{\parallel}{\ensuremath{\texttt{parallel}}\xspace }
 \newcommand{\asequence}{\ensuremath{\text{sequence}}\xspace }
 \newcommand{\basequence}{\ensuremath{\text{sequence}^{r,a}}\xspace }
 \newcommand{\aparallel}{\ensuremath{\text{parallel}}\xspace }
 \newcommand{\Sequence}{\ensuremath{\text{Sequence}}\xspace }
 \newcommand{\Parallel}{\ensuremath{\text{Parallel}}\xspace }
\newcommand{\amatsemantics}{\ensuremath{\text{Mat}}\xspace }
\newcommand{\asize}[1]{\ensuremath{\textbf{s}_{#1}}\xspace }
\newcommand{\basize}[1]{\ensuremath{\textbf{s}^{r,a}_{#1}}\xspace }
\newcommand{\ry}{\ensuremath{\text{Ry}}}
\newcommand{\bv}{\ensuremath{\text{bv}}}
\newcommand{\rxx}{\ensuremath{\text{rx}}}
\newcommand{\barxx}{\ensuremath{\text{rx}^{r,a}}}
\newcommand{\arxx}{\ensuremath{\text{rx}}}
\newcommand{\aryy}{\ensuremath{\text{ry}}}
\newcommand{\bryy}{\ensuremath{\text{ry}}}
\newcommand{\baryy}{\ensuremath{\text{ry}^{r,a}}}
\newcommand{\barzz}{\ensuremath{\text{rz}^{r,a}}}
\newcommand{\brxx}{\ensuremath{\text{rx}}}
\newcommand{\arzz}{\ensuremath{\text{rz}}}
\newcommand{\ryy}{\ensuremath{\text{ry}}}
\newcommand{\rzz}{\ensuremath{\text{ry}}}
\newcommand{\phase}{\ensuremath{\texttt{phase}}}
\newcommand{\Phase}{\ensuremath{\textit{Ph}}}
\newcommand{\Had}{\ensuremath{\textit{H}}}
\newcommand{\Rz}{\ensuremath{\textit{R}_{\textit{Z}}}}
\newcommand{\Post}{\ensuremath{\text{Post}}}
\newcommand{\phasee}{\ensuremath{\text{ph}}}
\newcommand{\aphasee}{\ensuremath{\text{ph}}}
\newcommand{\baphasee}{\ensuremath{\text{ph}^{r,a}}}
\newcommand{\cnott}{\ensuremath{\text{cnot}}}
\newcommand{\acnott}{\ensuremath{\text{cnot}}}
\newcommand{\sumrange}{\ensuremath{\textit{sumRange}}\ }
\newcommand{\ketpart}{\ensuremath{\textit{ketPart}}\ }
\newcommand{\sr}[1]{\ensuremath{\textbf{sR}_{#1}}\xspace }
\newcommand{\Rak}{\ensuremath{\textbf{Rak}}\xspace }
\newcommand{\rak}{\ensuremath{\textbf{rak}}\xspace }
\newcommand{\drak}{\ensuremath{\textbf{drak}}\xspace }
\renewcommand{\frak}{\ensuremath{\textbf{frak}}\xspace }
\newcommand{\hpz}{\ensuremath{\textbf{HPZ}}\xspace }
\newcommand{\deltaraks}{\ensuremath{\Delta-\textbf{raks}}\xspace }
\newcommand{\deltarak}{\ensuremath{\Delta-\textbf{rak}}\xspace }
\newcommand{\hpzraks}{\ensuremath{\textbf{HPZraks}}\xspace }
\newcommand{\hpzrak}{\ensuremath{\textbf{HPZrak}}\xspace }
\newcommand{\raks}{\ensuremath{\textbf{raks}}\xspace }
\newcommand{\scalp}[1]{\ensuremath{\textbf{a}_{#1}}\xspace }
\newcommand{\asr}[1]{\ensuremath{\textbf{sR}_{#1}}\xspace }
\newcommand{\ascalp}[1]{\ensuremath{\textbf{sC}_{#1}}\xspace }
\newcommand{\dscalp}[1]{\ensuremath{\textbf{dA}_{#1}}\xspace }
\newcommand{\fscalp}[1]{\ensuremath{\textbf{fA}_{#1}}\xspace }
\newcommand{\bk}[1]{\ensuremath{\textbf{bK}_{#1}}\xspace }
\newcommand{\abk}[1]{\ensuremath{\textbf{bK}_{#1}}\xspace }
\newcommand{\fbk}[1]{\ensuremath{\textbf{fBk}_{#1}}\xspace }
\newcommand{\acircuit}{\ensuremath{\text{circuit}}\ }
\newcommand{\bacircuit}{\ensuremath{\text{circuit}^{r,a}}\ }
\newcommand{\Circuit}{\ensuremath{\text{Circuit}}\ }
\newcommand{\Acircuit}{\ensuremath{\text{Circuit}}\ }
\newcommand{\Bcircuit}{\ensuremath{\text{Circuit}^b}\ }
\newcommand{\Bacircuit}{\ensuremath{\text{Circuit}^{r,a}}\ }
\newcommand{\am}{\ensuremath{^a}\xspace }
\newcommand{\afunc}{\ensuremath{\textit{f}^a}\ }
\newcommand{\reverse}[1]{\ensuremath{\texttt{inverse}(#1)}\ }
\newcommand{\reversen}{\ensuremath{\texttt{invert}}\xspace }
\newcommand{\areverse}[1]{\ensuremath{\textbf{Rev}(#1)}\ }
\newcommand{\areversen}{\ensuremath{\textbf{Rev}}\xspace }
\newcommand{\Mat}[1]{\ensuremath{\textbf{Mat}(#1)}\ }
\newcommand{\aMat}[1]{\ensuremath{\textbf{Mat}(#1)}\ }
\newcommand{\Mate}{\ensuremath{\textbf{Mat}}\ }
\newcommand{\aMate}{\ensuremath{\textbf{Mat}}\ }
\newcommand{\mat}{\ensuremath{\texttt{mat}}\ }
\newcommand{\pat}{\ensuremath{\textbf{Ps}}}
 \newcommand{\unitps}[1]{\ensuremath{\textbf{Ps}^{\textit{b}}_{#1}}}
\newcommand{\apat}[1]{\ensuremath{\textbf{Ps}_{#1}}\ }
\newcommand{\bapat}[1]{\ensuremath{\textbf{Ps}^{r,a}_{#1}}\ }
\newcommand{\aunitps}[1]{\ensuremath{\textbf{Ps}^{\textit{r,a}}_{#1}}}
\newcommand{\hadamard}{\ensuremath{\textit{hadamard}}}
\newcommand{\hadd}{\ensuremath{\textit{h}}}
\newcommand{\ahadd}{\ensuremath{\textit{h}}}
\newcommand{\bahadd}{\ensuremath{\textit{h}^{r,a}}}
\newcommand{\tcircuit}[1]{\begin{tabular}{c}\scalebox{.8}{\begin{tikzpicture}
\clip (-.3,-.24) rectangle (.3,.24);
    \node[draw,rectangle] at (0,0){$#1$};
  \end{tikzpicture}}\end{tabular}}
  
\newcommand{\bcircuit}[1]{\begin{tabular}{c}\begin{tikzpicture}
    \node[draw,rectangle] at (0,0){$#1$};
  \end{tikzpicture}\end{tabular}}

\newcommand{\mycomment}[1]{}

\newcommand{\pxy}{\ensuremath{\mathcal{P}_{\ket{Ox}{},\ket{Oy}{}}}}
\newcommand{\A}{\ensuremath{\mathcal{A}}}

\hyphenation{se-quence}
\hyphenation{che-cking}
\hyphenation{groun-ded}
\hyphenation{se-man-tic}
\hyphenation{o-bli-ga-tion}
\hyphenation{a-ri-thme-tic}
\hyphenation{se-quences}
\hyphenation{es-ti-ma-tion}
\hyphenation{pa-ra-me-tri-zed}
\hyphenation{in-te-res-ted}
\hyphenation{in-va-riance}
\hyphenation{cry-pto-gra-phy}
\hyphenation{spe-ci-fi-ca-tion}
\hyphenation{cer-ti-fi-ca-tion}
\hyphenation{spe-ci-fi-ca-tions}
\hyphenation{re-sul-ting}
\hyphenation{as-ser-tions}
\hyphenation{pro-ba-bi-lis-ti-cal-ly}
\hyphenation{in-va-ri-an-ce}
\hyphenation{su-per-po-si-tion}
\hyphenation{im-ple-ment}
\hyphenation{pro-gram}
\hyphenation{func-tion}
\hyphenation{func-tions}
\hyphenation{trans-for-ma-tion}
\hyphenation{pro-grams}
\hyphenation{al-go-ri-thm}
\hyphenation{ma-the-ma-ti-cal}
\hyphenation{par-ti-cu-lar-ly}
\hyphenation{par-ti-cu-lar}
\hyphenation{ab-so-lu-te}
\hyphenation{for-mal}
\hyphenation{spe-ci-fy}
\hyphenation{se-man-tics}
\hyphenation{im-ple-men-ta-tion}
\hyphenation{im-ple-men-ta-tions}
\hyphenation{in-va-riant}
\hyphenation{ope-ra-tion}
\hyphenation{ope-ra-tionq}
\hyphenation{Sec-tion}
\hyphenation{al-go-ri-thms}
\hyphenation{me-cha-nism}
\hyphenation{me-cha-nisms}
\hyphenation{re-cur-si-ve-ly}
\hyphenation{ve-ri-fi-ca-tion}
\hyphenation{com-po-si-tion}
\hyphenation{post-con-di-tion}
\hyphenation{pre-con-di-tion}
\hyphenation{post-con-di-tions}
\hyphenation{pre-con-di-tions}
\hyphenation{de-com-po-si-tion}
\hyphenation{de-fi-ni-tion}
\hyphenation{de-fi-ni-tions}
\hyphenation{com-plexed}
\hyphenation{coef-fi-cient}
\hyphenation{coef-fi-cients}
\hyphenation{pa-ra-lel-ly}
\hyphenation{ele-men-ta-ry}
\hyphenation{tran-sla-ting}
\hyphenation{com-pa-ri-son}
\hyphenation{au-to-ma-tion}
\hyphenation{uni-ver-sal-ly}
\hyphenation{au-to-ma-tion}
\hyphenation{pro-ba-bi-lis-tic-al-ly}
\hyphenation{se-quen-tial}
\hyphenation{se-quen-ce}
\hyphenation{pre-sen-ted}
\hyphenation{lit-te-ra-tu-re}
\hyphenation{pro-gram-ming}
\hyphenation{com-pu-ting}
\hyphenation{quan-tum}
\hyphenation{phe-no-me-na}
\hyphenation{op-ti-mi-za-tion}
\hyphenation{im-ple-men-ta-tion}
\hyphenation{in-va-riance}
\hyphenation{con-nec-ti-vi-ty}

\newcommand{\bor}{\mid}
\newcommand{\tuple}[1]{\langle#1\rangle}
\newcommand{\proj}[2]{\pi_{#1}#2}
\newcommand{\ttrue}{\texttt{t\!t}}
\newcommand{\ffalse}{\texttt{f\!f}}
\newcommand{\iftermx}[3]{\texttt{if}\,{#1}\,\texttt{then}\,{#2}\,\texttt{else}\,{#3}}
\newcommand{\letinterm}[3]{\texttt{let}\,{#1}\,{=}\,{#2}\,\texttt{in}\,{#3}}
\newcommand{\decl}[3]{\texttt{let}~{#1}(#2)~{=}~{#3}}
\newcommand{\bool}{\texttt{bool}}
\newcommand{\inttype}{\texttt{int}}
\newcommand{\settype}{\texttt{set}}
\newcommand{\realtype}{\texttt{real}}
\newcommand{\complextype}{\texttt{complex}}
\newcommand{\circtype}{\texttt{circ}}
\newcommand{\unittype}{\top}
\newcommand{\pps}{\texttt{pps}}
\newcommand{\Pps}{\texttt{Pps}}
\newcommand{\hop}{\texttt{hop}}
\newcommand{\probexecs}{\texttt{proba-execs}}
\newcommand{\probexec}{\texttt{proba-exec}}
\newcommand{\probvalue}{\texttt{proba-value}}
\newcommand{\angletype}{\texttt{angle}}
\newcommand{\bvconcat}{\texttt{concat}}
\newcommand{\concat}{\texttt{concat}}
\newcommand{\programtype}{\texttt{program}}
\newcommand{\gatetype}{\texttt{gate}}
\newcommand{\bitvector}{\texttt{bit\_vector}}
\newcommand{\bitvec}[1]{\ensuremath{\overrightarrow{#1}}}
\newcommand{\inttobv}{\texttt{int\_to\_bv}}
\newcommand{\mylist}{\ensuremath{\texttt{list}}\xspace}
\newcommand{\srak}{\ensuremath{\texttt{wrak}}\xspace}
\newcommand{\cond}[1]{\{{#1}\}}
\newcommand{\ahoare}[3]{\begin{array}{rcl}\cond{#1}&{#2}&\cond{#3}\end{array}}
\newcommand{\ahoarelr}[3]{\begin{array}{rcl}\left\{#1\right\}&#2&\left\{#3\right\}\end{array}}
\newcommand{\ahoarelrc}[4]{\begin{array}{rcl}
\multicolumn{1}{l}{#1}&&\\
\begin{array}{r}
    \left\{#2\right\}\end{array}&#3&\left\{#4\right\}\end{array}}

\newcommand{\ahoareBigr}[3]{\begin{array}{rcl}\cond{#1}&{#2}&\Big\{#3\Big\}\end{array}}
\newcommand{\ahoareBiggr}[3]{\begin{array}{rcl}\cond{#1}&{#2}&\Bigg\{#3\Bigg\}\end{array}}
\newcommand{\ahoareBig}[3]{\begin{array}{rcl}\Big\{#1 \Big\}&{#2}&\Big\{#3\Big\}\end{array}}
\newcommand{\ahoareBigg}[3]{\begin{array}{rcl} \Bigg\{#1 \Bigg\}&{#2}&\Bigg\{#3\Bigg\}\end{array}}
\newcommand{\hoare}[3]{\cond{#1}{#2}\cond{#3}}
\newcommand{\iterterm}[5]{\texttt{iter}\,{{#1}\mapsto{#2}}\,\texttt{from}\,{#3}\,\texttt{to}\,{#4}\,\texttt{with}\,{#5}}
\newcommand{\whilebterm}[6]{\texttt{cond\_iter}\,{{#1}\mapsto{#2}}\,\texttt{from}\,{#3}\,\texttt{to}\,{#4}\,\texttt{with}\,{#5},{#6}}

\newcommand{\denot}[1]{{[\!|{#1}|\!]}}
\newcommand{\interp}[2]{{\llparenthesis#1,#2\rrparenthesis}}

\newcommand{\lb}{\ensuremath{\mathit{lb}}}
\newcommand{\cip}{\ensuremath{\mathit{CIP}}}
\newcommand{\bcip}{\ensuremath{\mathit{BCIP}}}
\newcommand{\hgate}{\vspace{-.5cm}\begin{tikzpicture}\scalebox{.8}{\clip(-.25,-.25)rectangle(.25,.25);\node[draw,rectangle]at(0,0){H};}\end{tikzpicture}\vspace{.5cm}}
\newcommand{\xgate}{\vspace{-.5cm}\begin{tikzpicture}\scalebox{.8}{\clip(-.25,-.25)rectangle(.25,.25);\node[draw,rectangle]at(0,0){X};}\end{tikzpicture}\vspace{.5cm}}
\newcommand{\rzgate}[1]{\vspace{-.5cm}\begin{tikzpicture}\scalebox{.8}{\clip(-.28,-.27)rectangle(.28,.27);\node[draw,rectangle]at(0,0){$R_{#1}$};}\end{tikzpicture}\vspace{.5cm}}
\newcommand{\phgate}[1]{\vspace{-.5cm}\begin{tikzpicture}\scalebox{.8}{\clip(-.65,-.3)rectangle(.6,.32);\node[draw,rectangle]at(0,0){$\textit{Ph}_{#1}$};}\end{tikzpicture}\vspace{.5cm}}


\newcommand{\WHILE}{{\rm \textsc{While}}}
\newcommand{\SKIP}{\textbf{skip}}
\newcommand{\assnequal}{:=}
\newcommand{\starequal}{\; {\ast}{=}\; }
\newcommand{\ifStat}[1]{\textbf{if}\ #1\ \textbf{fi}}
\newcommand{\whileStat}[2]{\textbf{while}\ #1\ \textbf{do}\ #2\ \textbf{od}}

\newcommand{\lst}[1]{\bar{#1}}

\newcommand{\Boolean}{\textbf{Bool}}
\newcommand{\integer}{\textbf{Int}}
\newcommand{\tr}{\mathit{tr}}
\newcommand\Hh{H}
\newcommand{\deSem}[1]{\llbracket #1 \rrbracket}
\newcommand{\fwp}{\mathit{wp}}
\newcommand{\fwlp}{\mathit{wlp}}
\newcommand{\IF}{\mathbf{if}}
\newcommand{\bottom}{\mathbf{bottom}}
\newcommand{\Ee}{\mathbb{E}}

\newcommand{\outprod}[2]{\ket{#1}{}\bra{#2}{}}
\newcommand{\pcor}[3]{ \{ #1 \}\ #2\ \{ #3 \} }
\newcommand{\tcor}[3]{ \langle #1 \rangle #2 \langle #3 \rangle }

\newtheorem{definition}{Definition}[section]
\newtheorem{example}{Example}[section]

\newcommand{\stdinterp}[1]{\ensuremath{\left\llbracket #1 \right\rrbracket}}
\newcommand{\cat}[1]{\ensuremath{\mathbf{#1}}}
\newcommand{\ax}[1]{\texttt{#1}}

\usetikzlibrary{decorations.markings,arrows,decorations.pathmorphing}
\usetikzlibrary{shapes.misc,matrix,backgrounds,folding,positioning}
\usetikzlibrary{shapes.geometric}
\pgfdeclarelayer{edgelayer}
\pgfdeclarelayer{nodelayer}
\pgfsetlayers{background,edgelayer,nodelayer,main}
\tikzset{every path/.style={draw=black!80, line width=0.6pt}}
\tikzstyle{every picture}=[baseline=-0.25em]
\tikzstyle{none}=[inner sep=0mm]
\tikzstyle{zxnode}=[shape=circle, minimum width=.25cm, inner sep=0.5pt, font=\footnotesize, draw=black,thick]
\tikzstyle{gn}=[zxnode ,fill=green, draw=green!10!black]
\tikzstyle{rn}=[zxnode ,fill=red, draw=red!10!black]
\tikzstyle{H box}=[rectangle,fill=yellow, draw=yellow!10!black,thick,xscale=1,yscale=1,font=\footnotesize,inner sep=1.2pt,minimum width=0.15cm,minimum height=0.15cm]
\tikzstyle{ug}=[regular polygon, regular polygon sides=3, fill=red,draw=black,inner sep = 0pt,minimum width=0.8em]
\tikzstyle{black dot}=[inner sep=0.7mm,minimum width=0pt,minimum height=0pt,fill=black,draw=black,shape=circle]
\tikzstyle{dot}=[black dot]
\tikzstyle{white dot}=[dot,fill=white]
\tikzstyle{arrow}=[decoration={markings,mark=at position 1 with
    {\arrow[scale=1.2,>=stealth]{>}}},postaction={decorate}]
\newcommand{\edgetick}{{\arrow[black,scale=0.7,very thick]{|}}}
\tikzstyle{glabel}=[rounded corners=0.2em,fill=green!30,inner sep=0.1em,font=\scriptsize, anchor=west, xshift=-0.3em, yshift=0,opacity=1]
\tikzstyle{rlabel}=[rounded corners=0.2em,fill=red!30,inner sep=0.1em,font=\scriptsize, anchor=west, xshift=-0.3em, yshift=0,opacity=1]
\tikzstyle{box}=[rectangle, draw=black, fill=white, inner sep=2pt]
\tikzstyle{box-no-outline}=[rectangle, draw=white, fill=white, inner sep=2pt]
\tikzstyle{circle-no-outline}=[circle, draw=white, fill=white, inner sep=0pt]
\tikzstyle{squigglearrow}=[->, line join=round, decorate, decoration={zigzag, segment length=4, amplitude=0.8, post=lineto, post length=2pt}]
\tikzstyle{divide}=[regular polygon, regular polygon sides=3, draw=black, fill=gray!50, inner sep=1.6pt, rounded corners=0.8mm]
\tikzstyle{very thick}=[-, line width=1pt]
\tikzstyle{boxedge}=[draw=gray!50]
\tikzstyle{rule-box}=[dashed, orange]

\pgfdeclarelayer{edgelayer}
\pgfdeclarelayer{nodelayer}
\pgfsetlayers{background,edgelayer,nodelayer,main}
\tikzstyle{every loop}=[]

\usetikzlibrary{circuits.ee.IEC}
\newcommand{\ground}
{
\begin{tikzpicture}[circuit ee IEC,yscale=0.9,xscale=0.8]
\draw[solid,arrows=-] (0,1ex) to (0,0) node[anchor=center,ground,rotate=-90,xshift=.66ex] {};
\end{tikzpicture}} 

\newcommand{\sground}{\hspace*{-1pt}\scalebox{0.5}{\ground}}

\newcommand{\tikzfig}[1]{
\InputIfFileExists{./figures/#1.tikz}{}{{\color{red}\colorbox{pink}{missing file : #1}}}
}
\def\fig{}
\newcommand{\tikzfigc}[1]{\input{./figures/\fig/\fig_#1.tikz}}

\allowdisplaybreaks


\newcommand{\criscomment}[1]{\textcolor{blu}{
\newline
$\qquad$ \textbf{Christophe says:} #1} 
}
\newcommand{\bellzz}{\ensuremath{\ket{\beta_{00}}{}}}
\newcommand{\bellzo}{\ensuremath{\ket{\beta_{01}}{}}}
\newcommand{\belloz}{\ensuremath{\ket{\beta_{10}}{}}}
\newcommand{\belloo}{\ensuremath{\ket{\beta_{11}}{}}}
\newenvironment{psmallmatrix}
  {\left(\begin{smallmatrix}}
  {\end{smallmatrix}\right)}
\chapter[Formal methods for quantum algorithms]{Formal methods for quantum algorithms}

\section{Introduction}

 \label{sec:intro}
\paragraph*{Cryptography and quantum information.}
Quantum computing  dates back to 1982, when
Richard Feynman raised the idea~\cite{feynman:1982:spc} of simulating
quantum mechanics phenomena by storing information in particles
and controlling them according to the laws of  quantum mechanics.
In the brief history of quantum computing, the description  in 1994
by Peter Shor  of an algorithm~\cite{shor1994algorithms}, performing 
the decomposition of prime
integers in polynomial time on the size of the input, plays a major role. 
Indeed, it was the first-ever described quantum algorithm with
a practical utility--breaking the RSA public key
cryptosystems in a tractable manner.  

In an asymmetric  cryptosystem such as RSA, information is encrypted via a key that
is a solution for a given mathematical function--the decomposition
of a given integer into prime factors for the case of RSA. The
security of such a protocol is based on the fundamental assumption
that no potential eavesdropper has the means to compute this solution
efficiently.
Shor's algorithm is based on (1) a reduction of the prime factor
decomposition problem into the \emph{order-finding} problem and (2) an
adequate use of quantum parallelism to perform modular exponentiation
of integers over many different inputs in a single row, enabling a polynomial resolution of  the \emph{order-finding}. Thus, the
computation time for performing the prime decomposition is reduced
from exponential  to polynomial, and therefore breaks RSA's fundamental
assumption. 
Shor's original article ~\cite{shor1994algorithms} also presents a
variation of the order-finding resolution algorithm, solving the
discrete logarithm problem with similar performances. Doing so, it brings  a
procedure for breaking elliptic curve cryptosystems.

Symmetric-key cryptosystems are also challenged by quantum computing
\cite{santoli2016using}. As an example, Simon's quantum
algorithm~\cite{simon1997power} brings an exponential speedup for
computing the period of a function (given the promise that this period
indeed exists). Several applications in public-key cryptosystems were
described, providing exponential gain in, e.g., distinguishing a
three-round Feistel construction~\cite{kuwakado2010quantum,
  santoli2016using}, key recovering in the Evan Mansour encryption
scheme~\cite{kuwakado2012security} and attacking the CBC-MAC message
authentication scheme~\cite{santoli2016using}.

Finally, Grover search quantum algorithm~\cite{grover1996fast} brings
a quadratic speedup in the search for a distinguished element in
unstructured databases. Hence, while in this case the complexity gain is less
decisive  than for the procedures introduced above, its potential
for cryptography is significant as it weakens \textit{any}
symmetric-key encryption system.

Thus, quantum computing challenges current cryptographic uses and
practices.  Shor's algorithm opened a research program for
cryptographic solutions resisting the  power of quantum computation, called \emph{post-quantum
  cryptography}~\cite{chen2016report}.

\medskip Interestingly the induced challenge also
received answers from quantum information theory itself. Indeed, one of the
major distinctive features of quantum information is that
it cannot be read without being affected. This entails that an
eavesdropper trying to access a quantum information exchange cannot
help betraying her attempt.
Based on this feature, one can encode a cryptographic key in a quantum
message and, in case of eavesdropping, detect it \emph{a posteriori},
renounce this particular key and try another
sending. The study of \textit{Quantum key distribution protocols} is an active
research
area~\cite{scarani2009security,bennett2020quantum,liao2017satellite,%
lo2005decoy}.

\paragraph*{Quantum computing and quantum software.}
These cryptographic aspects are one of many applications studied in the
young research field of quantum computing. Others are, e.g., machine
learning~\cite{biamonte2017quantum,schuld2018supervised,lloyd2013quantum},
optimization~\cite{farhi2014quantum}, solving linear
systems~\cite{harrow2009quantum}, etc. In all these domains there are
quantum algorithms beating the best known classical algorithms by
quadratic or even exponential factors, complexity-wise.

These algorithms are based on laws and phenomena specific to quantum
mechanics (such as quantum superposition, entanglement, unitary
operations). Therefore, implementing them requires a framework consisting of both a
\textit{dedicated hardware} (quantum computers) and a
\textit{dedicated software} (quantum programming languages
and compilation toolchains).

In the last 20 years, several such languages have been proposed, such
as \qiskit~\cite{QiskitCommunity2017},
\liquid~\cite{wecker2014liqui}, Q\#~\cite{svore2018q},
\quipper~\cite{green2013quipper}, \projectq~\cite{projectq},
\textit{etc}.
Still, the field is in its infancy, and many questions still need to
be answered before we can reach the level of maturity observed for
classic programming languages. Standing questions include, for
example, introducing a foundational computing model and semantics for quantum
programming languages, adequate programming abstractions and type
systems, or the ability to interact with severely constrained hardware
in an efficient way (optimizing compilers).
  
\paragraph*{Verification of quantum programs.}
While testing and debugging are the common verification practice in
classical programming, they become extremely complicated in the
quantum case.  Indeed, debugging and assertion checking are \textit{a
  priori} very complicated due to the destructive aspects of quantum
measurement (see Section~\ref{sec:destructive-meas} below). Moreover,
the probabilistic nature of quantum algorithms seriously impedes
system-level quantum testing. Finally, classical emulation of quantum
algorithms is (strongly believed to be) intractable.

On the other hand, nothing prevents {\it a priori} the \textit{formal
  verification} \cite{clarkew96formal} of quantum programs,
i.e.~\textit{proving} by (more or less) automated reasoning methods
that a given quantum program behaves as expected for any input, or at
least that it is free from certain classes of bugs.

\smallskip 

\textit{Interestingly, while formal methods were first developed for
  the classical case where they are still used with parsimony--mainly
  for safety-critical domains--as testing remains the main
  validation methods, their application to quantum computing could
  become more mainstream, due to the inherent difficulties of testing
  quantum programs. }

\paragraph*{Goal of this survey} This survey introduces both the
requirements and challenges for formal methods in quantum programs specification and verification,
and the existing propositions to overcome these challenges.

The first sections give the general background.  In
Section~\ref{sec:gen-back-qc} we introduce the main concepts at stake
with quantum computing and quantum algorithms.  We provide a state
of the art introduction for formal methods, given in
Section~\ref{sec:gen-back-fm}. The specific requirements for formal
reasoning in the quantum case are then developed in
Section~\ref{sec:overview-fm}.
Then we come to concrete quantum programming and formal verification
material. In Section~\ref{sec:low-level-verif} we introduce several
existing solutions for the formal verification of quantum compilation
and the equivalence of quantum program runs. Generating such runs
requires specific programming languages. The formal interpretation of
quantum languages is introduced in Section~\ref{sec:formal_qpl}. Then
in Section~\ref{sec:high-and-lid-level-verif} we present the main
existing solutions for formally verified quantum programming languages.
In Section~\ref{sec:discussion} we introduce references for further usage
of formal methods linked with quantum information, and we conclude
this survey with a discussion in Section~\ref{sec:conclusion}.

\section{General Background in Quantum Computing}
\label{sec:gen-back-qc}

By many aspects, quantum computing constitutes a new paradigm.  Making
great use of quantum superposition and quantum entanglement, it
requires to define proper versions for such fundamental concepts as
data structures at stake in computation, or the elementary logical
operations at use. We introduce the well-known hybrid quantum computation model in
Section~\ref{sec:comp_mod}. 

Quantum computers are not intended to, and will
not, replace classical ones. One should better see the opening of a new
field, with possibilities to solve new
problems. Section~\ref{sec:algos} presents these new problems and
introduces quantum algorithms design.

As a new software technology, quantum computing comes with specific
challenges and difficulties. These specificities are closely related
to the particular needs for formal reasoning in quantum computing. They
are introduced in Section~\ref{sec:pro}.

\subsection{Hybrid Computational Model}
\label{sec:comp_mod}

Let us first introduce the main concepts at stake in quantum
programming. They concern the architecture of quantum computers, the
structure of quantum information and quantum programs, and their
formal interpretation.

\subsubsection{Hybrid Circuit Model}
\label{sec:hybrid-model}

\begin{figure}[tbh]
      \begin{center}
        \begin{tikzpicture}
        
\begin{scope}

  \node(cc)[draw,rectangle,minimum width = 1.6cm,minimum height =
        .9cm,ultra thick,rounded corners=3pt]at(0,0){{\tiny\begin{tabular}{l}Classical\\controller\end{tabular}}};
\draw[fill= black](-.6,-.7) to[bend right](cc.230) to (cc.310) to [bend right](.6,-.7); 
\draw[very thick,rounded corners=2pt](-.2,-.8)to(.7,-.8)to[very thick](.8,-1)to[very thick](-.8,-1)to(-.7,-.8)to(.2,-.8);
\draw[fill= black,rounded corners=2pt](.85,-.95) to (.8,-.8) to (1,-.8) to (1.1,-1) to (.9,-1) to (.85,-.95); 

\begin{scope}[xshift =-.2cm]
  \node(comp)[draw,fill=black,rectangle,minimum width = .6cm,minimum height = 1.15cm,
  ultra thick,rounded corners=3pt]at(-1.25,-.15){};
  \node(hole)[fill=white,rectangle,minimum width = .55cm,minimum height = .52cm,
  ultra thick,rounded corners=3pt]at(-1.25,.1){};
  
  \draw[thick](hole.0)to (hole.180);
  \draw[thick](hole.25)to (hole.155);
  \draw[thick](hole.-25)to (hole.205);
  \draw[thick](hole.-25)to (hole.205);
  \node[very thick,rectangle,draw, minimum width = .5cm, rounded corners=1pt]at(-1.25,-.7){};
  
\end{scope}
\begin{scope}[xshift =0cm]
  \node(s1) at(1.2,.5){};
  \node(s2) at(3,.5){};
  \node(s3) at(1.2,-1){};
  \node(s4) at(3,-1){};

\begin{scope}[yshift =2.7cm,xshift =3.5cm]
\node(qram)[draw,ultra thick,rounded corners=2pt,rectangle,minimum width=1cm,minimum height=1cm] at(0,-3){{\tiny\begin{tabular}{l}Quantum\\memory\end{tabular}}};
\draw[ultra thick](.4,-2.5)to  (.4,-2.3);
\draw[ultra thick](.2,-2.5)to  (.2,-2.3);
\draw[ultra thick](0,-2.5)to  (0,-2.3);
\draw[ultra thick](-.2,-2.5)to  (-.2,-2.3);
\draw[ultra thick](-.4,-2.5)to  (-.4,-2.3);
\draw[ultra thick](.4,-3.5)to  (.4,-3.7);
\draw[ultra thick](.2,-3.5)to  (.2,-3.7);
\draw[ultra thick](0,-3.5)to  (0,-3.7);
\draw[ultra thick](-.2,-3.5)to  (-.2,-3.7);
\draw[ultra thick](-.4,-3.5)to  (-.4,-3.7);
\end{scope}
\end{scope}
\end{scope}

\draw[<-,
          thick, bend right,color=blu] (s2) to
        node[above,color=blu]{Instructions}(s1);

        \draw[->, thick, bend left,color=re](s4) to
        node[below]{\begin{tabular}{l}Feedback\end{tabular}}(s3);

        \end{tikzpicture}
\vspace{-0.3cm}
\caption{Scheme of the hybrid model}
\label{qram}
    \end{center}
\end{figure}
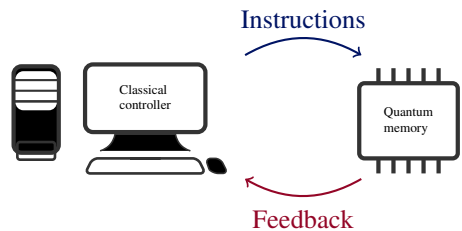

The vast majority of quantum algorithms are described within
the context of the
\emph{quantum co-processor model}~\cite{knill1996conventions}, i.e. an
hybrid model where a {\it classical} computer controls a {\it quantum}
co-processor holding a quantum memory, as shown in Figure~\ref{qram}. In particular, the classical computer performs control operations (\textbf{if \dots else} statements, loops, etc). The
co-processor can apply a fixed set of elementary operations
(buffered as {\it quantum circuits}) to update and query ({\it
  measure}) the quantum memory. Importantly, while { measurement}
allows retrieving classical (probabilistic) information from the
quantum memory, it also modifies it ({\it destructive effect}).

Major {\it quantum programming languages} such as \quipper
\cite{green2013quipper}, \liquid \cite{wecker2014liqui}, Q\#
\cite{svore2018q}, \projectq~\cite{projectq}, \silq \cite{silq}, and
the rich ecosystem of existing quantum programming frameworks
\cite{qecosystem} follow this hybrid model.

\subsubsection{Quantum Data Registers}
\label{sec:qd}

The following paragraphs introduce several definitions and notations
for quantum data registers. In particular, we follow the standard
\emph{Dirac notation}. For more details about this content, we refer
the reader to the standard literature~\cite{ DBLP:books/daglib/0046438}.

\paragraph*{Kets and basis-kets.}
While in classical computing the state of a bit is one between two
possible states ($0$ or $1$), in quantum computing the state of a
\emph{quantum bit} (or \emph{qubit}) is described by {\it amplitudes}
over the two elementary values $0$ and $1$ (denoted $\ket{0}{}$ and
$\ket{1}{}$), i.e.~linear combinations of vectors
\(\alpha_0 \ket{0}{} + \alpha_1\ket{1}{}\) where $\alpha_0$ and
$\alpha_1$ are any {\it complex values} satisfying
$\vert \alpha_0\vert^2 + \vert \alpha_1\vert^2 = 1$. In a sense,
amplitudes are generalization of probabilities.

More generally, quantum states are defined in complex finite-dimensional Hilbert spaces\footnote{In the finite-dimensional case, Hilbert spaces are vector spaces equipped with an \textit{inner} (scalar) product.}: the state of a {\it qubit register} of $n$ qubits
(called a \emph{ket} of length $n$--dimension $2^n$) is a column
vector with $2^n$ rows, formed as a {\it superposition} of the $2^n$
elementary basis vectors of length $n$ (the ``basis kets''), \ie a ket
is any linear combination of the form
\begin{equation}
  \ket{u}{n}= \sum_{k = 0}^{2^n-1}
  \alpha_k\ket{k}{n}\label{ket-eg}
\end{equation}
such that $\sum_{k = 0}^{2^n-1} \vert \alpha_k \vert^2 = 1 $.

\paragraph*{Bit-vectors and basis kets.}
Depending on the context, it may be more convenient to index the terms
in the sum above with bit vectors instead of integers.  We call
\emph{bit vector of length $n$} any sequence $x_0 x_1 \dots x_{n-1}$
of elements in $\{0,1\}$. Along this survey, we assume the implicit
casting of these values to/from booleans (with the least significant
bit on the right). For any positive $n$, we denote the set of bit
vectors of size $n$ by $\Bv_n$. We also surcharge notation
$\ket{j}{n}$ shown above with bit vector inputs. Formally, for any
bitvector $\vec{x}$ of length $n$,
$\ket{\vec{x}}{n}=\ket{\sum_{i=0}^{n-1} x_i* 2^{n-i-1} }{n}$.
Hence, one can write state $\ket{u}{n}$ from (\ref{ket-eg}) as
\[\ket{u}{n}= \sum_{\vec{x}\in\Bv_n} \alpha_{\sum_{i=0}^{n-1} x_i\cdot
    2^{n-i-1} )}\ket{\vec{x}}{n}\]
It may also be convenient to represent basis kets through their
index's binary writing.  For example, the two qubits kets basis is
equivalently given as $\{\ket02, \ket12, \ket22 , \ket32\}$ or as
$\{\ket{00}{}, \ket{01}{}, \ket{10}{}, \ket{11}{}\}$.

We omit the length index $n$ from notation $\ket{u}{n}$ when it is
either obvious from the context or irrelevant. We also adopt the
implicit convention of writing basis kets with either integer
indexes $k,i,j$ or bit-vector $\vec{x}$ and general kets with indexes
$u,v,w$. Hence, in the following $\ket{u}{},\ket{v}{},\ket{i}{}$ and
$\ket{\vec{x}}{}$ all designate kets, the last two having the
additional characteristics of being basis kets.

When considering two registers of respective size $m$ and $n$, the
state of the compound system lives in the \emph{Kronecker product}\footnote{Given two matrices $A$ (with $r$ rows and $c$ columns) and
  $B$, their Kronecker product is the matrix
  $A\otimes B = \begin{pmatrix} a_{11}B & \dots& a_{c} B\\
    \rotatebox{90}{\dots}& \rotatebox{135}{\dots}&
    \rotatebox{90}{\dots} \\ a_{r1}B & \dots &a_{rc} B\
  \end{pmatrix}$. This operation is central in quantum information
  representation. It enjoys a number of useful algebraic properties
  such as associativity, bilinearity or the equality
  $(A\otimes B)\cdot(C\otimes D) = (A\cdot C)\otimes (B\cdot D)$,
  where $\cdot$ denotes matrix multiplication.}
--or \emph{tensor product}--of the original state spaces: a general
state is then of the form
\[
  \sum_{\vec{x}\in\Bv_m,\vec{y}\in\Bv{n}}
  \alpha_{\vec{x},\vec{y}}\ket{\vec{x}}{m}\otimes\ket{\vec{y}}{n}.
\]
In particular, the state of a qubit register of $n$ qubits lives in
the tensor product of $n$ state-spaces of one single
qubit.

\paragraph*{Adjointness.}
In the following we also use the adjoint transformation for
matrices. The adjoint of matrix $M$ with $r$ rows and $c$ columns is
the matrix $M^\dagger$, with $c$ rows and $r$ columns and such that
for any indexes $j,k \in \tofset{c}\times \tofset{r}$\footnote{where, for any two integers $i,j$ with $i<j, \llbracket i,j\llbracket$ denote the induced interval, that is the set of integers k such that $i\leq k<j$.}, cell
$M^\dagger(j,k)$ holds the conjugate value $M(k,j)^*$ of $M(k,j)$ (for
any complex number $c$, its conjugate $c^*$ is the complex number with
the same real part and the opposite imaginary part as $c$). The
adjoint of a ket $\ket{u}{n}$ is called a \emph{bra}. It is a row
vector with $2^n$ columns denoted $\bra{u}{n}$--or simply
$\bra{u}{}$--and with indices the conjugates of those of
$\ket{u}{n}$.  This bra-ket notation is particularly convenient for
representing operations over  vectors.  Given a ket $\ket{u}{}$
and a bra $\bra{v}{}$, $\ketbra{u}{v}$ denotes their Kronecker product
--or \emph{outer
  product}. Furthermore, if $\ket{u}{}$ and $\bra{v}{}$ have the
same length, then $\braket{v}{u}$ denotes their scalar product--also
called \emph{inner product}. In particular, in the case of basis
states $\ket{i}{}$ and $\bra{j}{}$, $\braket{i}{j} = 1$ if $i=j$ and
$0$ otherwise and $\ketbra{i}{j}$ is the square matrix of width $2^n$
with null coefficient everywhere except for cell $(i,j)$ with
coefficient $1$. If $i = j,$ then $\ketbra{i}{j}$ operates as the
projector upon $\ket{i}{}$.

\paragraph*{Quantum measurement and Born rule.}
The probabilistic law for measurement of kets  is given by the
so-called \emph{Born rule}:
for any $k \in \llbracket 0 , 2^n \llbracket$, measuring
state $\ket{u}{n}$ from Equation~\ref{ket-eg} results in $k$ with
probability $\vert \alpha_k\vert ^2$. The measurement is destructive:
if the result were $k$, the state of the register is now $\ket{k}{n}$
(with amplitude $1$).

\subsubsection{Separable and Entangled States}
\label{quantum_state}

From Section~\ref{sec:qd}, a quantum state vector of length $n$ is a
superposition of basis elements with coefficients whose squared moduli
sum to one. Then, tensoring $n$ quantum states $\ket{u_j}{1}$ of
length $1$ results in a state
$\ket{u}{n} = \bigotimes_{j\in\tofset{n}}\ket{u_j}{1}$ of length
$n$. One can decompose back $\ket{u}{n}$ into the family
$\{\ket{u_j}{}\}_{j\in\tofset{n}}$: we say that $\ket{u}{n}$ is a
\emph{separable} state. Note that the structure of quantum information
introduced above contains states missing the property of being
separable. As an example, the state
\[
  \bellzz =\frac{1}{\sqrt{2}} (\ket{00}{} + \ket{11}{})
\]
cannot be written as a tensor product of two single-qubit states.
This phenomenon is called \emph{entanglement}, and $\bellzz$ is an
\emph{entangled state}.  It induces that one can store more quantum
information in $n$ qubits altogether than separately.

\begin{example}[Bell states]
  \label{ex:bell-states}
  State \bellzz is a construction of particular interest in quantum
  mechanics and quantum computing. In their famous 1935
  article~\cite{einstein1935can}, Einstein, Podolsky and Rosen argued
  for the incompleteness of quantum mechanics, based on considerations
  upon \bellzz. In 1964~\cite{bell1964einstein}, J.S. Bell proposed an
  experiment to test the argument. It was based on statistics over
  experiments on the four following states:
  \[\begin{array}{rclrcl}
      \bellzz & =  \frac{1}{\sqrt{2}} (\ket{00}{2} + \ket{11}{2})
      & \qquad
        \bellzo & =  \frac{1}{\sqrt{2}} (\ket{01}{2} + \ket{10}{2}) \\
      \belloz & =  \frac{1}{\sqrt{2}} (\ket{00}{2} - \ket{11}{2})
      & \qquad
        \belloo & =  \frac{1}{\sqrt{2}} (\ket{01}{2} - \ket{10}{2}) \\
    \end{array}
  \]
  These four states are now known as the \emph{Bell states} (notation
  $\beta$ stands for the initial $B$) and are used in many quantum
  protocols, such as teleportation or superdense coding~(see
  Section~\ref{sec:zx}). We use them and their generation as a running
  example in the rest of this survey.
\end{example}

\subsubsection{Quantum Circuits}
\label{quantum circuits}

\begin{figure}[tbh]
  \centering
  \scalebox{1}{
    \begin{tikzpicture}
      \foreach \i in {1,...,2}{
        \draw[very thick] (0,\i+.2)--(0,\i-.2);
        \draw (0,\i)--(3.5,\i);
        \node[draw,rectangle,fill=white]at (3.5,\i){\Large $\meassymb$};
    \node at (0,3.3 -\i){$i_{\i}$};
      }
      \node at (.1,.4)[rotate =90]{$\dots$};
      \node at (3.2,.4)[rotate =90]{$\dots$};
        \draw (0,-.2)--(4,-.2);
        \draw (0,-.4)--(4,-.4);
        \draw (0,0)--(4,0);
      \node[draw,rectangle,dashed, minimum width =2.3cm, minimum height =3.7cm] at(1.5,.8){};
      \node[draw,rectangle,dotted, minimum width =.8cm, minimum height =3.2cm] at(1,.8){};
      \node[draw,rectangle,dotted, minimum width =.8cm, minimum height =3.2cm] at(2,.8){};
      \node at (1,-.6){\small a.};
      \node at (2,-.6){\small b.};
      \node[draw,rectangle,fill=white] at (1,2){\Large H};
      \node(oplus) at (2,1){\huge $\oplus$};
      \node(bullet) at (2,2){\Large $\bullet$};
      \draw(oplus.center)to(bullet.center);
    \end{tikzpicture}
  }
  \caption{Generalized circuit to create and measure Bell states}
  \label{fig:circuit_bell_states}
\end{figure}
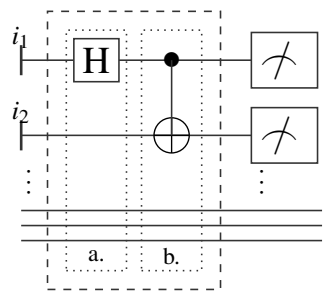

Three kinds of operations may be applied to quantum memory,
exemplified in Figure~\ref{fig:circuit_bell_states} with the circuit
generating and measuring Bell states:
\begin{itemize}
\item the \emph{initialization} phase allocates and initializes
  quantum registers (arrays of qubits) from classical data. In
  Figure~\ref{fig:circuit_bell_states} it is represented on the two first qubit
  wires  as
  \begin{tikzpicture}\draw[very thick](-.2,.15)to(-.2,-.15);\draw
    (-.2,0) to (0,0); \end{tikzpicture}, indexed by value $i_w$.
It \emph{creates} a quantum register in one of the basis
  states (that is, in the case of a two-qubit register, in one of the four basis
  states $\ket{00}{},\ket{01}{},\ket{10}{},\ket{11}{}$),

\item the actual quantum computing part consists in transforming an
  initialized state. This is performed by applying a sequence of
  proper quantum operations, structured in a so-called \emph{quantum
    circuit}. In Figure~\ref{fig:circuit_bell_states} this part is
  identified with a dashed box (itself sequenced with dotted boxes
  a. and b.),
  
\item the extraction of useful information from a quantum computation
  is performed through the \emph{measurement} operation, by which one
  probabilistically gets classical data from the quantum memory
  register. Measurement is represented, on each qubit it is applied to,
  as
  \scalebox{.45}{%
    \begin{tikzpicture}
      \node[draw] at (0,0){\Huge $\meassymb$} ;
    \end{tikzpicture}%
  }.
  
  \item In generalized circuits, not all wires in a register need to be initialized and measured. Hence in Figure~\ref{fig:circuit_bell_states} the transformations are performed over the  two first wires of a wider register, and the additional wires are left untouched.
\end{itemize}

\paragraph*{}
\textit{Note that we reserve the term \emph{circuit} for the pure
  quantum part (the dashed box in
  Figure~\ref{fig:circuit_bell_states}). We call \emph{generalized
    circuit} a process made of a circuit together with, possibly,
  initialization and measurements.}

\paragraph*{}
Quantum circuits are built by \textit{combining}, either
in \textit{sequence} or in
\textit{parallel}, a given set of elementary operations
called \emph{quantum
  gates}.
In addition to sequence and parallelism, derived circuit combinators
(controls, reversion, ancillas, etc) are often used in quantum circuit
design (See Figure~\ref{fig:hl-circ-comb} in
Section~\ref{sec:struct-quantum-algo} for details).
The circuit part of Figure~\ref{fig:circuit_bell_states}
uses two
different quantum gates, drawn in dotted boxes:

\begin{itemize}
\item the Hadamard gate \Had\ (a.), which induces a state
  superposition on a given qubit,
\item the \emph{control not} gate, often written  $\Cnot$ (b.) and
  represented as
  \scalebox{.7}{
  \begin{tikzpicture}
    \node(bullet) at (.5,.3){$\bullet$};
    \node(oplus) at (.5,0){\large $\oplus$};
    \draw(bullet.center)to(oplus.center);
  \end{tikzpicture}}.
  It is a binary gate, flipping the \emph{target qubit} (in wire 2 in
  our case) when the \emph{control qubit} (wire 1) has value 1.
\end{itemize}

\subsubsection{Quantum Matrix Semantics and density operators}
\label{sec:quant-circ-semant}

The transformation operated by a quantum circuit on a quantum register is
commonly interpreted as a matrix. In this setting, the parallel
combination of circuits is interpreted by the \textit{Kronecker product} and
the sequential combination by the \textit{matrix multiplication}.

\paragraph*{Quantum circuits.}
Quantum circuits happen to operate as \emph{unitary} operators
(preserving the inner product between vectors).  A set of elementary
gates is (pseudo-) \emph{universal} if, by a combination of parallel and
sequential composition, one can synthesize (or approximate) all
\emph{unitary} operations.  Examples for elementary gates are given in
Table~\ref{elemgate}, with their matrix semantics
interpretation. Apart from the already encountered gates \Had\ and
\Cnot, it features two additional families of gates, $\Phase(\theta)$
and $\Rz(\theta)$, where $\theta$ is an angle inducing a so-called \emph{phase factor} $e^{i\theta}$.  $\Phase(\theta)$
operates a simple scalar multiplication by a phase factor, while
$\Rz(\theta)$ operates as a rotation. Table~\ref{elemgate} is given
with indexes ranging over any angle $\theta$, making the set of gates
universal. Usually, we restrict it to angles of measure
$\frac{\pi}{2^n}$, with $n$ ranging over integers. This restriction
makes the resulting set of gates pseudo-universal.

\begin{table}[tbh]
\caption{Elementary gates and their matrix semantics}
\label{elemgate}
\begin{minipage}{1.0\linewidth}
\centering\scalebox{1}{\begin{tabular}{|c|c|c|c|}
\hline  \Had & \Cnot&$\Phase(\theta)$ & $\Rz(\theta)$ \\
  \hline
 $\frac{1}{\sqrt{2}} \begin{pmatrix}1 &1 \\ 1 &-1\end{pmatrix} $
&  $\begin{pmatrix} 1&0&0&0 \\ 0&1&0&0\\ 0&0&0&1\\ 0&0&1&0  \end{pmatrix}$&
$\begin{pmatrix} e^{i\theta} &  0 \\ 0 & e^{i\theta}  \end{pmatrix}$ &
$\begin{pmatrix} e^{-i\theta} & 0 \\ 0 &
  e^{i\theta} \end{pmatrix}$
\\\hline

\end{tabular}}
\end{minipage}

\end{table}

\begin{example}[Semantics for the Bell generating circuit]
  Let us look at Figure~\ref{fig:circuit_bell_states} again. First, an
  Hadamard gate is applied to the first wire and nothing happens to
  the second wire (it stays untouched, which is represented by the
  identity matrix). The matrix for the first \emph{column}--the
  dotted box indexed with a.--of Figure~\ref{fig:circuit_bell_states}
  is
  \[
    \frac{1}{\sqrt{2}}
    \begin{pmatrix}1 &1 \\ 1 &-1\end{pmatrix}
    \otimes
    \begin{pmatrix}1 &0 \\ 0 &1\end{pmatrix}
    = 
    \frac{1}{\sqrt{2}}
    \begin{psmallmatrix}
      1&0&1&0 \\
      0 &1&0&1 \\
      1 &0&-1&0 \\
      0 &1&0&-1 
    \end{psmallmatrix}
  \]
  Then gate \Cnot\ is applied, with matrix
  $\begin{psmallmatrix} 1&0&0&0 \\ 0&1&0&0\\ 0&0&0&1\\
    0&0&1&0 \end{psmallmatrix}$ and the sequential combination of the
  two subcircuits translates, in the matrix semantics, as their usual
  product (mind the reverse ordering, \wrt the figure):
  \[
    \Mat{\textit{Bell-circuit}} =
    \begin{psmallmatrix}
      1&0&0&0 \\ 0&1&0&0\\ 0&0&0&1\\ 0&0&1&0  \end{psmallmatrix}
    \cdot \frac{1}{\sqrt{2}}
    \begin{psmallmatrix}
      1&0&1&0 \\
      0 &1&0&1 \\
      1 &0&-1&0 \\
      0 &1&0&-1 
    \end{psmallmatrix} = 
    \frac{1}{\sqrt{2}}
    \begin{psmallmatrix}
      1&0&1&0 \\
      0 &1&0&1 \\
      0 &1&0&-1 \\
      1 &0&-1&0 
    \end{psmallmatrix}
  \]
\end{example}

\paragraph*{Application to input initialized kets.}
In the matrix formalism, we interpret $\ket{0}{}$ as the column
vector $\begin{psmallmatrix} 1\\0
\end{psmallmatrix}$, $\ket{1}{}$ as$\begin{psmallmatrix} 0\\1
\end{psmallmatrix}$ and the concatenation $\ket{i j}{}$, where
$i$ and $j$ both are sequences of $0$ or $1$, as the Kronecker product
$\ket{i}{}\otimes\ket{j}{}$. For example, the two qubits basis kets
$\ket{00}{},\ket{01}{},\ket{10}{},\ket{11}{}$ are represented,
respectively, as
\[
  \begin{psmallmatrix} 1\\0\\0\\0
  \end{psmallmatrix},
  \begin{psmallmatrix}
    0\\1\\0\\0
  \end{psmallmatrix},
  \begin{psmallmatrix}
    0\\0\\1\\0
  \end{psmallmatrix} \text{ and }
  \begin{psmallmatrix}
    0\\0\\0\\1
  \end{psmallmatrix}.
\]
The transformation performed by a quantum circuit $C$ upon a quantum
state $\ket{\psi}{}$ is interpreted as the matrix product
$\Mat{C}\cdot\ket{\psi}{}$ of the matrix for this circuit by the
column vector for this quantum state. By notation abuse, we also
simply write it $C\ket{\psi}{}$.

\begin{example}
  One can now directly verify that the Bell generating circuit from
  Figure~\ref{fig:circuit_bell_states} generates the Bell states from
  Example~\ref{ex:bell-states}: for any $a,b\in \{0,1\}$,
\[\textit{Bell-circuit} \cdot \ket{ab}{} = \ket{\beta_{ab}}{}\]
\end{example}

\paragraph*{Measurement.}
Last, measurement is performed over an orthonormal basis of the Hilbert space. For sake of simplicity, we only consider the case of measurements in the computational basis. Hence, measuring a quantum register results in a basis state, with
probabilities following the Born rule introduced in
Section~\ref{sec:qd}: measuring any state $\ket{u}{n}$ results in
basis state $\ket{k}{n}$ with probability (written
$\probmeas(\ket{u}{n},\ket{k}{n})$) $\mid \alpha_k\mid^2$, where
$\alpha_k$ is the amplitude of $\ket{k}{n}$ in $\ket{u}{n}$. Applying
this rule to the Bell state, one easily state that for any
$a,b,i,j\in \{0,1\},$
\[
  \probmeas(\ket{\beta_{ab}}{},\ket{ij}{}) = \frac{1}{2} (\texttt{if }
  b = 0 \texttt{ then }i\oplus j \texttt{ else } 1-(i\oplus j))
\]
where $\oplus$ denotes addition modulo $2$. Note, from   Example~\ref{ex:bell-states}, that index $a$ in notation $\ket{\beta_{ab}}{}$ only accounts for a $-1$ factor in the second term of the state superposition. Hence, since measurement is ruled by the Born rule and since this rule ignores negation (see Section~\ref{sec:qd}), then index $a$ does not appear in the expression of $\probmeas(\ket{\beta_{ab}}{},\ket{ij}{})$.

\paragraph*{Discussion over the matrix semantics.}
Matrix semantics is the usual standard formalism for quantum computing
(see~\cite{ DBLP:books/daglib/0046438} for example). Still, the size of
matrices grows exponentially with the width (number of qubits) of circuits, so it is
often cumbersome when addressing circuits from non-trivial algorithm
instances. Furthermore, algorithms usually manipulate parametrized families
of circuits. The resulting parametrized families of matrices may not
be conveniently writable.

Hence, a more compact interpretation for quantum circuits may be
helpful.  In particular, path-sum
semantics~\cite{amy2018towards,dblp:phd/basesearch/amy19} directly
interprets quantum circuits by the input/output function they induce
over kets--corresponding, in matrix terms, for any circuit $C$ of
width $n$, to the function $\ket{u}{n}\mapsto
\Mat{C}\cdot\ket{u}{n}$. To do so, it exhibits a generic form for
quantum registers description, which is generated by a restricted
number of parameters and composes nicely with sequence and parallel compositions.  Path-sum semantics plays a growing role in formal
specification and verification. It is introduced with further details
in Section~\ref{sec:path-sum-circuit}.

\paragraph*{Density operators.}

In the preceding paragraph, we introduced measurement as a non-deter\-ministic operation over quantum states.
Another strategy consists in dealing with a notion of states featuring this non-determinism. A \emph{mixed state} (as opposed to a~\emph{pure state}) is a probability distribution over several states. Alternatively, it can be seen as an incomplete description of a state, featuring the incomplete knowledge one may have about it. Then, measurement can be characterized as a simple transition between mixed states.

In quantum processes, this view is formalized by  density operators, that extends matrices formalism with the characterization of probabilistic states.
For sake of brevity, in this paragraph we give only a short introduction to the density operator formalism. Our aim here is only to provide the  required definitions and notations for  this review. For further detail about density operators and for the related soundness proofs, we refer the interested reader either to~\cite{selinger2004} or~\cite{nielsen2002quantum} (Section 2.4).

Basically,  the density operator for a pure state  $\ket{x}{}$ is the reflexive outer product $\ketbra{x}{x}$. 
 Given a set $S$ of indices and a distribution of states $\{\ket{x_k}{}\}_{k\in S}$, each occurring with probability $p_k$, we represent the overall mixed state as the density operator 
$$\rho :=\sum_{k\in S}p_k\ketbra{x_k}{x_k}$$

By linearity, the result of applying a unitary $U$ to $\ketbra{x}{x}$ is given by the product $U\ketbra{x}{x}U^{\dagger}$.
A measurement of a quantum register $q$ of size $n$ may be described by the collection of possible projectors it realizes, that is the set $M=\{M_k:=\ket{k}{n}\bra{k}{n}\}_{k\in\tofset{2^n-1}}$. In the density operators formalism, the action of $M$ over a state $\rho$ may result in any
state $M_k \rho M_k^{\dagger}$, with  probability $\tr(M_k M_k^{\dagger}\rho )$, where the trace $\tr(M)$ of a square
  matrix $M$ with $n$ rows and columns is defined as the sum
  $\sum_{j\in \tofset{n}} M(j,j)$ of its diagonal cell values.
Then, the overall action of a measurement $M$ over a density operator $\rho$ 
results in $\rho' = \sum_{k\in\tofset{2^n-1}} M_k\rho M_k^\dagger$.

Now, measurement description generalizes to the case of \emph{partial measurements},  where only a sub-register is measured. Let us consider the case of a quantum register $q$ of size $n=n_1+n_2$. We write $q_1$ and $q_2$ for the concatenated sub-registers and $H, H_1, H_2$ for the respectively induced  Hilbert spaces. To simplify the notations we consider the case of measuring the first  $n$ qubits. For any density operator $\rho$, if it is separable as $\rho = \ket{x_1}{n_1}\bra{y_1}{n_1}\otimes\ket{x_2}{n_2}\bra{y_2}{n_2}$, then the partial trace of $\rho$ over $H_2$  is defined as $\tr_2(\rho) = \braket{y_1}{x_1}\ketbra{x_2}{y_2}$ and the definition generalizes by linearity to any density operator $\rho$. Then,  $\tr_2(\rho)$ equivalently represents the result of:
\begin{itemize}
    \item (1) measuring register $q_2$ from the mixed state $\rho$ and (2) forgetting the measured qubits while conserving memory of the unmeasured subregister  $q_1$ state; 
    \item or just forgetting  about (\emph{discarding}) register $q_2$ in the description of $\rho$. Then,  $\tr_2(\rho)$ is the description of the sub-system held by $q_1$. We call  $\tr_2(\rho)$  a  \emph{partial density operator} on $H_1$.
    \end{itemize}

\subsubsection{Other Models for Quantum Computations}

Many alternatives are currently explored for physical implementations
of quantum computing machines and worth mentioning. Some of them (such as Measurement
Based Quantum Computing~\cite{raussendorf2003measurement,Briegel2009}, topological
quantum computations~\cite{freedman2003topological}, linear optical
networks~\cite{aaronson2011computational}, adiabatic quantum
computing~\cite{farhi2001quantum}, \textit{etc}.) differing on rather
fundamental aspects (like, e.g., the elementary
operations constituting computations). Nevertheless, currently,  formal methods developments mainly address the standard circuit model introduced above. 

The ZX-Calculus~\cite{coecke2017picturing} also provides an
alternative graphical formalism to reason about quantum processes.
Basically, in this setting, quantum operations are represented by
diagrams and their composition through sequence or parallelism
corresponds to graphical compositions in the calculus. This language
comes with a series of enabled transformations over graphs, preserving
computational equivalence. ZX-Calculus is presented
in Section~\ref{sec:zx}.

\subsection{Algorithms}
\label{sec:algos}

As previously introduced, quantum computers are meant to perform calculations
that classical computers are  \emph{a priori} not able to perform in a
\emph{reasonable} time.  We give the formal complexity theory
characterization for this point in
Section~\ref{sec:quantum_algo_complexity}, then
Section~\ref{sec:quantum-algorithm-design} discusses the usual
conventions for quantum algorithms descriptions.

\subsubsection{Quantum Algorithms and Complexity}
\label{sec:quantum_algo_complexity}

\begin{figure}[tbh]
  \centering
  \begin{tikzpicture}
    \node(bqp)[draw, very thick,ellipse, minimum width = 10cm, minimum height = 2.5cm,fill= gray!75] at (2,1.05){};
    \node(np)[draw,very thick, rounded corners=10pt,fill= white, minimum width = 8cm, minimum height = 3.5cm] at (2,1.25){};
    \node(bqp1)[draw, very thick,ellipse, minimum width = 10cm, minimum height = 2.5cm,fill= gray!35,opacity=.8] at (2,1.05){};
    
    \node(p)[draw,rectangle,very thick, rounded corners=10pt,fill=white, minimum width = 5cm, minimum height = .9cm] at (2,.5){};
    \node at (2,2){\textbf{BQP}};
    \node at (2,.5){\textbf{P}};
    \node at (2,2.8){\textbf{NP}};
  \end{tikzpicture}
  \caption{Complexity classes P, NP, BQP}
  \label{fig:BQP}
\end{figure}
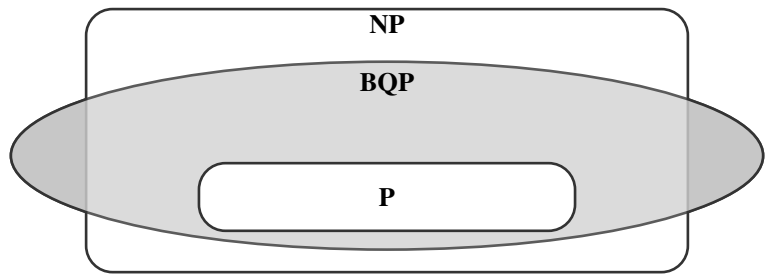

It is commonly assumed that formal problems are \emph{tractable} by a
computer if there exists an algorithm to solve this problem in time
(measured by the number of elementary operations it requires) that is
bounded by a polynomial over the size of the input parameters.
Formal problems satisfying this
criterion for classical computers form a \emph{complexity class}
usually referred to as \textbf{P}. It is schematically
represented in Figure~\ref{fig:BQP}. 

As introduced in Section~\ref{sec:qd}, extracting useful information from a quantum register requires a measurement, ruled by the Born law. Therefore, a quantum computation is an alternation of non-deterministic (measurement) and deterministic (circuit unitary application, classical post-treatment, etc) operations.
Since such computations are probabilistic, the tractability
criterion from above needs to be slightly adapted. Instead of
considering problems for which a polynomial algorithm brings a
solution with certainty, we consider those for which a polynomial
algorithm brings a solution with an error probability of at most
$\frac{1}{3}$.  The
corresponding class of problems for quantum computers is called
\emph{bounded error quantum polynomial time} (\textbf{BQP}).

In addition to \textbf{P} and \textbf{BQP}, Figure~\ref{fig:BQP}
represents the \emph{non-deterministic polynomial time} class
\textbf{NP}. It gathers formal problems $\mathcal{P}$ for which there
is an algorithm that, given a candidate solution, checks whether this
candidate is an actual solution for $\mathcal{P}$ in polynomial
time. It is trivial that $\textbf{P}$ is included in $\textbf{NP}$ and
it is also proved that $\textbf{P}\subseteq \textbf{BQP}$. There are
good reasons to believe that these inclusions are
strict. Nevertheless, strictness is not formally proved and there are a
variety of problems that belong to $\textbf{NP}$ without a known
tractable resolution algorithm.

Hence, quantum algorithm performance is not to be evaluated against
the  best \emph{possible} performance of any classical
computation (which depends on whether
$\textbf{BQP} = \textbf{P}$)  but, more pragmatically,
against the best classical \emph{known} equivalent.
      
Then, quantum computing is relevant for problems that are polynomially
solvable by a quantum computer (with a given probability of success)
but \textit{intractable} by a classical one. They appear in the gray zone in Figure~\ref{fig:BQP}. The
question whether the dark gray part (\textbf{BQP}$\backslash$
\textbf{NP}) is empty or not depends on whether
$\textbf{BQP} \subseteq \textbf{NP}$, which is unknown, but several
\textbf{BQP}-complete problems have been described through the
literature ~\cite{DBLP:reference/nc/Zhang12b}.  These are neither
easily computable nor verifiable with known classical means, but may
be computed with quantum means: since quantum computers output a right
solution for them with probability $>\frac{2}{3}$, after several runs one can select the best
represented output as the sought solution.

Let us stress that the correspondence between polynomial solvability
and tractability is not strict. For a problem, not belonging to a
polynomial class bounds the size of input parameters concretely
tractable by a computer, but does not absolutely forbid computation
for any instance of it. Hence, for some problems, the quantum
advantage does not consist in providing a polynomial resolution, but
in reducing computation time to extend the set of tractable
inputs. A typical example is the Grover search
algorithm~\cite{grover1996fast}, searching for a distinguished element  in
 an unstructured data set, providing a quadratic acceleration against
classical  procedures.

\subsubsection{Quantum Algorithm Design}
\label{sec:quantum-algorithm-design}

\begin{figure}[tbh]
  \begin{center}
    \framebox{\begin{tikzpicture}
        \node{
          \scalebox{0.9}{
            \begin{minipage}{\textwidth}
              \begin{tabular}{l}
                {\normalsize         \textbf{Inputs:} (1) A black-box $U_{x,N}$ which performs the transformation}\\ {\normalsize
                $\ket{j}{}\ket{k}{}\rightarrow\ket{j}{}\ket{x^j k\ \text{mod } N}{}$, for $x$ co-prime to the $L-$bit number $N$,}\\ {\normalsize (2)
                $t=2L+1 + \big\lceil \text{log}(2 + \frac{1}{2\epsilon})\big\rceil$ qubits initialized to $\ket{0}{}$, and}\\ {\normalsize(3) $L$ qubits initialized to the state $\ket{1}{}$.}
                \\[1ex]
                {\normalsize\textbf{Outputs:} The least integer $r>0$ such that $x^r = 1$ (mod $N$).}
                \\[1ex]
                {\normalsize\textbf{Runtime:} $O(L^3)$ operations. Succeeds with probability $O(1)$.}
                \\[1ex]
                {\normalsize\textbf{Procedure:}}
              \end{tabular}
              \begin{alignat*}{100}
                &1. &\quad& \ket{0}{}\ket{u}{}
                && \text{\hspace{2ex}initial state}
                \\
                &2.&& \to{} \frac1{\sqrt{2^t}}\sum_{j=0}^{2^t-1}\ket{j}{}\ket{1}{}
                && \text{\hspace{2ex}create superposition}
                \\
                &3. && \to{}
                \frac1{\sqrt{2^t}}\sum_{j=0}^{2^t-1}\ket{j}{}\ket{x^j \text{mod } N}{}
                && \text{\hspace{2ex}apply $U_{x, N}$}
                \\
                & && \approx{}
                \frac1{\sqrt{r2^t}}\sum_{s=0}^{r-1}\sum_{j=0}^{2^t-1}e^{2\pi i s j
                  /r}\ket{j}{}\ket{u_s}{}
                \\
                &4. && \to \frac{1}{\sqrt{r}}\sum_{s=0}^{r-1}\widetilde{\ket{s/r}{}}\ket{u_s}{}
                &&
                \begin{array}{@{}l}\text{\hspace{2ex}apply inverse Fourier transform 
                to the first register}\end{array}
                \\
                &5. && \to{} \widetilde{\ket{s/r}{}}&& \text{\hspace{2ex}measure first register}\\
                &6. &&\to{}  r&&\hspace{-0ex}
                \begin{array}{@{}l}\text{\hspace{2ex}apply continued fraction 
                algorithm}\end{array}
              \end{alignat*}
          \end{minipage}
          }
        };
        \node[draw, rectangle, color=blu,thick]at
        (-3.7,.95){\phantom{\tiny le petit chat est vivant le petit chat est vivant}};
        \node[draw, rectangle, color=re,thick,minimum height=1cm]at
        (-3.7,0.15){\phantom{\tiny le petit chat est vivant le petit chat est vivant}};
    \end{tikzpicture}}
  \end{center}
  \caption{Bird-eye view of the circuit for Shor's factoring algorithm~\cite{shor1994algorithms} 
    (as presented in \cite[p.\,232]{ DBLP:books/daglib/0046438})}
  \label{ncpe}
\end{figure}
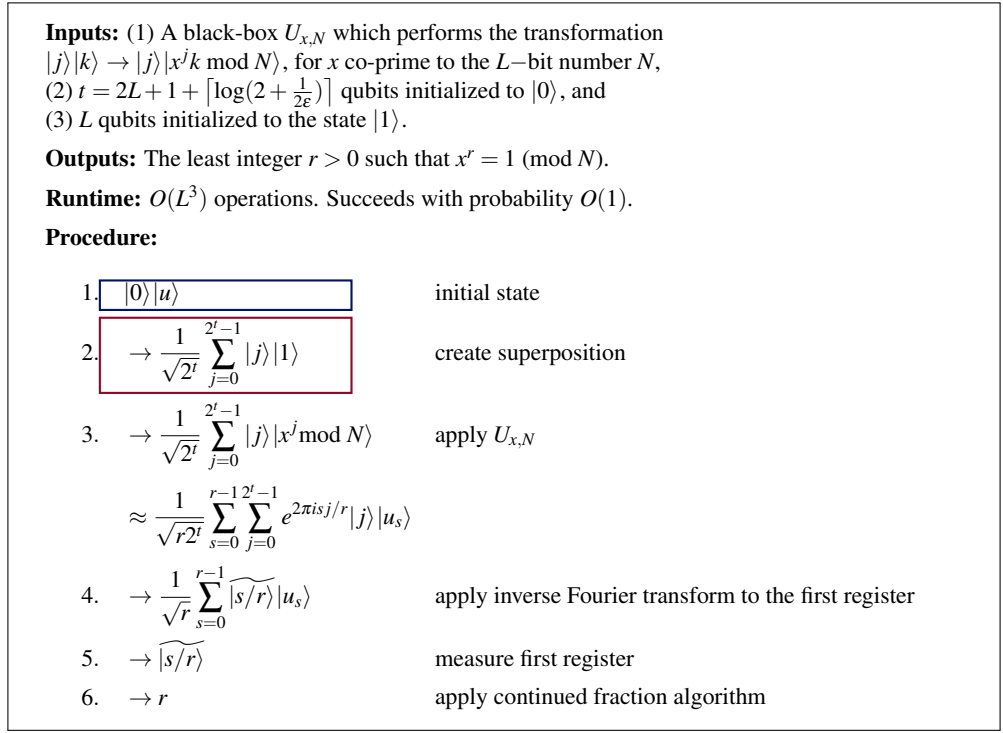

Before introducing the challenges at stake with implementations of
quantum algorithms and their formal solutions, we make a few
observations on the usual format used to describe quantum algorithms,
based on an example: Figure~\ref{ncpe} reproduces the core quantum
part for Shor's algorithm~\cite[p.\,232]{ DBLP:books/daglib/0046438}--
certainly the most emblematic of all
quantum algorithms.

The following observations generally hold for other
quantum algorithms in the literature.
We give them together with illustrations (within parenthesis) from the
example of Figure~\ref{ncpe}.
\begin{enumerate}
\item The algorithm is structured in two main parts: a specification
  preamble and a \textbf{Procedure} description. The preamble
  indicates the minimal specification an implementation should
  satisfy. Note that it figures as an actual part of the
  algorithm itself. It contains three types of entries:
  \begin{itemize}
  \item a description of the \textbf{Inputs}, giving a signature for
    the parameters (a black-box circuit $U$, integers $x,N,L$ and two
    quantum registers of sizes $t$ and $L$) and some preconditions for
    these elements (e.g. $x$ co-prime to $L$, $L$ being $N$ bits long,
    \textit{etc});
  \item a description of the \textbf{Outputs} of the algorithm. It
    contains, again, a signature (an integer) and a success condition
    for these \textbf{Outputs} (to be equal to the sought modular
    order);
  \item a \textbf{Runtime} specification, containing: (1) a
    probability of success for each run of the \textbf{Procedure}
    ($O(1)$), (2) resources specifications. In the example, the latter consists
    in bounding the number of required elementary operations. Further
    metrics are also often used (the maximal \emph{width} of required
    quantum circuits--the number of qubits a circuits
    requires, the maximal depth of a circuit--the maximal number
    of operations performed on a given qubit, \textit{etc}).
  \end{itemize}
    
\item The \textbf{Procedure} itself consists of a sequence of declared
  operations, interspersed with formal descriptions of the state of
  the system along with the performance of these operations. (in
  Figure~\ref{ncpe} these elements are given in parallel, declarations
  of operations constitute the right-hand side and intermediate
  formal assertions are on the left-hand side). These
  assertions serve as specifications for the declared operations.  For
  instance, operation ``create superposition'' has precondition the
  formal expression of Line 1, left (framed in blue) and postcondition
  the one of Line 2, left (framed in red). They serve as arguments to
  convince the reader that the algorithm \textbf{Outputs} conditions
  are met at the end of the \textbf{Procedure} (notice that the
  ultimate such postcondition--the measured state being $r$--
  corresponds to the success condition for the overall algorithm). But
  we can also interpret them as \emph{contracts} for the programmer,
  committing her to implement each function in any way provided that
  whenever its inputs satisfy the preconditions, then its outputs
  satisfy the postconditions.

\item The algorithm description is parametric, and so should be any
  program implementing it. Hence, the quantum programming paradigm is
  higher-order: a quantum program is a function from (classical data)
  input parameters to quantum circuits. Then, each instance of a
  quantum circuit behaves as a function from its (quantum data)
  inputs to its (quantum data) outputs.
\end{enumerate}

Most of the current quantum
programs~\cite{QiskitCommunity2017,green2013quipper,%
wecker2014liqui,svore2018q,projectq,silq,CirqDevelopers2018}
proceed as implementations for functions such as those declared in
Figure~\ref{ncpe} (right part of the \textbf{Procedure} part),
providing no guarantee over the algorithm specifications, be it about
their functional behavior or their resource specifications.

Based on the preceding comments, the purpose of formal verification
can be summarized as completing algorithm implementations with their proved 
specifications. In other words:

\paragraph*{}
\textit{Formal verification in quantum programming aims at providing
  solutions to furnish, in addition to quantum programs, evidence that
  these programs meet their specifications, in terms of both success
  probability and resource usage.}

\subsection{Challenges for Quantum Computation}
\label{sec:pro}

Let us now introduce some particularities of quantum programming with
regards to classical computing. They raise design challenges that are
particular to this programming paradigm.

\subsubsection{Destructive Measurement and Non-Determinism}
\label{sec:destructive-meas}
One of the main particularities of quantum programming is that the
output produced by the quantum memory device follows the probabilistic
Born rule (see Section~\ref{sec:qd}). So, in the general case, the
result of quantum computation is non-deterministic. 

Furthermore, a computation in the model from Figure~\ref{qram} contains
both probabilistic quantum computations and classical control
structures, performed by the classical controller. Hence, control
itself may depend on the probabilistic data received from the quantum
device and the execution flow itself is probabilistic.

\subsubsection{Quantum Noise}
\label{sec:quantum-noise}

Another particularity comes from the difficulty to maintain big quantum
systems in a given state and to control the evolution of this state
through time. Along with a quantum computation,
uncontrolled modifications (bit or phase flip, amplitude damping,
\textit{etc}.) of the quantum state may occur. 

To overcome this phenomenon, one solution consists in integrating
 error correction mechanisms into the
compilation.
Error
correction design is an active research
fields~\cite{chiaverini2004realization,lidar2013quantum,%
gottesman1997stabilizer, Fowler2012surfacecodes}. Many
propositions have been developed. They mainly consist in designing
redundant quantum circuits (one logical qubit is implemented by many
different physical qubits). Then the main challenge is to design
a solution for testing the reliability along with a computation without losing
the state of the register due to destructive measurement.

Since error correction requires bigger quantum registers (due to redundancy), its possible implementation is conditioned by the design, elaboration and availability of large quantum processors.

An alternative strategy is to not correct quantum errors, but design computations to limit their effect. John Preskill introduced the notion of \emph{Noisy-Intermediate Scale Quantum (NISQ)} technologies~\cite{nisq}.
The formal analysis of error propagation requires the
identification of possible errors together with rules specifying how their
probability of occurrences propagates along with quantum
circuits~\cite{hung2019quantitative, rand2019formal}.  

\subsubsection{Efficient Compilation on Constrained Hardware }
\label{sec:compilation}

Languages such as \liquid, Q\#, \quipper, \textit{etc}. enable
the description and building of quantum circuits for so-called
\emph{logical qubits}. In practice, realizing a quantum circuit on an
actual quantum machine (\emph{physical qubits}) requires several
compilation passes, in addition to the error correction mentioned
in the preceding paragraph. Among others:
\begin{itemize}
\item the physical realization should respect the physical constraints
  of its target architecture, which concerns, e.g. connectivity of
  qubits or register size limits. Considering this point requires
  qubit reordering intermediary operations and an adequate mapping
  between theoretical and physical qubits;
\item the set of possible quantum operations over physical qubits may
  not correspond to the set of elementary gates from the logical
  circuit description, which would require an adequate gate synthesis and
  circuit rewriting;
\item last but not least, physical realization of an algorithm should be
  as resource frugal as possible, requiring the development of circuit
  optimization techniques.
\end{itemize}
Each of these steps consists of low-level operations over circuits.
They must all preserve functional equivalence while reaching their
proper purpose. For these low-level developing layers, one requires
tools and languages formalizing functionally equivalent circuit
transformation operations. ZX-Calculus~\cite{coecke2017picturing}
(Section~\ref{sec:zx}) is particularly well-fitted for such design,
other propositions include the proof of path-sum semantics
equivalence~\cite{amy2018towards,dblp:phd/basesearch/amy19}
(Section~\ref{sec:path-sum-circuit}) and formally verified circuit
optimization~\cite{hietala19:verif_optim_quant_circuit}
(Section~\ref{sec:voqc}).

\section{General Background on Formal Methods}
\label{sec:gen-back-fm}

We now present a brief overview of formal methods. While the domain is old and has led to rich literature, we try to highlight the underlying main principles and to quickly describe the most popular classes of techniques so far.

\subsection{Introduction}

Formal methods and formal verification \cite{clarkew96formal} denote a
wide range of techniques aiming at {\it proving} the correctness of a
a system with a \textit{mathematical guarantee}—reasoning over
\textit{all} possible inputs and paths of the system, with methods
drawn from logic, automated reasoning and program analysis.
The last two decades have seen an extraordinary blooming of the field,
with significant case studies ranging from pure mathematics
\cite{Gonthier2008} to complete software architectures
\cite{Klein2010,Leroy2009} and industrial systems
\cite{Behm1999,KKP15}.
In addition to offering an alternative to testing, formal verification
has in principle the decisive additional advantages to both enable
parametric proof certificates and offer once-for-all absolute
guarantees for the correctness of programs.

\subsection{Principles}

Formal methods' main principles were laid mostly in the 1970s. Pioneers
include Floyd \cite{Flo67}, Hoare \cite{Hoa69}, Dijkstra \cite{Dij76},
Cousot \cite{CC77} and Clark \cite{CE81}. While there is a wide
diversity of approaches in the field, any formal method builds upon
the following three key ingredients:

\begin{itemize}
\item a formal semantics $M$ representing the possible behaviors of a
  system or program--$M$ is typically equipped with an operational
  semantics and behaviors are often represented as a set of traces
  $L(M)$;

\item a formal specification $\varphi$ of the acceptable or correct
  behaviors--$\varphi$ is typically a logical formula or an
  automaton representing a set of traces $L_{\varphi}$;

\item a (semi-)decision procedure verifying that possible behaviors
  are indeed correct, denoted $M \models \varphi$--typically a
  semi-algorithm to check whether $L(M) \subseteq L_{\varphi}$ holds
  or not.

\end{itemize}

Regarding the complexity of realistic systems and programs, the
verification problem is usually undecidable, hence the impossibility
to have a fully automated and perfectly precise decision procedure for
it.
The different formal method communities bring different responses to
get around this fundamental limitation, yielding different trade-offs
in the design space, favoring either restriction of the classes of
systems under analysis, restrictions of the classes of properties,
human guidance or one-sided answer (over-approximations or
under-approximations).

Overall, after two decades of maturation, formal methods have made
enough progress to be successfully applied to (mostly safety-critical)
software~\cite{HJMS03, VPK04, BCLR04,BGM13,CCF05, KT14,KKP15}.

\subsection{The Formal Method Zoo}
\label{sec:fm_zoo}
We present now in more detail the main classes of formal
methods. While recent techniques tend to blur the lines and combine
aspects from several main approaches, this classification is still
useful to understand the trade-off at stake in the field.

\begin{itemize}
\item Type checking and unification: at the crossroad of programming
  language design and formal methods, type systems
  \cite{pierce02types} allow forbidding by design certain classes of
  errors or bad code patterns (such as trying to  add together a number and a boolean in Java or, in a quantum setting, trying to apply a unitary operation over a classical data register). Traditionally, type systems focus on
  simple ``well-formedness" properties (good typing), but they scale
  very well (modular reasoning) and require only a little manual
  annotation effort (type inference).  While first type systems were
  based on basic unification
  \cite{hindley1969principal,milner1978theory}, advanced type systems
  with dependent types or flow-sensitivity come closer and closer to
  full-fledged verification techniques;
\item Model checking and its many variants: while initially focused on
  finite-state systems \cite{CE81} (typically, idealized protocols or
  hardware models) and complex temporal properties--with essentially
  graph-based and automata-based decision procedures, model checking
  \cite{ModelCheckingBook2018} has notably evolved along the year to
  cope with infinite-state systems, either through specific decidable
  classes (e.g., Petri Nets or Timed Automata) or through
  abstractions. The current approaches to software model checking
  include notably symbolic bounded verification \cite{KT14,Cadar2013}
  for bug finding and counter-example guided abstraction refinement
  \cite{HJMS03} for bug finding and proof of invariants (but it may
  loop forever). Usually, model checking relies on specifications expressed in a variant of modal logic such as temporal logic~\cite{CE81,pnueli1977temporal}, dynamic logic~\cite{harel2001dynamic,lange2006model,fischer1979propositional} or mu-calculus~\cite{kozen1983results,scott1969theory};

\item Abstract Interpretation: Generally speaking, Abstract
  Interpretation  \cite{CC77} is a general theory of abstraction
  for fixpoint computations. Abstract
  Interpretation-based static
  analysis~\cite{RivalBook2020} builds over Abstract Interpretation to
  effectively compute sound (i.e.~overapproximated) abstractions of
  all reachable states of a program. Hence, these techniques are well
  suited for proving invariants. More precisely, Abstract
  Interpretation  provides a systematic recipe to design sound abstract computation over sets of program states,
by connecting the concrete domain (e.g., sets of states) to a given abstract domain (e.g., interval constraints) through a
Galois Connexion between an abstraction and a concretization functions. In practice, Abstract
  Interpretation  comes down to computing
the fixpoint over the abstract domain, ensuring termination but losing precision.  Different abstract domains provide different trade-offs between cost and precision.
Historically speaking, the approach targets full automation and implicit properties (e.g., runtime error);

\item Deductive verification and first-order reasoning: Deductive
  program verification \cite{Barnett2011, Filliatre2011, Hoa69,swamy2016dependent} is
  probably the oldest formal method technique, dating back to 1969
  \cite{Hoa69} and the development of Hoare logic. In this approach, programs are annotated with logical
  assertions, such as pre- and postconditions for operations or loop
  invariants, then so-called proof obligations are automatically
  generated (e.g., by the weakest precondition algorithm) in such a
  way that proving (a.k.a. discharging) them ensures that the logical
  assertions hold along any execution of the program. These proof
  obligations are commonly expressed in first-order or separation logic~\cite{reynolds2002separation} and proven by the help of proof assistants \cite{paulin2015introduction,nipkow2002isabelle} or
  automatic solvers lying on Satisfiability Modulo Theory  ~\cite{barrett2018satisfiability} or Automated Theorem Proving~\cite{fitting2012first};

\item Interactive proof and second-order reasoning: some techniques
  completely drop the hope for automation in favor of expressivity,
  relying on 2nd order or even higher-order specification and proofs
  languages--typically in \coq \cite{paulin2015introduction} or
  \isabelle \cite{nipkow2002isabelle}. Once programmed and proved
  in the language, a certified functional program can then often be
  extracted. This family of approaches is very versatile and almost any
  problem or specification can be encoded, yet it requires lots of
  manual effort, both for the specification and the proofs--
  higher-order proofs can be automatically checked but not
  found. Still, the technique has been for example used for certified
  compilers or operating systems \cite{Leroy2009,Klein2010}.

\end{itemize}

\section{Overview of Formal Methods in Quantum Computing}
\label{sec:overview-fm}

As presented in Section~\ref{sec:gen-back-fm}, formal methods for
proving properties of classical algorithms, programs and systems are
well-developed and versatile. In this section, we present the needs
for formal methods in the realm of quantum computation, while the
later sections are devoted to answering them.

\subsection{The Need for Formal Methods in Quantum Computing}
\label{sec:need-formal-methods}

As introduced in Section~\ref{sec:gen-back-qc}, the data structures at
stake in quantum computing make the computations hard to represent for
developers.  Intermediary formal languages are of great help for
understanding what quantum programs do and describing their
functional behavior.

Furthermore, as introduced in Section~\ref{sec:intro}, directly
importing the testing and debugging practices at use in classical
programming is extremely difficult in the
quantum case\footnote{It requires major adaptations and redefinitions,
  see Section~\ref{sec:assertion_checking} for details.}, due to the
destructive aspect of quantum measurement.
Moreover, the probabilistic nature of quantum algorithms
seriously impedes system-level quantum testing. As a consequence,
test-based programming strategies do not seem adequate in the quantum case
and quantum computing needs alternative debugging strategies and
methodologies.

So far, existing quantum processors were small enough that their
behavior could be entirely simulated on a classical device. Hence, a
short-term solution for overcoming the debugging challenge relied on
classical simulations of quantum programs. Since quantum computer 
prototypes are now reaching the size limit over which this simulation will not be
possible anymore (among others, \cite{arute2019quantum} is often
referred to as the milestone for this context change, referred to as
\emph{quantum supremacy}), more robust solutions must be developed.

On the other hand, nothing prevents \emph{a priori} the formal
verification of quantum programs.  In addition to constituting
alternative debugging strategies, formal methods have several
additional decisive advantages. In particular:

\begin{itemize}
\item They enable parametrized reasoning and certification in the
  higher-order quantum programming context introduced in
  Section~\ref{sec:quantum-algorithm-design}: formal certification of
  a parametrized program holds for any circuit generated by this
  program, whatever the value of its parameters. In contrast, testing
  certification holds for the particular values of these parameters
  that  are used in a test.  Formal certification
  is not limited by the size of the parameters; 

\item It provides once for all an absolute, mathematically proven,
  certification of a program's specifications, whereas testing
  furnishes at best only statistical arguments based on bounded-size
  input samples.
\end{itemize}

\subsection{Typology of Properties to Verify}
\label{sec:typology_properies}

In this section we detail the different properties one has to mind for
developing correct quantum programs. The goal of formal certification
is to provide solutions for their verification.

\subsubsection{Functional Specifications}
\label{sec:func_spec}

A major challenge is to give assurance on the input/output
relationship computed by a given program, that is verifying whether a
given program implements an intended function $f$.  Functional
specifications are two-layered :

\paragraph*{High-level specifications.}
We give a circuit an overall specification independent from the
concrete implementation. It is made of a success condition and a
minimum probability, for any run of this circuit, to result in an
output satisfying the success condition.

In the hybrid model, circuits are run on a quantum co-processor but
controlled by a classical computer, performing control operations
(such as \textbf{if} and \textbf{while} instructions, simple sequence,
\textit{etc}). As a simple example, an algorithm such as the one from
Figure~\ref{ncpe} outputs a success with probability $p$. It can be
included in a control structure including $k$ iterations of it. This
higher-level procedure has probability $[1-(1-p)^k]$ to output a success
at least once, which can be made arbitrarily close to $1$.

More generally, a high-level quantum verification
framework~\cite{10.1007/978-3-030-25543-5_12,dblp:journals/toplas/ying11}
considers an algorithm as a controlled sequence of quantum
functions. There, one considers quantum operations as primitives and
composes them together via controlled sequence operations. These
operations are interpreted as functions in the semantical formalism. For
example,~\cite{selinger2004,10.1007/978-3-030-25543-5_12,dblp:journals/toplas/ying11}
formalize quantum programs in the \emph{density operators
  formalism}. This view is introduced in detail in
Section~\ref{sec:high-level-algorithm} together with the Quantum Hoare
Logic
(QHL)~\cite{10.1007/978-3-030-25543-5_12,dblp:journals/toplas/ying11}.

\paragraph*{Intermediate specifications.} In the \textbf{Procedure}
section from Figure~\ref{ncpe}, each mid-level step of the algorithm
is given a formal specification, that is a description of the state of
the system. These intermediate specifications are deterministic and
concern quantum data.

In a lower-level verification approach view, instead of inputting
quantum operations as primitive functions, one builds quantum circuit
implementations of these operations, by adequately combining quantum
gates. Such a framework~\cite{hietala19:verif_optim_quant_circuit,
  dblp:journals/corr/abs-1803-00699,chareton2021automated} relies on a
circuit description language such as \quipper or \qwire. Then, an
adequate semantics characterization for the built circuits enables to
reason about the quantum data received as inputs and delivered as
outputs. A certification solution for quantum
circuits enables us to reason compositionally about their
semantics. This programming view is explored in
Sections~\ref{sec:qbricks}
 and\ref{sec:sqir}.

\subsubsection{Complexity specifications}
\label{sec:complexity-req}

The major reason for developing quantum computers and quantum
algorithms is to lower computing complexity specifications,
\textit{w.r.t.}~classical computing solutions (see
Section~\ref{sec:quantum_algo_complexity} for precisions).  Therefore,
the relevance of a quantum implementation relies on the fact it satisfies low
complexity specifications. As introduced in
Section~\ref{sec:quantum-algorithm-design}, they may be formulated
through different metrics, such as the width and/or depth of quantum
circuits, their number of elementary gates or more complex metrics
such as \emph{quantum volume}~\cite{lewandowski1997volume}.

The complexity specification is also crucial for another reason: remind
from Section~\ref{sec:quantum-noise} that quantum computation is
subject to noise: the bigger a quantum circuit is, the most prone to
error it is. Functional specifications introduced so far reason about
the theoretical output of quantum computations, in the absence of
errors. The risk of error in a circuit is closely related to
the structural characteristics of this circuit, among which are the
different measures of complexity. Therefore, the information provided
by these measures is also crucial to appreciate the functional
trustfulness of an implementation.

\subsubsection{Structural Constraints}
\label{sec:build-quant-circ}
Quantum circuit design must also consider various \emph{structural
  constraints} that discriminate through several criteria:

\begin{enumerate}
\item They can be either relative to a target architecture or absolute
  (induced by quantum physics laws). The first category comprises, for
  example, the number of available qubits in a processor, the
  connectivity between physical qubits, the set of available
  elementary gates, \textit{etc}. The second category mainly deals
  with aspects induced by quantum Calculus unitarity (no cloning
  theorem, ancilla management, quantum control, \textit{etc}.);

\item Now, depending on the programming language at stake, absolute
  structural constraints may either be taken into charge by the
  language design or left to the user's responsibility.  For example,
  the \emph{no-cloning rule} is derived from the unitarity of quantum
  processes. It forbids using the same quantum data register twice:

  \begin{itemize} 

  \item in languages where quantum data registers are full right
    objects (eg: \quipper, \qwire, \textit{etc}), caring for the respect of
    no-cloning is left to the user. In this case, formal verification
    may help her to do so. Solutions like
    \protoquipper~\cite{ross2015algebraic} or
    \qwire~\cite{paykin2017qwire,dblp:journals/corr/abs-1803-00699}
    tackle this problem through linear type systems (see
    Section~\ref{sec:type-system}); 

  \item another possibility is to reduce the expressivity of the
    language (eg: QFC/QPL~\cite{selinger2004}, \sqir~\cite{hietala19:verif_optim_quant_circuit}, \qbrick~\cite{chareton2021automated}), to prevent any possible violation of
    no-cloning. In \sqir or
    \qbrick, quantum data registers are
    addressed via integer indexes, but the quantum data they hold are not directly accessible from the programming language itself. These data 
    concern the semantics of the language and they are formalized only in  the specification 
    language.  Hence, the respecting conditions for the no-cloning
    theorem are reduced to simple indexing rules for quantum circuits.
  \end{itemize}
  
\item Last, structural program constraints can be either syntactic
  or semantic. The first category contains, for example, all 
  constraints that are linked with qubit identification (eg. :
  \emph{do not control an operation by the value of a qubit it is
    acting on}). The most representative example for semantic 
  constraints concerns a particular aspect of quantum computing that
  we do not detail in this survey: the management of ancilla
  qubits. Ancilla qubits provide additional memory for some
  sections of quantum circuits, the content of which is then
  discharged at some stage of the computation. Discharging a part of a
  register is possible (without affecting the rest of the memory) only
  if there is no interaction between the memory to discharge and the
  rest of the memory (See \cite{ DBLP:books/daglib/0046438} for further
  details). Hence, ancilla management is possible modulo some
  non-entanglement specifications, regarding the semantics of quantum
  circuits.
\end{enumerate}

\subsubsection{Circuit Equivalence}
\label{sec:circuit-equiv-req}

Compilation of quantum programs contains many circuit rewriting
operations (see Section ~\ref{sec:compilation}).  They concern the
implementation of logical qubits in a physical framework and require 
certification for functional behavior preservation. Concretely, given
a logical circuit $C$, compiling $C$ traces as a chain of circuits,
starting from $C$ and each obtained from the precedent by applying a
circuit rewriting operation.  Each such rewriting must preserve the
input/output relation, to ensure that, provided $C$ fits its
functional specifications, then so does the final physical qubits
circuit.  In Section~\ref{sec:low-level-verif} we present two tools
enabling the verification of circuit equivalence: the ZX-Calculus
(Section~\ref{sec:zx}) and the path-sum equivalence verification
(Section~\ref{sec:path-sum-circuit}).

\paragraph*{Further formal comparisons between quantum processes.} 
Different notions of equivalence between quantum processes are also at
stake with further uses of quantum information, such as communication
protocols.  Recent
developments~\cite{DBLP:journals/pacmpl/Unruh19,BartheHYYZ20}
generalize the equivalence specification to further comparison
predicates between quantum processes. Since they are not designed for
the formalization of algorithms, which is the scope of the present
survey, we do not detail these propositions in the present survey.

\section{Low-Level Verification: Compilation and Equivalence}
\label{sec:low-level-verif}

Realizing logical circuits into physical devices (\textit{circuit
  compilation}) requires to deal with severe constraints: the number
of available qubits, their connectivity, the set of elementary
operations, the instability of quantum information--requiring the
insertion of error correction mechanisms, \textit{etc}.  As mentioned
in Section~\ref{sec:circuit-equiv-req}, the underlying circuit
transformations must preserve functional equivalence with the initial
circuit representation, all along the compilation process. In the
present section we introduce formal tools for checking such
equivalences and certifying compilation correctness.

\subsection{ZX-Calculus and Quantomatic/PyZX}
\label{sec:zx}

ZX-Calculus \cite{Coecke2011interacting} is a powerful graphical
language for representing and manipulating quantum information. This
language historically stems from category theory applied to quantum
mechanics, through the program Categorical Quantum Mechanics initiated
by Samson Abramsky and Bob Coecke \cite{Abramsky2009categorical}.

For our purposes, it is interesting to see ZX-diagrams as a
lax version of quantum circuits. This laxness on the one hand implies
that not all ZX-diagrams are implementable with physical qubits, but
on the other hand, it allows formalism to get powerful results on
the underlying equational theory (rewriting rules, pseudo-normal forms).  

The level of abstraction provided by the language allows the user to
reason about quantum programs or protocols while significantly
alleviating the "bureaucracy checks" typically coming with
circuit-level reasoning, in particular, checking sub-circuit
equivalence in the presence of ancillas. It also allows unifying
different models of quantum computation (circuits, measurement-based
quantum computing, lattice surgery, \textit{etc}.), as well as to
provide optimization strategies for these models. Last but not least,
it can be used to formally (yet, graphically) verify properties on
protocols or programs--all that based on simple graph-based
manipulations.

\subsubsection{Semantical Model}

The ZX-diagrams are generated from a set of primitives:
\[
  \left\lbrace 
\begin{tikzpicture}
	\begin{pgfonlayer}{nodelayer}
		\node [style=none] (0) at (0, 0.25) {};
		\node [style=none] (1) at (0, -0.25) {};
	\end{pgfonlayer}
	\begin{pgfonlayer}{edgelayer}
		\draw (0.center) to (1.center);
	\end{pgfonlayer}
\end{tikzpicture}
\colorbox{pink}{missing file : id}}}
, 
\begin{tikzpicture}
	\begin{pgfonlayer}{nodelayer}
		\node [style=none] (0) at (0.25, 0.25) {};
		\node [style=none] (1) at (-0.25, -0.25) {};
		\node [style=none] (2) at (-0.25, 0.25) {};
		\node [style=none] (3) at (0.25, -0.25) {};
	\end{pgfonlayer}
	\begin{pgfonlayer}{edgelayer}
		\draw [in=90, out=-90] (0.center) to (1.center);
		\draw [in=90, out=-90] (2.center) to (3.center);
	\end{pgfonlayer}
\end{tikzpicture}\colorbox{pink}{missing file : swap}}}
, 
\begin{tikzpicture}
	\begin{pgfonlayer}{nodelayer}
		\node [style=none] (0) at (-0.25, 0.125) {};
		\node [style=none] (1) at (0.25, 0.125) {};
		\node [style=none] (2) at (0, -0.125) {};
	\end{pgfonlayer}
	\begin{pgfonlayer}{edgelayer}
		\draw [bend right=90, looseness=1.75] (0.center) to (1.center);
	\end{pgfonlayer}
\end{tikzpicture}
\colorbox{pink}{missing file : cup}}}
,
    
\begin{tikzpicture}
	\begin{pgfonlayer}{nodelayer}
		\node [style=none] (0) at (-0.25, -0.125) {};
		\node [style=none] (1) at (0.25, -0.125) {};
		\node [style=none] (2) at (0, 0.125) {};
	\end{pgfonlayer}
	\begin{pgfonlayer}{edgelayer}
		\draw [bend left=90, looseness=1.75] (0.center) to (1.center);
	\end{pgfonlayer}
\end{tikzpicture}
\colorbox{pink}{missing file : cap}}}
, 
\begin{tikzpicture}
	\begin{pgfonlayer}{nodelayer}
		\node [style=gn] (0) at (0, 0) {};
		\node [style=glabel] (1) at (0.25, 0) {$\alpha$};
		\node [style=none] (2) at (0.25, 0.375) {};
		\node [style=none] (3) at (-0.25, -0.375) {};
		\node [style=none] (4) at (-0.25, 0.375) {};
		\node [style=none] (5) at (0.25, -0.375) {};
		\node [style=none] (6) at (0, 0.375) {$\overset{n}{...}$};
		\node [style=none] (7) at (0, -0.375) {$\underset{m}{...}$};
	\end{pgfonlayer}
	\begin{pgfonlayer}{edgelayer}
		\draw [in=90, out=-90] (2.center) to (3.center);
		\draw [in=90, out=-90] (4.center) to (5.center);
	\end{pgfonlayer}
\end{tikzpicture}\colorbox{pink}{missing file : gn-alpha}}}
, 
\begin{tikzpicture}
	\begin{pgfonlayer}{nodelayer}
		\node [style=rn] (0) at (0, 0) {};
		\node [style=rlabel] (1) at (0.25, 0) {$\alpha$};
		\node [style=none] (2) at (0.25, 0.375) {};
		\node [style=none] (3) at (-0.25, -0.375) {};
		\node [style=none] (4) at (-0.25, 0.375) {};
		\node [style=none] (5) at (0.25, -0.375) {};
		\node [style=none] (6) at (0, 0.375) {$\overset{n}{...}$};
		\node [style=none] (7) at (0, -0.375) {$\underset{m}{...}$};
	\end{pgfonlayer}
	\begin{pgfonlayer}{edgelayer}
		\draw [in=90, out=-90] (2.center) to (3.center);
		\draw [in=90, out=-90] (4.center) to (5.center);
	\end{pgfonlayer}
\end{tikzpicture}\colorbox{pink}{missing file : rn-alpha}}}
,
    
\begin{tikzpicture}
	\begin{pgfonlayer}{nodelayer}
		\node [style=none] (0) at (0, 0.25) {};
		\node [style=none] (1) at (0, -0.25) {};
		\node [style={{H box}}] (2) at (0, 0) {};
	\end{pgfonlayer}
	\begin{pgfonlayer}{edgelayer}
		\draw (0.center) to (1.center);
	\end{pgfonlayer}
\end{tikzpicture}
\colorbox{pink}{missing file : H}}}
\right\rbrace_{\substack{n,m\in\mathbb
      N\\\alpha\in\mathbb R}}
\]
which can be composed either:
\begin{itemize}
\item sequentially: 
\begin{tikzpicture}
	\begin{pgfonlayer}{nodelayer}
		\node [style=none] (0) at (-0.375, -0.125) {};
		\node [style=none] (1) at (-0.375, -0.625) {};
		\node [style=none] (2) at (0.375, -0.125) {};
		\node [style=none] (3) at (0.375, -0.625) {};
		\node [style=none] (4) at (-0.375, 0.625) {};
		\node [style=none] (5) at (-0.375, 0.125) {};
		\node [style=none] (6) at (0.375, 0.625) {};
		\node [style=none] (7) at (0.375, 0.125) {};
		\node [style=none] (8) at (-0.25, 0.875) {};
		\node [style=none] (9) at (-0.25, 0.625) {};
		\node [style=none] (10) at (0.25, 0.875) {};
		\node [style=none] (11) at (0.25, 0.625) {};
		\node [style=none] (12) at (0, 0.75) {...};
		\node [style=none] (13) at (-0.25, 0.125) {};
		\node [style=none] (14) at (-0.25, -0.125) {};
		\node [style=none] (15) at (0.25, 0.125) {};
		\node [style=none] (16) at (0.25, -0.125) {};
		\node [style=none] (17) at (0, 0) {...};
		\node [style=none] (18) at (-0.25, -0.625) {};
		\node [style=none] (19) at (-0.25, -0.875) {};
		\node [style=none] (20) at (0.25, -0.625) {};
		\node [style=none] (21) at (0.25, -0.875) {};
		\node [style=none] (22) at (0, -0.75) {...};
		\node [style=none] (23) at (0, -0.375) {$D_2$};
		\node [style=none] (24) at (0, 0.375) {$D_1$};
	\end{pgfonlayer}
	\begin{pgfonlayer}{edgelayer}
		\draw (0.center) to (1.center);
		\draw (2.center) to (3.center);
		\draw (1.center) to (3.center);
		\draw (2.center) to (0.center);
		\draw (4.center) to (5.center);
		\draw (6.center) to (7.center);
		\draw (5.center) to (7.center);
		\draw (6.center) to (4.center);
		\draw (8.center) to (9.center);
		\draw (10.center) to (11.center);
		\draw (13.center) to (14.center);
		\draw (15.center) to (16.center);
		\draw (18.center) to (19.center);
		\draw (20.center) to (21.center);
	\end{pgfonlayer}
\end{tikzpicture}\colorbox{pink}{missing file : compo}}}

\item or in parallel: 
\begin{tikzpicture}
	\begin{pgfonlayer}{nodelayer}
		\node [style=none] (2) at (-0.625, 0.25) {};
		\node [style=none] (3) at (-0.625, -0.25) {};
		\node [style=none] (4) at (0.125, 0.25) {};
		\node [style=none] (5) at (0.125, -0.25) {};
		\node [style=none] (6) at (-0.5, 0.5) {};
		\node [style=none] (7) at (-0.5, 0.25) {};
		\node [style=none] (8) at (0, 0.5) {};
		\node [style=none] (9) at (0, 0.25) {};
		\node [style=none] (10) at (-0.25, 0.375) {...};
		\node [style=none] (11) at (-0.5, -0.25) {};
		\node [style=none] (12) at (-0.5, -0.5) {};
		\node [style=none] (13) at (0, -0.25) {};
		\node [style=none] (14) at (0, -0.5) {};
		\node [style=none] (15) at (-0.25, -0.375) {...};
		\node [style=none] (16) at (-0.25, 0) {$D_1$};
		\node [style=none] (19) at (0.375, 0.25) {};
		\node [style=none] (20) at (0.375, -0.25) {};
		\node [style=none] (21) at (1.125, 0.25) {};
		\node [style=none] (22) at (1.125, -0.25) {};
		\node [style=none] (23) at (0.5, 0.5) {};
		\node [style=none] (24) at (0.5, 0.25) {};
		\node [style=none] (25) at (1, 0.5) {};
		\node [style=none] (26) at (1, 0.25) {};
		\node [style=none] (27) at (0.75, 0.375) {...};
		\node [style=none] (28) at (0.5, -0.25) {};
		\node [style=none] (29) at (0.5, -0.5) {};
		\node [style=none] (30) at (1, -0.25) {};
		\node [style=none] (31) at (1, -0.5) {};
		\node [style=none] (32) at (0.75, -0.375) {...};
		\node [style=none] (33) at (0.75, 0) {$D_2$};
	\end{pgfonlayer}
	\begin{pgfonlayer}{edgelayer}
		\draw (2.center) to (3.center);
		\draw (4.center) to (5.center);
		\draw (3.center) to (5.center);
		\draw (4.center) to (2.center);
		\draw (6.center) to (7.center);
		\draw (8.center) to (9.center);
		\draw (11.center) to (12.center);
		\draw (13.center) to (14.center);
		\draw (19.center) to (20.center);
		\draw (21.center) to (22.center);
		\draw (20.center) to (22.center);
		\draw (21.center) to (19.center);
		\draw (23.center) to (24.center);
		\draw (25.center) to (26.center);
		\draw (28.center) to (29.center);
		\draw (30.center) to (31.center);
	\end{pgfonlayer}
\end{tikzpicture}\colorbox{pink}{missing file : tensor}}}

\end{itemize}
where $D_1$ and $D_2$ are both ZX-diagrams (themselves composed of the above primitives).
We denote by $\cat{ZX}$ the set of ZX-diagrams. In these, 
information flows from top to bottom, which is in
contrast with quantum circuits  where it flows from left to
right. This is only a matter of convention, as string diagrams, on
which the ZX-Calculus formalism relies upon, are oriented vertically.

These diagrams are used to represent linear maps, thanks to the
so-called \emph{standard interpretation} of ZX-diagrams as complex number matrices
$\stdinterp{.}:\cat{ZX}\to \mathcal M(\mathbb C)$\footnote{To be more precise, the standard interpretation associates to any ZX-diagram in $\cat{ZX}[n,m]$ (i.e. with $n$ inputs and $m$ outputs) a complex matrix of dimension $2^m\times2^n$ i.e. in $\mathcal M_{2^m\times2^n}(\mathbb C)$.}. It is
inductively defined as:

~\hfill
$\stdinterp{
\begin{tikzpicture}
	\begin{pgfonlayer}{nodelayer}
		\node [style=none] (0) at (-0.375, -0.125) {};
		\node [style=none] (1) at (-0.375, -0.625) {};
		\node [style=none] (2) at (0.375, -0.125) {};
		\node [style=none] (3) at (0.375, -0.625) {};
		\node [style=none] (4) at (-0.375, 0.625) {};
		\node [style=none] (5) at (-0.375, 0.125) {};
		\node [style=none] (6) at (0.375, 0.625) {};
		\node [style=none] (7) at (0.375, 0.125) {};
		\node [style=none] (8) at (-0.25, 0.875) {};
		\node [style=none] (9) at (-0.25, 0.625) {};
		\node [style=none] (10) at (0.25, 0.875) {};
		\node [style=none] (11) at (0.25, 0.625) {};
		\node [style=none] (12) at (0, 0.75) {...};
		\node [style=none] (13) at (-0.25, 0.125) {};
		\node [style=none] (14) at (-0.25, -0.125) {};
		\node [style=none] (15) at (0.25, 0.125) {};
		\node [style=none] (16) at (0.25, -0.125) {};
		\node [style=none] (17) at (0, 0) {...};
		\node [style=none] (18) at (-0.25, -0.625) {};
		\node [style=none] (19) at (-0.25, -0.875) {};
		\node [style=none] (20) at (0.25, -0.625) {};
		\node [style=none] (21) at (0.25, -0.875) {};
		\node [style=none] (22) at (0, -0.75) {...};
		\node [style=none] (23) at (0, -0.375) {$D_2$};
		\node [style=none] (24) at (0, 0.375) {$D_1$};
	\end{pgfonlayer}
	\begin{pgfonlayer}{edgelayer}
		\draw (0.center) to (1.center);
		\draw (2.center) to (3.center);
		\draw (1.center) to (3.center);
		\draw (2.center) to (0.center);
		\draw (4.center) to (5.center);
		\draw (6.center) to (7.center);
		\draw (5.center) to (7.center);
		\draw (6.center) to (4.center);
		\draw (8.center) to (9.center);
		\draw (10.center) to (11.center);
		\draw (13.center) to (14.center);
		\draw (15.center) to (16.center);
		\draw (18.center) to (19.center);
		\draw (20.center) to (21.center);
	\end{pgfonlayer}
\end{tikzpicture}\colorbox{pink}{missing file : compo}}}
}=\stdinterp{
\begin{tikzpicture}
	\begin{pgfonlayer}{nodelayer}
		\node [style=none] (0) at (-0.375, 0.25) {};
		\node [style=none] (1) at (-0.375, -0.25) {};
		\node [style=none] (2) at (0.375, 0.25) {};
		\node [style=none] (3) at (0.375, -0.25) {};
		\node [style=none] (4) at (-0.25, 0.5) {};
		\node [style=none] (5) at (-0.25, 0.25) {};
		\node [style=none] (6) at (0.25, 0.5) {};
		\node [style=none] (7) at (0.25, 0.25) {};
		\node [style=none] (8) at (0, 0.375) {...};
		\node [style=none] (9) at (-0.25, -0.25) {};
		\node [style=none] (10) at (-0.25, -0.5) {};
		\node [style=none] (11) at (0.25, -0.25) {};
		\node [style=none] (12) at (0.25, -0.5) {};
		\node [style=none] (13) at (0, -0.375) {...};
		\node [style=none] (14) at (0, 0) {$D_2$};
	\end{pgfonlayer}
	\begin{pgfonlayer}{edgelayer}
		\draw (0.center) to (1.center);
		\draw (2.center) to (3.center);
		\draw (1.center) to (3.center);
		\draw (2.center) to (0.center);
		\draw (4.center) to (5.center);
		\draw (6.center) to (7.center);
		\draw (9.center) to (10.center);
		\draw (11.center) to (12.center);
	\end{pgfonlayer}
\end{tikzpicture}\colorbox{pink}{missing file : D2}}}
}\circ
\stdinterp{
\begin{tikzpicture}
	\begin{pgfonlayer}{nodelayer}
		\node [style=none] (0) at (-0.5, 0.25) {};
		\node [style=none] (1) at (-0.5, -0.25) {};
		\node [style=none] (2) at (0.25, 0.25) {};
		\node [style=none] (3) at (0.25, -0.25) {};
		\node [style=none] (4) at (-0.375, 0.5) {};
		\node [style=none] (5) at (-0.375, 0.25) {};
		\node [style=none] (6) at (0.125, 0.5) {};
		\node [style=none] (7) at (0.125, 0.25) {};
		\node [style=none] (8) at (-0.125, 0.375) {...};
		\node [style=none] (9) at (-0.375, -0.25) {};
		\node [style=none] (10) at (-0.375, -0.5) {};
		\node [style=none] (11) at (0.125, -0.25) {};
		\node [style=none] (12) at (0.125, -0.5) {};
		\node [style=none] (13) at (-0.125, -0.375) {...};
		\node [style=none] (14) at (-0.125, 0) {$D_1$};
	\end{pgfonlayer}
	\begin{pgfonlayer}{edgelayer}
		\draw (0.center) to (1.center);
		\draw (2.center) to (3.center);
		\draw (1.center) to (3.center);
		\draw (2.center) to (0.center);
		\draw (4.center) to (5.center);
		\draw (6.center) to (7.center);
		\draw (9.center) to (10.center);
		\draw (11.center) to (12.center);
	\end{pgfonlayer}
\end{tikzpicture}\colorbox{pink}{missing file : D1}}}
}$
\hfill
$\stdinterp{
\begin{tikzpicture}
	\begin{pgfonlayer}{nodelayer}
		\node [style=none] (2) at (-0.625, 0.25) {};
		\node [style=none] (3) at (-0.625, -0.25) {};
		\node [style=none] (4) at (0.125, 0.25) {};
		\node [style=none] (5) at (0.125, -0.25) {};
		\node [style=none] (6) at (-0.5, 0.5) {};
		\node [style=none] (7) at (-0.5, 0.25) {};
		\node [style=none] (8) at (0, 0.5) {};
		\node [style=none] (9) at (0, 0.25) {};
		\node [style=none] (10) at (-0.25, 0.375) {...};
		\node [style=none] (11) at (-0.5, -0.25) {};
		\node [style=none] (12) at (-0.5, -0.5) {};
		\node [style=none] (13) at (0, -0.25) {};
		\node [style=none] (14) at (0, -0.5) {};
		\node [style=none] (15) at (-0.25, -0.375) {...};
		\node [style=none] (16) at (-0.25, 0) {$D_1$};
		\node [style=none] (19) at (0.375, 0.25) {};
		\node [style=none] (20) at (0.375, -0.25) {};
		\node [style=none] (21) at (1.125, 0.25) {};
		\node [style=none] (22) at (1.125, -0.25) {};
		\node [style=none] (23) at (0.5, 0.5) {};
		\node [style=none] (24) at (0.5, 0.25) {};
		\node [style=none] (25) at (1, 0.5) {};
		\node [style=none] (26) at (1, 0.25) {};
		\node [style=none] (27) at (0.75, 0.375) {...};
		\node [style=none] (28) at (0.5, -0.25) {};
		\node [style=none] (29) at (0.5, -0.5) {};
		\node [style=none] (30) at (1, -0.25) {};
		\node [style=none] (31) at (1, -0.5) {};
		\node [style=none] (32) at (0.75, -0.375) {...};
		\node [style=none] (33) at (0.75, 0) {$D_2$};
	\end{pgfonlayer}
	\begin{pgfonlayer}{edgelayer}
		\draw (2.center) to (3.center);
		\draw (4.center) to (5.center);
		\draw (3.center) to (5.center);
		\draw (4.center) to (2.center);
		\draw (6.center) to (7.center);
		\draw (8.center) to (9.center);
		\draw (11.center) to (12.center);
		\draw (13.center) to (14.center);
		\draw (19.center) to (20.center);
		\draw (21.center) to (22.center);
		\draw (20.center) to (22.center);
		\draw (21.center) to (19.center);
		\draw (23.center) to (24.center);
		\draw (25.center) to (26.center);
		\draw (28.center) to (29.center);
		\draw (30.center) to (31.center);
	\end{pgfonlayer}
\end{tikzpicture}\colorbox{pink}{missing file : tensor}}}
}=\stdinterp{
\begin{tikzpicture}
	\begin{pgfonlayer}{nodelayer}
		\node [style=none] (0) at (-0.5, 0.25) {};
		\node [style=none] (1) at (-0.5, -0.25) {};
		\node [style=none] (2) at (0.25, 0.25) {};
		\node [style=none] (3) at (0.25, -0.25) {};
		\node [style=none] (4) at (-0.375, 0.5) {};
		\node [style=none] (5) at (-0.375, 0.25) {};
		\node [style=none] (6) at (0.125, 0.5) {};
		\node [style=none] (7) at (0.125, 0.25) {};
		\node [style=none] (8) at (-0.125, 0.375) {...};
		\node [style=none] (9) at (-0.375, -0.25) {};
		\node [style=none] (10) at (-0.375, -0.5) {};
		\node [style=none] (11) at (0.125, -0.25) {};
		\node [style=none] (12) at (0.125, -0.5) {};
		\node [style=none] (13) at (-0.125, -0.375) {...};
		\node [style=none] (14) at (-0.125, 0) {$D_1$};
	\end{pgfonlayer}
	\begin{pgfonlayer}{edgelayer}
		\draw (0.center) to (1.center);
		\draw (2.center) to (3.center);
		\draw (1.center) to (3.center);
		\draw (2.center) to (0.center);
		\draw (4.center) to (5.center);
		\draw (6.center) to (7.center);
		\draw (9.center) to (10.center);
		\draw (11.center) to (12.center);
	\end{pgfonlayer}
\end{tikzpicture}\colorbox{pink}{missing file : D1}}}
}\otimes
\stdinterp{
\begin{tikzpicture}
	\begin{pgfonlayer}{nodelayer}
		\node [style=none] (0) at (-0.375, 0.25) {};
		\node [style=none] (1) at (-0.375, -0.25) {};
		\node [style=none] (2) at (0.375, 0.25) {};
		\node [style=none] (3) at (0.375, -0.25) {};
		\node [style=none] (4) at (-0.25, 0.5) {};
		\node [style=none] (5) at (-0.25, 0.25) {};
		\node [style=none] (6) at (0.25, 0.5) {};
		\node [style=none] (7) at (0.25, 0.25) {};
		\node [style=none] (8) at (0, 0.375) {...};
		\node [style=none] (9) at (-0.25, -0.25) {};
		\node [style=none] (10) at (-0.25, -0.5) {};
		\node [style=none] (11) at (0.25, -0.25) {};
		\node [style=none] (12) at (0.25, -0.5) {};
		\node [style=none] (13) at (0, -0.375) {...};
		\node [style=none] (14) at (0, 0) {$D_2$};
	\end{pgfonlayer}
	\begin{pgfonlayer}{edgelayer}
		\draw (0.center) to (1.center);
		\draw (2.center) to (3.center);
		\draw (1.center) to (3.center);
		\draw (2.center) to (0.center);
		\draw (4.center) to (5.center);
		\draw (6.center) to (7.center);
		\draw (9.center) to (10.center);
		\draw (11.center) to (12.center);
	\end{pgfonlayer}
\end{tikzpicture}\colorbox{pink}{missing file : D2}}}
}$
\hfill~

~\hfill
$\stdinterp{~
\begin{tikzpicture}
	\begin{pgfonlayer}{nodelayer}
		\node [style=none] (0) at (0, 0.25) {};
		\node [style=none] (1) at (0, -0.25) {};
	\end{pgfonlayer}
	\begin{pgfonlayer}{edgelayer}
		\draw (0.center) to (1.center);
	\end{pgfonlayer}
\end{tikzpicture}
\colorbox{pink}{missing file : id}}}
~}=id_{\mathbb C^2}=\ketbra00+\ketbra11$
\hfill
$\displaystyle\stdinterp{~
\begin{tikzpicture}
	\begin{pgfonlayer}{nodelayer}
		\node [style=none] (0) at (0.25, 0.25) {};
		\node [style=none] (1) at (-0.25, -0.25) {};
		\node [style=none] (2) at (-0.25, 0.25) {};
		\node [style=none] (3) at (0.25, -0.25) {};
	\end{pgfonlayer}
	\begin{pgfonlayer}{edgelayer}
		\draw [in=90, out=-90] (0.center) to (1.center);
		\draw [in=90, out=-90] (2.center) to (3.center);
	\end{pgfonlayer}
\end{tikzpicture}\colorbox{pink}{missing file : swap}}}
~}=
\sum_{i,j\in\{0,1\}}\ketbra{ji}{ij}$
\hfill~

~\hfill
$\stdinterp{~
\begin{tikzpicture}
	\begin{pgfonlayer}{nodelayer}
		\node [style=none] (0) at (-0.25, -0.125) {};
		\node [style=none] (1) at (0.25, -0.125) {};
		\node [style=none] (2) at (0, 0.125) {};
	\end{pgfonlayer}
	\begin{pgfonlayer}{edgelayer}
		\draw [bend left=90, looseness=1.75] (0.center) to (1.center);
	\end{pgfonlayer}
\end{tikzpicture}
\colorbox{pink}{missing file : cap}}}
~}=\stdinterp{~
\begin{tikzpicture}
	\begin{pgfonlayer}{nodelayer}
		\node [style=none] (0) at (-0.25, 0.125) {};
		\node [style=none] (1) at (0.25, 0.125) {};
		\node [style=none] (2) at (0, -0.125) {};
	\end{pgfonlayer}
	\begin{pgfonlayer}{edgelayer}
		\draw [bend right=90, looseness=1.75] (0.center) to (1.center);
	\end{pgfonlayer}
\end{tikzpicture}
\colorbox{pink}{missing file : cup}}}
~}^\dagger=
\ket{00}{}+\ket{11}{}$
\hfill $\stdinterp{~
\begin{tikzpicture}
	\begin{pgfonlayer}{nodelayer}
		\node [style=none] (0) at (0, 0.25) {};
		\node [style=none] (1) at (0, -0.25) {};
		\node [style={{H box}}] (2) at (0, 0) {};
	\end{pgfonlayer}
	\begin{pgfonlayer}{edgelayer}
		\draw (0.center) to (1.center);
	\end{pgfonlayer}
\end{tikzpicture}
\colorbox{pink}{missing file : H}}}
~}=\ketbra+0+\ketbra-1$ \hfill~

~\hfill
$\stdinterp{
\begin{tikzpicture}
	\begin{pgfonlayer}{nodelayer}
		\node [style=gn] (0) at (0, 0) {};
		\node [style=glabel] (1) at (0.25, 0) {$\alpha$};
		\node [style=none] (2) at (0.25, 0.375) {};
		\node [style=none] (3) at (-0.25, -0.375) {};
		\node [style=none] (4) at (-0.25, 0.375) {};
		\node [style=none] (5) at (0.25, -0.375) {};
		\node [style=none] (6) at (0, 0.375) {$\overset{n}{...}$};
		\node [style=none] (7) at (0, -0.375) {$\underset{m}{...}$};
	\end{pgfonlayer}
	\begin{pgfonlayer}{edgelayer}
		\draw [in=90, out=-90] (2.center) to (3.center);
		\draw [in=90, out=-90] (4.center) to (5.center);
	\end{pgfonlayer}
\end{tikzpicture}\colorbox{pink}{missing file : gn-alpha}}}
}=\ketbra{0^m}{0^n}
+e^{i\alpha}\ketbra{1^m}{1^n}$
\hfill
$\stdinterp{
\begin{tikzpicture}
	\begin{pgfonlayer}{nodelayer}
		\node [style=rn] (0) at (0, 0) {};
		\node [style=rlabel] (1) at (0.25, 0) {$\alpha$};
		\node [style=none] (2) at (0.25, 0.375) {};
		\node [style=none] (3) at (-0.25, -0.375) {};
		\node [style=none] (4) at (-0.25, 0.375) {};
		\node [style=none] (5) at (0.25, -0.375) {};
		\node [style=none] (6) at (0, 0.375) {$\overset{n}{...}$};
		\node [style=none] (7) at (0, -0.375) {$\underset{m}{...}$};
	\end{pgfonlayer}
	\begin{pgfonlayer}{edgelayer}
		\draw [in=90, out=-90] (2.center) to (3.center);
		\draw [in=90, out=-90] (4.center) to (5.center);
	\end{pgfonlayer}
\end{tikzpicture}\colorbox{pink}{missing file : rn-alpha}}}
}=\ketbra{+^m}{+^n}
+e^{i\alpha}\ketbra{-^m}{-^n}$
\hfill~

\noindent
where $\ket+{}:=\frac{\ket0{}+\ket1{}}{\sqrt2}$ and
$\ket-{}:=\frac{\ket0{}-\ket1{}}{\sqrt2}$ and $\ketbra uv$ is
the ket bra outer product from Section~\ref{sec:qd}. For example,
$id_{\mathbb C^2} = \ketbra00+\ketbra11
= \begin{psmallmatrix}1&0\end{psmallmatrix}\otimes 
\begin{psmallmatrix}1\\0\end{psmallmatrix}
+ \begin{psmallmatrix}0&1\end{psmallmatrix}\otimes 
\begin{psmallmatrix}0\\1\end{psmallmatrix}
=
\begin{psmallmatrix}1&0\\0&0\end{psmallmatrix}+ 
\begin{psmallmatrix}0&0\\0&1\end{psmallmatrix} = 
\begin{psmallmatrix}1&0\\0&1\end{psmallmatrix}$.

Notice that the green (light) and red (dark) nodes only differ from the basis in which
they are defined (as $(\ket+{},\ket-{})$ defines an orthonormal basis
of $\mathbb C^2$) and that they can have an arbitrary number of
inputs and outputs.  It often happens that a green or red node has a
parameter of value $0$. In this case, by convention, this angle $0$ is
omitted.
Finally, notice that $
\begin{tikzpicture}
	\begin{pgfonlayer}{nodelayer}
		\node [style=none] (0) at (0, 0.25) {};
		\node [style=none] (1) at (0, -0.25) {};
		\node [style={{H box}}] (2) at (0, 0) {};
	\end{pgfonlayer}
	\begin{pgfonlayer}{edgelayer}
		\draw (0.center) to (1.center);
	\end{pgfonlayer}
\end{tikzpicture}
\colorbox{pink}{missing file : H}}}
$ represents exactly the Hadamard
gate of quantum circuits. This is not a coincidence, as 
ZX-diagrams can be seen as a generalization of quantum
circuits. In particular, we can map any quantum circuit to a
ZX-diagram that represents exactly the same quantum operator:
\begin{center}
  $\begin{array}{c@{~\mapsto~}c@{\qquad\qquad}c@{~\mapsto~}c}
     \Phase(\theta) & 
\begin{tikzpicture}
	\begin{pgfonlayer}{nodelayer}
		\node [style=none] (0) at (0, 0.375) {};
		\node [style=none] (1) at (0, -0.375) {};
		\node [style=rn] (2) at (0.75, 0.2) {};
		\node [style=gn] (3) at (0.75, -0.2) {};
		\node [style=rlabel] (4) at (1, 0.2) {$\pi$};
		\node [style=glabel] (5) at (1, -0.2) {$2\theta$};
		\node [style=rn] (6) at (0.35, 0.2) {};
		\node [style=gn] (7) at (0.35, -0.2) {};
	\end{pgfonlayer}
	\begin{pgfonlayer}{edgelayer}
		\draw (0.center) to (1.center);
		\draw (2) to (3);
		\draw (6) to (7);
		\draw [bend left=45, looseness=1.25] (6) to (7);
		\draw [bend left=45, looseness=1.25] (7) to (6);
	\end{pgfonlayer}
\end{tikzpicture}\colorbox{pink}{missing file : phase}}}

    & \Rz(\theta) & 
\begin{tikzpicture}
	\begin{pgfonlayer}{nodelayer}
		\node [style=gn] (0) at (0, 0) {};
		\node [style=glabel] (1) at (0.25, 0) {$4\theta$};
		\node [style=none] (2) at (0, 0.375) {};
		\node [style=none] (3) at (0, -0.375) {};
		\node [style=rn] (4) at (1.15, 0.2) {};
		\node [style=gn] (5) at (1.15, -0.2) {};
		\node [style=rlabel] (6) at (1.4, 0.2) {$\pi$};
		\node [style=glabel] (7) at (1.4, -0.2) {-$2\theta$};
		\node [style=rn] (8) at (0.75, 0.2) {};
		\node [style=gn] (9) at (0.75, -0.2) {};
	\end{pgfonlayer}
	\begin{pgfonlayer}{edgelayer}
		\draw [in=90, out=-90] (2.center) to (3.center);
		\draw (4) to (5);
		\draw (8) to (9);
		\draw [bend left=45, looseness=1.25] (8) to (9);
		\draw [bend left=45, looseness=1.25] (9) to (8);
	\end{pgfonlayer}
\end{tikzpicture}\colorbox{pink}{missing file : rz-theta}}}
\\[1.5em]
     \Had & 
\begin{tikzpicture}
	\begin{pgfonlayer}{nodelayer}
		\node [style=none] (0) at (0, 0.25) {};
		\node [style=none] (1) at (0, -0.25) {};
		\node [style={{H box}}] (2) at (0, 0) {};
	\end{pgfonlayer}
	\begin{pgfonlayer}{edgelayer}
		\draw (0.center) to (1.center);
	\end{pgfonlayer}
\end{tikzpicture}
\colorbox{pink}{missing file : H}}}
 & \Cnot & 
\begin{tikzpicture}
	\begin{pgfonlayer}{nodelayer}
		\node [style=rn] (0) at (0, 0.2) {};
		\node [style=gn] (1) at (0, -0.2) {};
		\node [style=none] (2) at (-1, 0.375) {};
		\node [style=none] (3) at (-1, -0.375) {};
		\node [style=none] (4) at (-0.5, 0.375) {};
		\node [style=none] (5) at (-0.5, -0.375) {};
		\node [style=gn] (6) at (-1, 0.125) {};
		\node [style=rn] (7) at (-0.5, -0.125) {};
	\end{pgfonlayer}
	\begin{pgfonlayer}{edgelayer}
		\draw (0) to (1);
		\draw (2.center) to (6);
		\draw (6) to (3.center);
		\draw (4.center) to (7);
		\draw (7) to (5.center);
		\draw (6) to (7);
	\end{pgfonlayer}
\end{tikzpicture}\colorbox{pink}{missing file : CNot}}}

   \end{array}$
\end{center}
and that preserves sequential and parallel compositions. The
elementary gates given above are the ones detailed in Table
\ref{elemgate}.

We can actually map any \emph{generalized} quantum circuit
(i.e. circuit including measure) into a ZX-diagram. Indeed,
initializations of qubits are easy to represent:
$\ket0{}\mapsto 
\begin{tikzpicture}
	\begin{pgfonlayer}{nodelayer}
		\node [style=rn] (0) at (0, 0) {};
		\node [style=none] (3) at (0, -0.375) {};
		\node [style=rn] (4) at (0.425, 0.2) {};
		\node [style=gn] (5) at (0.425, -0.2) {};
	\end{pgfonlayer}
	\begin{pgfonlayer}{edgelayer}
		\draw (4) to (5);
		\draw [bend left=45, looseness=1.25] (4) to (5);
		\draw [bend left=45, looseness=1.25] (5) to (4);
		\draw (0) to (3.center);
	\end{pgfonlayer}
\end{tikzpicture}\colorbox{pink}{missing file : ket0}}}
$, and there exists an extension of the
ZX-Calculus \cite{Coecke2012environment,Carette2019completeness} that
allows the language to represent measurements.
In this extension, we represent the environment as $\ground$, which
becomes an additional generator of the diagrams (we denote by
$\cat{ZX}^{\sground}$ this updated set of ZX-diagrams). This generator
can also be understood as discarding a qubit. However, contrary to
classical data, this action affects the rest of the system. 
Introducing $\ground$ forces us to change the codomain of the standard
interpretation, but we will not give the details here. Simply keep in
mind that the measurement in the computational basis
$(\ket0{},\ket1{})$ is represented by $
\begin{tikzpicture}
	\begin{pgfonlayer}{nodelayer}
		\node [style=gn] (0) at (0, 0) {};
		\node [style=none] (1) at (0.325, -0.275) {};
		\node [style=none] (2) at (0, 0.5) {};
		\node [style=none] (3) at (0, -0.5) {};
		\node [style=none] (4) at (0.325, -0.45) {$\ground$};
	\end{pgfonlayer}
	\begin{pgfonlayer}{edgelayer}
		\draw (2.center) to (3.center);
		\draw [in=90, out=-15] (0) to (1.center);
		\draw (1.center) to (4.center);
	\end{pgfonlayer}
\end{tikzpicture}\colorbox{pink}{missing file : measurement}}}
$.

In this way, we can (fairly) easily represent any generalized quantum
circuit as a ZX-diagram. But we can actually represent more, and this
is an active field of research to try and characterize diagrams that
can be put in circuit form (we talk about ``extracting a
circuit''). First was introduced the notion of causal flow
\cite{Danos2006determinism} which was then extended to that of
``gflow'' (for generalized flow)
\cite{Browne2007generalizedflow}. Some other variations
exist~\cite{Backens2020ThereAB}.

Quantum circuits, however, are not the only computational model one
might want ZX-diagrams to compile to. Indeed, it so happens that the
primitives of the ZX-Calculus quite naturally match those of 
\emph{lattice surgery} \cite{deBeaudrap2020zxlattice}, 
a scheme for error 
correction \cite{Fowler2012surfacecodes,Horsman2012surfacecode}. 
In particular, ZX-diagrams implementing a (physical) lattice surgery
procedure features a special notion of flow, the \emph{PF flow} (for
Pauli fusion flow) \cite{deBeaudrap2019paulifusion}.

\subsubsection{Verified Properties}

The strength of ZX-Calculus comes from its powerful equational
theory. This equational theory allows to define equivalence classes
of ZX-diagrams and to conveniently decide whether two different
diagrams represent the same quantum operator.

This question can be asked for quantum circuits as well, as two
different circuits may represent the same operator
(e.g.~$\Had^2=\Id$). Some such equational theories exist for quantum
circuits \cite{Selinger2015generators,Amy2018cnotdihedral}, but none,
as of today, for a universal fragment of quantum mechanics (notice that 
in \cite{10.1145/2775051.2676999} a completeness theorem is given for 
the interaction between measurements and pure parts of the circuit, 
but crucially one for the pure part itself is not provided but only assumed).

The main difference between the two formalisms  is that the equational
theory of the ZX-Calculus allows for a powerful result in this
language, aggregated under the paradigm \textit{``only connectivity
matters''}. This result states that we can treat any ZX-diagram as an
undirected open graph, where the Hadamard box and the green and red
nodes are considered as vertices. In particular, any (open) graph
isomorphism is an allowed transformation.

\begin{example}
  $
\begin{tikzpicture}
	\begin{pgfonlayer}{nodelayer}
		\node [style=gn] (0) at (0, 0.4) {};
		\node [style=glabel] (000000) at (-0.35, 0.4) {$\frac\pi 2$};
		\node [style=rn] (1) at (0.5, 0) {};
		\node [style={{H box}}] (2) at (-0.125, -0.05) {};
		\node [style=none] (3) at (-0.125, -0.65) {};
		\node [style=none] (4) at (0.5, -0.65) {};
		\node [style=none] (5) at (0.5, 0.7) {};
		\node [style=none] (6) at (0, 0.7) {};
	\end{pgfonlayer}
	\begin{pgfonlayer}{edgelayer}
		\draw (0) to (1);
		\draw [in=90, out=-90] (1) to (3.center);
		\draw [in=90, out=-90] (2) to (4.center);
		\draw [in=90, out=-116] (0) to (2);
		\draw (5.center) to (1);
		\draw (6.center) to (0);
	\end{pgfonlayer}
\end{tikzpicture}\colorbox{pink}{missing file : example-composition-zx-no-box}}}
~~=~~
\begin{tikzpicture}
	\begin{pgfonlayer}{nodelayer}
		\node [style=gn] (0) at (-0.15, 0.15) {};
		\node [style=glabel] (1) at (0.1, 0.15) {$\frac\pi 2$};
		\node [style=rn] (2) at (-0.775, 0.475) {};
		\node [style={{H box}}] (3) at (-0.15, -0.325) {};
		\node [style=none] (4) at (-0.775, -0.65) {};
		\node [style=none] (5) at (-0.15, -0.65) {};
		\node [style=none] (6) at (-0.15, 0.95) {};
		\node [style=none] (7) at (-0.775, 0.95) {};
	\end{pgfonlayer}
	\begin{pgfonlayer}{edgelayer}
		\draw (0) to (2);
		\draw (2) to (4.center);
		\draw (3) to (5.center);
		\draw (0) to (3);
		\draw [in=90, out=-90, looseness=0.75] (7.center) to (0);
		\draw [in=-90, out=60] (2) to (6.center);
	\end{pgfonlayer}
\end{tikzpicture}\colorbox{pink}{missing file : example-isom}}}
$
  because the two diagrams can be obtained from one another by simply
  ``moving their nodes around'' (while keeping inputs and outputs
  fixed).
\end{example}

This result also allows us to unambiguously represent a horizontal
wire. For instance:
\begin{center}

\begin{tikzpicture}
	\begin{pgfonlayer}{nodelayer}
		\node [style=none] (0) at (0.5, 0.375) {};
		\node [style=none] (1) at (0.5, -0.375) {};
		\node [style=none] (2) at (1, 0.375) {};
		\node [style=none] (3) at (1, -0.375) {};
		\node [style=gn] (4) at (0.5, 0.125) {};
		\node [style=rn] (5) at (1, -0.125) {};
		\node [style=none] (6) at (2, -0.375) {};
		\node [style=none] (7) at (2, 0.375) {};
		\node [style=none] (8) at (2.5, -0.375) {};
		\node [style=none] (9) at (2.5, 0.375) {};
		\node [style=gn] (10) at (2, -0.125) {};
		\node [style=rn] (11) at (2.5, 0.125) {};
		\node [style=none] (12) at (1.5, 0) {$=$};
		\node [style=none] (13) at (-1, 0.375) {};
		\node [style=none] (14) at (-1, -0.375) {};
		\node [style=none] (15) at (-0.5, 0.375) {};
		\node [style=none] (16) at (-0.5, -0.375) {};
		\node [style=gn] (17) at (-1, 0) {};
		\node [style=rn] (18) at (-0.5, 0) {};
		\node [style=none] (19) at (0, 0) {$:=$};
	\end{pgfonlayer}
	\begin{pgfonlayer}{edgelayer}
		\draw (0.center) to (4);
		\draw (4) to (1.center);
		\draw (2.center) to (5);
		\draw (5) to (3.center);
		\draw (4) to (5);
		\draw (6.center) to (10);
		\draw (10) to (7.center);
		\draw (8.center) to (11);
		\draw (11) to (9.center);
		\draw (10) to (11);
		\draw (13.center) to (17);
		\draw (17) to (14.center);
		\draw (15.center) to (18);
		\draw (18) to (16.center);
		\draw (17) to (18);
	\end{pgfonlayer}
\end{tikzpicture}\colorbox{pink}{missing file : CNot-horizontal-wire}}}
.
\end{center}
This ``meta''-rule, that all isomorphisms of open graphs are allowed,
constitutes the backbone of the ZX-Calculus. In what follows,
different sets of axioms, that satisfy different needs, will be
presented, but this meta-rule will always be there implicitly.

When two diagrams $D_1$ and $D_2$ are proven to be equal using the
equational theory $\mathcal{T}$, we write $\mathcal{T}\vdash D_1=D_2$.
The axiomatization \ax{zx} for the ZX-Calculus can be found in
Figure~\ref{fig:ZX}, and it was recently proven to be complete 
for the standard interpretation $\stdinterp{.}$ 
\cite{Vilmart2019nearminimal}:

\begin{figure}[tbh]
  \centering
  \begin{tabular}{|@{~}c@{$\qquad~$}c@{$\qquad~$}c@{~~}|}
    \hline
    && \\
    
\begin{tikzpicture}
	\begin{pgfonlayer}{nodelayer}
		\node [style=none] (0) at (-0.875, 0) {\rotatebox[origin=c]{63.43}{...}};
		\node [style=none] (1) at (0.5, 0) {$=$};
		\node [style=gn] (2) at (1.25, 0) {};
		\node [style=glabel] (2b) at (1.5, 0) {$\alpha{+}\beta$};
		\node [style=gn] (3) at (-0.5, -0.125) {};
		\node [style=glabel] (3b) at (-0.25, -0.125) {$\beta$};
		\node [style=none] (4) at (1.5, -0.625) {};
		\node [style=none] (5) at (-0.75, -0.625) {};
		\node [style=none] (6) at (-0.25, -0.625) {};
		\node [style=none] (7) at (1, -0.625) {};
		\node [style=none] (8) at (-1.5, -0.625) {};
		\node [style=none] (9) at (-1, -0.625) {};
		\node [style=none] (10) at (1.25, 0.475) {...};
		\node [style=none] (11) at (1, 0.625) {};
		\node [style=none] (12) at (-1.5, 0.625) {};
		\node [style=none] (13) at (-0.25, 0.625) {};
		\node [style=none] (14) at (-1, 0.625) {};
		\node [style=none] (15) at (1.5, 0.625) {};
		\node [style=gn] (16) at (-1.25, 0.125) {};
		\node [style=glabel] (16b) at (-1.55, 0.125) {$\alpha$};
		\node [style=none] (17) at (-0.75, 0.625) {};
		\node [style=none, yshift=-3.5pt] (18) at (-1.25, 0.625) {...};
		\node [style=none] (19) at (0.5, -0.25) {(S)};
		\node [style=none, yshift=2.5pt] (20) at (-0.5, -0.6) {...};
		\node [style=none, yshift=2.5pt] (21) at (-1.25, -0.6) {...};
		\node [style=none, yshift=-3.5pt] (22) at (-0.5, 0.625) {...};
		\node [style=none, yshift=2.5pt] (23) at (1.25, -0.55) {...};
	\end{pgfonlayer}
	\begin{pgfonlayer}{edgelayer}
		\draw [in=-90, out=60] (3) to (13.center);
		\draw [in=90, out=-135] (3) to (5.center);
		\draw [in=90, out=-45] (3) to (6.center);
		\draw [in=90, out=-120] (16) to (8.center);
		\draw [in=90, out=-60] (16) to (9.center);
		\draw [bend right] (16) to (3);
		\draw [bend left] (16) to (3);
		\draw [in=135, out=-90] (11.center) to (2);
		\draw [in=90, out=-135] (2) to (7.center);
		\draw [in=-45, out=90] (4.center) to (2);
		\draw [in=-90, out=45] (2) to (15.center);
		\draw [in=-90, out=135] (16) to (12.center);
		\draw [in=-90, out=45] (16) to (14.center);
		\draw [in=-90, out=120] (3) to (17.center);
	\end{pgfonlayer}
\end{tikzpicture}\colorbox{pink}{missing file : spider}}}

    &
\begin{tikzpicture}
	\begin{pgfonlayer}{nodelayer}
		\node [style=gn] (0) at (-0.75, 0) {};
		\node [style=none] (1) at (0, 0) {$=$};
		\node [style=none] (2) at (-0.75, 0.5) {};
		\node [style=none] (3) at (-0.75, -0.5) {};
		\node [style=none] (4) at (0.5, 0.5) {};
		\node [style=none] (5) at (0.5, -0.5) {};
		\node [style=none] (6) at (0, -0.25) {(I$_g$)};
		\node [style=none] (7) at (1, 0) {$=$};
		\node [style=none] (8) at (1, -0.25) {(I$_r$)};
		\node [style=none] (9) at (1.75, 0.5) {};
		\node [style=none] (10) at (1.75, -0.5) {};
		\node [style=rn] (11) at (1.75, 0) {};
	\end{pgfonlayer}
	\begin{pgfonlayer}{edgelayer}
		\draw (2.center) to (0);
		\draw (0) to (3.center);
		\draw (4.center) to (5.center);
		\draw (9.center) to (10.center);
	\end{pgfonlayer}
\end{tikzpicture}\colorbox{pink}{missing file : id-green-red}}}

    &
\begin{tikzpicture}
	\begin{pgfonlayer}{nodelayer}
		\node [style=rn] (0) at (-1, -0.25) {};
		\node [style=rlabel] (0b) at (-0.75, -0.25) {-$\frac{\pi}{4}$};
		\node [style=gn] (1) at (-1, 0.25) {};
		\node [style=glabel] (1b) at (-0.75, 0.25) {$\frac{\pi}{4}$};
		\node [style=none] (2) at (0, 0) {$=$};
		\node [style=none] (3) at (0.625, 0.25) {};
		\node [style=none] (4) at (0.625, -0.25) {};
		\node [style=none] (5) at (1.125, 0.25) {};
		\node [style=none] (6) at (1.125, -0.25) {};
		\node [style=none] (7) at (0, -0.25) {(E)};
	\end{pgfonlayer}
	\begin{pgfonlayer}{edgelayer}
		\draw (0) to (1);
		\draw [style=dashed] (3.center) to (5.center);
		\draw [style=dashed] (5.center) to (6.center);
		\draw [style=dashed] (6.center) to (4.center);
		\draw [style=dashed] (4.center) to (3.center);
	\end{pgfonlayer}
\end{tikzpicture}\colorbox{pink}{missing file : bicolor_pi_4_eq_empty}}}
\\
    && \\
    
\begin{tikzpicture}
	\begin{pgfonlayer}{nodelayer}
		\node [style=gn] (0) at (0.75, 0) {};
		\node [style=none] (1) at (2.25, -0.25) {};
		\node [style=none] (2) at (0.5, -0.35) {};
		\node [style=rn] (3) at (2.25, 0.15) {};
		\node [style=none] (4) at (1, -0.35) {};
		\node [style=rn] (5) at (0.75, 0.4) {};
		\node [style=rn] (6) at (2.75, 0.15) {};
		\node [style=none] (7) at (2.75, -0.25) {};
		\node [style=none] (8) at (1.5, 0) {$=$};
		\node [style=rn] (9) at (0.275, 0.35) {};
		\node [style=gn] (10) at (0.275, -0.05) {};
		\node [style=none] (11) at (1.5, -0.25) {(CP)};
	\end{pgfonlayer}
	\begin{pgfonlayer}{edgelayer}
		\draw [style=none] (5) to (0);
		\draw [style=none, bend right=23] (0) to (2.center);
		\draw [style=none, bend left=23] (0) to (4.center);
		\draw [style=none] (3) to (1.center);
		\draw [style=none] (6) to (7.center);
		\draw (9) to (10);
	\end{pgfonlayer}
\end{tikzpicture}\colorbox{pink}{missing file : copy}}}

    &
\begin{tikzpicture}
	\begin{pgfonlayer}{nodelayer}
		\node [style=none] (0) at (1, 0.55) {};
		\node [style=rn] (1) at (-1.275, -0.275) {};
		\node [style=none] (2) at (0.5, -0.575) {};
		\node [style=none] (3) at (-0.725, 0.575) {};
		\node [style=none] (4) at (0.5, 0.55) {};
		\node [style=none] (5) at (-1.275, -0.575) {};
		\node [style=none] (6) at (-1.275, 0.575) {};
		\node [style=gn] (7) at (-0.725, 0.275) {};
		\node [style=none] (8) at (0, 0) {$=$};
		\node [style=rn] (9) at (-0.725, -0.275) {};
		\node [style=gn] (10) at (0.75, -0.225) {};
		\node [style=gn] (11) at (-1.275, 0.275) {};
		\node [style=none] (12) at (-0.725, -0.575) {};
		\node [style=none] (13) at (1, -0.575) {};
		\node [style=rn] (14) at (0.75, 0.225) {};
		\node [style=rn] (15) at (-1.75, 0.2) {};
		\node [style=gn] (16) at (-1.75, -0.2) {};
		\node [style=none] (17) at (0, -0.25) {(B)};
	\end{pgfonlayer}
	\begin{pgfonlayer}{edgelayer}
		\draw [style=none] (12.center) to (9);
		\draw [style=none] (5.center) to (1);
		\draw [style=none] (7) to (3.center);
		\draw [style=none, bend right=23] (9) to (7);
		\draw [style=none] (11) to (6.center);
		\draw [style=none, bend left=23] (1) to (11);
		\draw [style=none, bend right=23] (13.center) to (10);
		\draw [style=none] (10) to (14);
		\draw [style=none, bend left=23] (14) to (4.center);
		\draw [style=none, bend right=23] (14) to (0.center);
		\draw [bend right=23] (10) to (2.center);
		\draw (11) to (9);
		\draw (7) to (1);
		\draw (15) to (16);
	\end{pgfonlayer}
\end{tikzpicture}\colorbox{pink}{missing file : bialgebra}}}

    &
\begin{tikzpicture}
	\begin{pgfonlayer}{nodelayer}
		\node [style=none] (0) at (0, 0) {$=$};
		\node [style=gn] (1) at (0.75, 0) {};
		\node [style=glabel] (1b) at (1, 0) {$\beta_2$};
		\node [style=rn] (2) at (0.75, 0.425) {};
		\node [style=rlabel] (2b) at (1, 0.425) {$\beta_1$};
		\node [style=rn] (3) at (0.75, -0.425) {};
		\node [style=rlabel] (3b) at (1, -0.425) {$\beta_3$};
		\node [style=rn] (4) at (-1, 0) {};
		\node [style=rlabel] (4b) at (-0.75, 0) {$\alpha_2$};
		\node [style=gn] (5) at (-1, 0.425) {};
		\node [style=glabel] (5b) at (-0.75, 0.425) {$\alpha_1$};
		\node [style=gn] (6) at (-1, -0.425) {};
		\node [style=glabel] (6b) at (-0.75, -0.425) {$\alpha_3$};
		\node [style=none] (7) at (0.75, 0.75) {};
		\node [style=none] (8) at (0.75, -0.75) {};
		\node [style=none] (9) at (-1, 0.75) {};
		\node [style=none] (10) at (-1, -0.75) {};
		\node [style=rn] (11) at (1.5, 0.225) {};
		\node [style=rlabel] (11b) at (1.75, 0.225) {$\pi$};
		\node [style=gn] (12) at (1.5, -0.225) {};
		\node [style=glabel] (12b) at (1.75, -0.225) {$\gamma$};
		\node [style=rn] (13) at (2.25, -0.225) {};
		\node [style=gn] (14) at (2.25, 0.225) {};
		\node [style=none] (15) at (0, -0.25) {(EU)};
	\end{pgfonlayer}
	\begin{pgfonlayer}{edgelayer}
		\draw [style=none] (7.center) to (8.center);
		\draw [style=none] (9.center) to (10.center);
		\draw [style=none] (12) to (11);
		\draw [style=none, bend right=45, looseness=1.25] (14) to (13);
		\draw [style=none, bend left=45, looseness=1.25] (14) to (13);
		\draw [style=none] (13) to (14);
	\end{pgfonlayer}
\end{tikzpicture}\colorbox{pink}{missing file : Euler}}}
\\
    
\begin{tikzpicture}
	\begin{pgfonlayer}{nodelayer}
		\node [style=gn] (0) at (0.5, -0.45) {};
		\node [style=glabel] (0b) at (0.75, -0.45) {$\frac{\pi}{2}$};
		\node [style=rn] (1) at (0.5, 0) {};
		\node [style=gn] (2) at (0.5, 0.45) {};
		\node [style=glabel] (2b) at (0.15, 0.45) {$\frac{\pi}{2}$};
		\node [style=none] (3) at (0.5, 0.75) {};
		\node [style=gn] (4) at (0.85, 0.225) {};
		\node [style=glabel] (4b) at (1.1, 0.225) {-$\frac{\pi}{2}$};
		\node [style=none] (5) at (0.5, -0.75) {};
		\node [style=none] (6) at (-1, 0.75) {};
		\node [style=none] (7) at (-1, -0.75) {};
		\node [style=none] (8) at (-0.25, 0) {$=$};
		\node [style={{H box}}] (9) at (-1, 0) {};
		\node [style=none] (10) at (-0.25, -0.25) {(HD)};
	\end{pgfonlayer}
	\begin{pgfonlayer}{edgelayer}
		\draw (3.center) to (2);
		\draw (2) to (1);
		\draw (1) to (0);
		\draw (0) to (5.center);
		\draw (1) to (4);
		\draw (6.center) to (7.center);
	\end{pgfonlayer}
\end{tikzpicture}\colorbox{pink}{missing file : euler-decomp-scalar-free}}}

    &
\begin{tikzpicture}
	\begin{pgfonlayer}{nodelayer}
		\node [style=none] (0) at (0.75, 0.5) {};
		\node [style=none] (1) at (-0.9, -0.65) {};
		\node [style=H box] (2) at (-0.9, 0.35) {};
		\node [style=rn] (3) at (-1.25, 0) {};
		\node [style=rlabel] (3b) at (-0.925, 0) {$\alpha$};
		\node [style=H box] (4) at (-1.6, -0.35) {};
		\node [style=none] (5) at (-0.9, 0.65) {};
		\node [style=H box] (6) at (-0.9, -0.35) {};
		\node [style=none] (7) at (-1.25, 0.5) {...};
		\node [style=none] (8) at (0, 0) {$=$};
		\node [style=gn] (9) at (1, 0) {};
		\node [style=glabel] (9b) at (1.25, 0) {$\alpha$};
		\node [style=none] (10) at (1.25, -0.5) {};
		\node [style=none] (11) at (1.25, 0.5) {};
		\node [style=none] (12) at (-1.6, 0.65) {};
		\node [style=H box] (13) at (-1.6, 0.35) {};
		\node [style=none] (14) at (-1.6, -0.65) {};
		\node [style=none] (15) at (0.75, -0.5) {};
		\node [style=none] (16) at (-1.25, -0.5) {...};
		\node [style=none] (17) at (1, 0.4) {...};
		\node [style=none] (18) at (1, -0.375) {...};
		\node [style=none] (19) at (0, -0.25) {(H)};
	\end{pgfonlayer}
	\begin{pgfonlayer}{edgelayer}
		\draw [bend right] (3) to (4);
		\draw [bend left] (3) to (6);
		\draw (6) to (1.center);
		\draw (4) to (14.center);
		\draw [bend left] (3) to (13);
		\draw [bend right] (3) to (2);
		\draw (2) to (5.center);
		\draw (13) to (12.center);
		\draw [bend left] (9) to (0.center);
		\draw [bend right] (9) to (11.center);
		\draw [bend right] (9) to (15.center);
		\draw [bend left] (9) to (10.center);
	\end{pgfonlayer}
\end{tikzpicture}\colorbox{pink}{missing file : colour-change}}}

    &$\overbrace{\begin{array}{@{}l@{}}
  x^+:=\frac{\alpha_1+\alpha_3}{2}\qquad x^-:=x^+-\alpha_3\\
  z := \cos{\frac{\alpha_2}{2}}\cos{x^+}+
               i\sin{\frac{\alpha_2}{2}}\cos{x^-}\\
  z' := \cos{\frac{\alpha_2}{2}}\sin{x^+}-
               i\sin{\frac{\alpha_2}{2}}\sin{x^-}\\
    \beta_1 = \arg z + \arg z\\
    \beta_2 = 2\arg\left(i+\left|\frac{z}{z'}\right|\right)\\
    \beta_3 = \arg z - \arg z'\\
    \gamma = x^+-\arg(z)+\frac{\alpha_2-\beta_2}{2}
 \end{array}}$\\
    && \\
    \hline
  \end{tabular}
  \caption[The equational theory $\ax{zx}$]{The equational theory $\ax{zx}$\footnotemark. All rules -- provably -- hold in their upside-down and color-swapped (between green and red) versions.}
  \label{fig:ZX}
\end{figure}
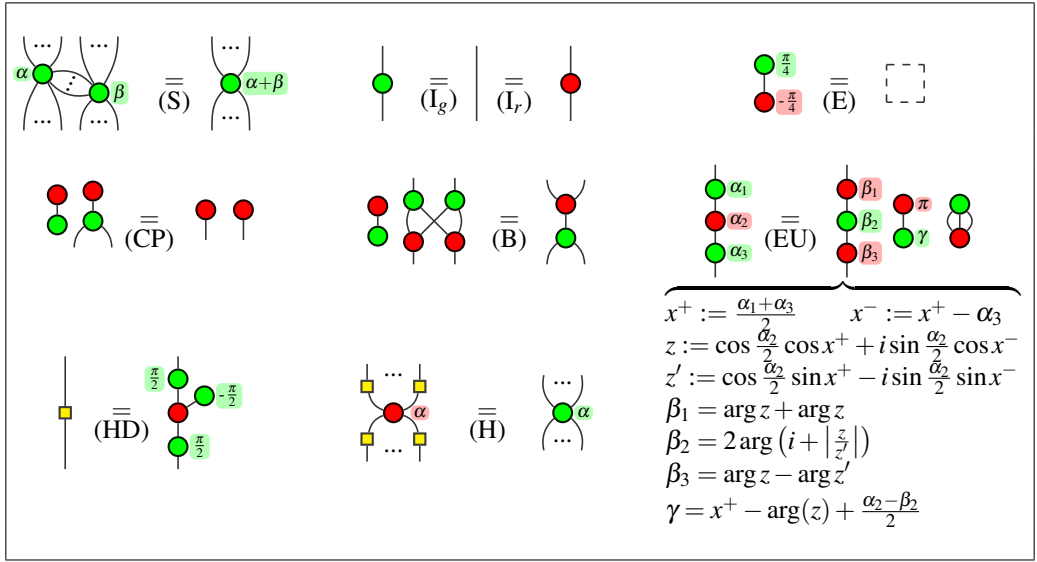

\begin{theorem}
  $\cat{ZX}/\ax{zx}$ is complete with respect to $\stdinterp{.}$:
  \[
    \forall D_1, D_2\in \cat{ZX},~~
    \stdinterp{D_1}=\stdinterp{D_2}\iff \ax{zx}\vdash D_1 = D_2
  \]
\end{theorem}

\noindent
Here $\cat{ZX}/\ax{zx}$ represents the quotient of $\cat{ZX}$ by the
equational theory $\ax{zx}$.  The completeness property is
fundamental. It allows us to reason on quantum processes through
diagrammatic transformations rather than by matrix computations. In
particular, it tells us that whenever two diagrams represent the same
quantum operator, they can be turned into one another using only the
rules of \ax{zx}.

It is customary in quantum computing to work with particular
(restricted) sets of gates. For a lot of such restrictions, there
exist complete axiomatizations
\cite{Hadzihasanovic2018twocomplete, Jeandel2018complete,%
Jeandel2018diagrammatic,jeandel2018rational, Jeandel2019normalform,%
Jeandel2020completeness}. The
ZX-Calculus with measurements has a similar completeness result for
$\cat{ZX}^{\sground}/\ax{zx}^{\sground}$
\cite{Carette2019completeness} with an updated set of rules
$\ax{zx}^{\sground}$ which we will not give here for conciseness
purposes.

\footnotetext{The way we denote the equations relates to their names in broader literature if such exists, and more informal names if not. (S) is the \emph{spider} rule, (I$_g$) and (I$_r$) are the green and red \emph{identity} rules, (E) is the \emph{empty diagram} rule, (CP) the \emph{copy} rule, (B) the \emph{bialgebra} rule, (EU) the \emph{Euler angles} rule, (HD) the \emph{Hadamard decomposition} and (H) the \emph{Hadamard colour change} rule.}

Some properties of quantum protocols or algorithms can then be
verified by diagram transformations.  To give the reader the flavor of
such verifications, we detail the example of superdense coding.

\begin{example}[Superdense Coding]
  The idea of the superdense coding protocol is to transmit two
  classical bits using a single qubit. This is not possible in
  general, but it is when the two parties initially share an entangled
  pair of qubits. The protocol goes as follows:
  \begin{itemize}
  \item Alice and Bob initially share the (previously defined) EPR
    pair $\beta_{00}=\frac{\ket{00}{}+\ket{11}{}}{\sqrt2}$, and Alice
    moreover has two bits she wants to send to Bob
  \item Alice applies
    $\sigma_X = \begin{pmatrix}0&1\\1&0\end{pmatrix}$ to her qubit if
    her first bit is $1$, then
    $\sigma_Z=\begin{pmatrix}1&0\\0&-1\end{pmatrix}$ if her second bit
    is $1$
  \item Alice sends her qubit to Bob
  \item Bob applies a $\Cnot$ between his qubit and the one he
    received from Alice, then a $\Had$ gate on his qubit, and finally
    measures his two qubits
  \end{itemize}
  This protocol can be represented with a ZX-diagram as follows.
  \[
    
\begin{tikzpicture}
	\begin{pgfonlayer}{nodelayer}
		\node [style=none] (0) at (-0.425, 1) {};
		\node [style=none] (1) at (-1.925, 1) {};
		\node [style=rn] (2) at (-1.925, 0.75) {};
		\node [style=gn] (3) at (-1.925, 0.25) {};
		\node [style={{H box}}] (4) at (-3.175, 0.675) {};
		\node [style=gn] (5) at (-0.925, -0.5) {};
		\node [style=rn] (6) at (-0.425, -0.5) {};
		\node [style={{H box}}] (7) at (-0.925, -0.875) {};
		\node [style=gn] (8) at (-0.425, -1.25) {};
		\node [style=none] (9) at (-0.1, -1.525) {};
		\node [style=none] (10) at (-0.425, -1.75) {};
		\node [style=none] (11) at (-0.1, -1.7) {$\ground$};
		\node [style=gn] (12) at (-0.925, -1.25) {};
		\node [style=none] (13) at (-1.25, -1.525) {};
		\node [style=none] (14) at (-0.925, -1.75) {};
		\node [style=none] (15) at (-1.25, -1.7) {$\ground$};
		\node [style=none] (16) at (-2.425, 1.75) {};
		\node [style=none] (17) at (-3.175, 1.75) {};
		\node [style=none] (18) at (-1.425, 2) {};
		\node [style=none] (19) at (-1.425, -1) {};
		\node [style=none, anchor=west] (20) at (-1.175, 2) {Bob};
		\node [style=none, anchor=east] (21) at (-1.675, 2) {Alice};
		\node [style=gn] (22) at (-2.425, 1.375) {};
		\node [style=none] (23) at (-2.75, 1.2) {};
		\node [style=none] (24) at (-2.75, 1.025) {$\ground$};
		\node [style=gn] (25) at (-3.175, 1.375) {};
		\node [style=none] (26) at (-3.5, 1.2) {};
		\node [style=none] (27) at (-3.5, 1.025) {$\ground$};
		\node [style=none] (28) at (-0.35, 1) {};
		\node [style=none] (29) at (-2, 1) {};
		\node [style=none] (30) at (-2, 1.5) {};
		\node [style=none] (31) at (-0.35, 1.5) {};
		\node [style=none] (32) at (-3.75, 1.575) {};
		\node [style=none] (33) at (-2.25, 1.575) {};
		\node [style=none] (34) at (-3.75, 0.825) {};
		\node [style=none] (35) at (-2.25, 0.825) {};
		\node [style=none] (36) at (-0.25, -0.325) {};
		\node [style=none] (37) at (-0.25, -1.075) {};
		\node [style=none] (38) at (-0.125, -0.325) {};
		\node [style=none] (39) at (-0.125, -1.075) {};
		\node [style=none] (40) at (0.15, -1.075) {};
		\node [style=none] (41) at (0.15, -1.825) {};
		\node [style=none] (42) at (0.275, -1.075) {};
		\node [style=none] (43) at (0.275, -1.825) {};
		\node [style=none, color=orange, font={\scriptsize}] (44) at (-1, 0.75) {};
		\node [style=none, color=orange, font={\scriptsize}] (45) at (-2.225, -0.5) {};
		\node [style=none] (47) at (-2.025, 0.025) {};
		\node [style=none] (48) at (-1.65, 0.75) {};
		\node [style=none, color=orange, font={\scriptsize}, anchor=west] (49) at (0.025, -0.675) {\Cnot and \Had};
		\node [style=none, color=orange, font={\scriptsize}, anchor=west] (50) at (0.525, -1.425) {measurements};
		\node [style=none, color=orange, font={\scriptsize}, anchor=west] (51) at (-0.225, 1.325) {EPR pair};
		\node [style=none, color=orange, font={\scriptsize}, anchor=east] (52) at (-3.975, 1.325) {classical data};
		\node [style=none, color=orange, font={\scriptsize}] (53) at (0.025, 0.75) {application of $\sigma_X$};
		\node [style=none, color=orange, font={\scriptsize}] (54) at (-2.725, -0.75) {application of $\sigma_Z$};
		\node [style=none, color=orange, font={\scriptsize}] (55) at (-3.1, -0.5) {};
		\node [style=none] (56) at (-3.15, 0.25) {};
	\end{pgfonlayer}
	\begin{pgfonlayer}{edgelayer}
		\draw [in=90, out=-15] (8) to (9.center);
		\draw (9.center) to (11.center);
		\draw [in=90, out=-165] (12) to (13.center);
		\draw (13.center) to (15.center);
		\draw [bend left=90] (1.center) to (0.center);
		\draw (17.center) to (4);
		\draw [in=180, out=-90] (4) to (3);
		\draw [in=-180, out=-90, looseness=1.25] (16.center) to (2);
		\draw (1.center) to (3);
		\draw [in=120, out=-75, looseness=0.75] (3) to (5);
		\draw (0.center) to (6);
		\draw (5) to (14.center);
		\draw (6) to (10.center);
		\draw (6) to (5);
		\draw [style=dashed] (18.center) to (19.center);
		\draw [in=90, out=-165] (22) to (23.center);
		\draw (23.center) to (24.center);
		\draw [in=90, out=-165] (25) to (26.center);
		\draw (26.center) to (27.center);
		\draw [style=rule-box] (30.center) to (29.center);
		\draw [style=rule-box] (30.center) to (31.center);
		\draw [style=rule-box] (31.center) to (28.center);
		\draw [style=rule-box] (28.center) to (29.center);
		\draw [style=rule-box] (33.center) to (32.center);
		\draw [style=rule-box] (35.center) to (34.center);
		\draw [style=rule-box] (32.center) to (34.center);
		\draw [style=rule-box] (33.center) to (35.center);
		\draw [style=rule-box] (38.center) to (39.center);
		\draw [style=rule-box] (38.center) to (36.center);
		\draw [style=rule-box] (39.center) to (37.center);
		\draw [style=rule-box] (42.center) to (43.center);
		\draw [style=rule-box] (42.center) to (40.center);
		\draw [style=rule-box] (43.center) to (41.center);
		\draw [style={arrows={->[]}}, color=orange] (45.center) to (47.center);
		\draw [style={arrows={->[]}}, color=orange] (44.center) to (48.center);
		\draw [style={arrows={->[]}}, color=orange] (55.center) to (56.center);
	\end{pgfonlayer}
\end{tikzpicture}\colorbox{pink}{missing file : superdense-coding-details}}}

  \]
  It is then possible to verify that Bob eventually does get (copies
  of) Alice's bits, using the equational theory (although the whole
  derivation is not given here): \def\fig{superdense-coding}
  \begin{align*}
    \input{./figures/\fig/\fig_00.tikz}
    &~=~\input{./figures/\fig/\fig_01.tikz}
        ~=~\input{./figures/\fig/\fig_04.tikz}
  \end{align*}
  We can hence see that data is transmitted from Alice to Bob, without
  any loss.  Interestingly, this protocol can be extended for secure
  communication between the two parties \cite{Wang2005quantum}. The
  larger protocol uses instances of the smaller one to also check
  whether an eavesdropper has tried intercepting or copying data.

  If Bob is aware that the data he received was compromised, he can
  abort everything by simply discarding his qubits, so that the
  eavesdropper (Eve) gets absolutely no information:
  \def\fig{superdense-coding-attack}
  \begin{align*}
    \input{./figures/\fig/\fig_00.tikz}
    ~=~\input{./figures/\fig/\fig_03.tikz}
  \end{align*}
  where $U$ denotes an unknown operator applied by Eve to the qubit
  she intercepted.  Notice here how no information can pass from Alice
  to Eve. No information is retrieved by the latter.
\end{example}

A plethora of quantum protocols have been verified with ZX-Calculus
in a similar manner 
\cite{Hillebrand2011protocols}. Note however that the theory is ever
evolving, and in particular, $\ground$ was not introduced in the
language at that time, so the author had to use a trick to make up for
the absence of measurement (namely case-based reasoning).

\subsubsection{Algorithms and Tools}

It is possible to manipulate ZX-diagrams in a computer-verified
way. For instance, Quantomatic
\cite{kissinger2015quantomatic} allows the users to define
at the same time diagrams in graphical form and equational
theories. It is also possible to work with
user-defined nodes in the diagram so that  even though $\ground$ is
not part of the ``vanilla'' ZX-Calculus, it can be defined  as a
new node. It is then possible in the tool to manipulate diagrams in a
way that satisfies the equational theory, and even to define rewriting
strategies that can be then applied in an automated way. 

The verification of protocols and programs using the ZX-Calculus
relies on diagrammatic equivalence. This problem, in general, is at
least QMA-complete,
\cite{Bookatz2014QMA,Janzing2005non-identity-check} (the quantum
counterpart of NP-completeness).  This problem is linked to the one of
simplification/optimization, which asks how a quantum operator can be
simplified, given a particular metric (e.g.~the number of non-Clifford
gates). Indeed, for instance if $D_1$ and $D_2$ are two diagrams 
representing the same \emph{unitary}
(i.e.~$\stdinterp{D_1}=\stdinterp{D_2}$), then simplifying
$D_2^\dagger\circ D_1$ should ideally get us to the identity.

In the case of the Clifford fragment (obtained when the angles in

\begin{tikzpicture}
	\begin{pgfonlayer}{nodelayer}
		\node [style=gn] (0) at (0, 0) {};
		\node [style=glabel] (1) at (0.25, 0) {$\alpha$};
		\node [style=none] (2) at (0.25, 0.375) {};
		\node [style=none] (3) at (-0.25, -0.375) {};
		\node [style=none] (4) at (-0.25, 0.375) {};
		\node [style=none] (5) at (0.25, -0.375) {};
		\node [style=none] (6) at (0, 0.375) {$\overset{n}{...}$};
		\node [style=none] (7) at (0, -0.375) {$\underset{m}{...}$};
	\end{pgfonlayer}
	\begin{pgfonlayer}{edgelayer}
		\draw [in=90, out=-90] (2.center) to (3.center);
		\draw [in=90, out=-90] (4.center) to (5.center);
	\end{pgfonlayer}
\end{tikzpicture}\colorbox{pink}{missing file : gn-alpha}}}
 and 
\begin{tikzpicture}
	\begin{pgfonlayer}{nodelayer}
		\node [style=rn] (0) at (0, 0) {};
		\node [style=rlabel] (1) at (0.25, 0) {$\alpha$};
		\node [style=none] (2) at (0.25, 0.375) {};
		\node [style=none] (3) at (-0.25, -0.375) {};
		\node [style=none] (4) at (-0.25, 0.375) {};
		\node [style=none] (5) at (0.25, -0.375) {};
		\node [style=none] (6) at (0, 0.375) {$\overset{n}{...}$};
		\node [style=none] (7) at (0, -0.375) {$\underset{m}{...}$};
	\end{pgfonlayer}
	\begin{pgfonlayer}{edgelayer}
		\draw [in=90, out=-90] (2.center) to (3.center);
		\draw [in=90, out=-90] (4.center) to (5.center);
	\end{pgfonlayer}
\end{tikzpicture}\colorbox{pink}{missing file : rn-alpha}}}
 are restricted to multiples
of $\frac\pi2$), there exists a strategy that reduces the diagram in
(pseudo-)normal form \cite{Backens2014complete}. When this algorithm
terminates, the resulting diagram is of size $O(n^2)$ where $n$ is the
number of inputs and outputs in the diagram. The algorithm is
polynomial in the overall size of the diagram it is applied on.

Turning an arbitrary diagram into a normal form can be done in
principle \cite{Jeandel2019normalform}, however, the complexity of this
algorithm is EXPSPACE for universal fragments.  So this approach is
obviously not preferred in general.
However, one can use the ideas of the algorithm for the Clifford
fragment as a starting point to get a rewriting strategy for the
general case. Applications to quantum circuits and improvements on
this strategy can be found in the literature
\cite{Kissinger2020reducing,beaudrap2020fast,%
duncan2019graph,Backens2020ThereAB},
and implementations in the PyZX tool \cite{kissinger2019pyzx}. 
This tool can in particular be used to tackle circuit equivalence 
verification, using a different but related approach to that of 
Section~\ref{sec:path-sum-circuit} below.
The formalism used later is that of \emph{path-sums}, where morphisms 
were showed in \cite{Lemonnier2020linking,Vilmart2020SOP} to be 
essentially equivalent to ZX-diagrams, allowing us to apply strategies for
path-sums to the ZX-Calculus and vice-versa.
Next Section is devoted to we introducing the path-sum formalism and its use for circuits verification.

\subsection{Path-Sum circuit Equivalence Verification}
\label{sec:path-sum-circuit}

Path-sums are a recent alternative direction for verifying the equivalence between quantum
circuits ~\cite{dblp:phd/basesearch/amy19,amy2018towards}.  In this
section we briefly present it, together with the main verification
related achievements.

Note that a generalization of path-sum semantics is introduced in
Section~\ref{sec:qbricks}, for parametrized families of circuits.  For
sake of readability, conciseness and coherence with this further
content, in the coming paragraphs we slightly simplify path-sums
related notations.  We refer the reader
to the original definitions~\cite{dblp:phd/basesearch/amy19,amy2018towards} for the full formalism and underlying
mathematical structures.

\subsubsection{Semantical Model}

The standard semantics for quantum circuits is the matrix formalism,
introduced in Section~\ref{sec:quant-circ-semant}.  It
associates to each quantum circuit $C$ a matrix $\mathbf{Mat}(C)$
and it interprets the behavior of this circuit as a function
$\ket{x}{}\to \mathbf{Mat}(C)\cdot \ket{x}{}$ from kets to kets, where
$\cdot$ stands for the usual matrix product.

Notice that this standard semantics builds on an intermediary object--
the matrix--to derive and interprets the functional behavior of
circuits. And it does so by use of a higher-order function--the
matrix product. Contrarily, path-sums are a straight construction of
the \emph{input/output} function performed by circuits, enabling
compositional reasoning.

Concretely, a path-sum $\textit{PS}(x)$ is a quantum register state (a
ket), parametrized by an input basis ket $\ket{x}{}$ and defined as
the sum of kets
\large
\begin{equation}
  \label{ps:def}\textit{PS}(x) ~~{:}{:}{=}~
  \frac{1}{{\sqrt2}^{n}}\sum_{k=0}^{2^{n}-1}
  e^{\frac{2\cdot\pi\cdot{i}\cdot{P_k(x)}}{2^m}}\ket{\phi_k(x)}{}
\end{equation}
\normalsize
where the $P_k(x)$ are called \emph{phase polynomials} while the
$\ket{\phi_k(x)}{}$ are \emph{basis-kets}.  This representation is
{\it closed} under functional composition and Kronecker product. For
instance, if \large
\[\begin{array}{l@{~:~}l@{{}\mapsto{}}l}
    C & \ket{x}{} & PS(x) = \frac{1}{{\sqrt2}^{n}} \sum_{k=0}^{2^{n}-1}
                    e^{ 
                    \frac{ 2\cdot\pi\cdot{i}\cdot{P_k(x)} }{ 2^{m} }
                    } \ket{\phi_k(x)}{},
    \\
    C' & \ket{y}{} & PS'(y) = \frac{1}{{\sqrt2}^{n'}}
                     \sum_{k=0}^{2^{n'}-1} e^{ \frac{
                     2\cdot\pi\cdot{i}\cdot{P'_k(y)} }{ 2^{m'} } }
                     \ket{\phi'_k(y)}{},
  \end{array}
\]
\normalsize
then their parallel combination
\texttt{parallel}$(C, C')$ sends $\ket{x}{}\otimes\ket{y}{}$ to:
\large\begin{equation*}
  \frac{1}{{\sqrt2}^{n+n'}}
  \sum_{j=0}^{2^{n + n'}-1}
  e^{
    \frac{2\cdot\pi\cdot{i}\left(2^{m'}\cdot{P_{j/2^n}(x)} +
      2^m\cdot{P'_{j\%2^n}(y)}\right)}{2^{m+m'}}
  }
  \ket{\phi_{j/2^n}(x)}{}\otimes\ket{\phi'_{j\%2^n}(y)}{}
\end{equation*}
\normalsize
The sequential combination of  quantum circuits $C$ and $C'$ 
receives a similar compositional definition, parametrized by path-sums
components for circuits $C$ and $C'$.

\subsubsection{Path-Sums Reduction}

While path-sums compose nicely, a given linear map (eg. the
input/output function for a quantum circuit) does not have a unique
representative path-sum. Hence, given two different path-sum, how to decide whether they both encode the behavior of a given circuit? To tackle this problem, an equivalence
relation is defined with a few, simple rules that can be oriented. As an example, the \emph{HH} rule enables to simplify a path-sum expression over the reduction from  a sequence of two consecutive Hadamard gates to the identity--see Figure~\ref{fig:hh_rule}. All
these rules transform a path-sum into an equivalent one, with a lower
number of \emph{path variables} (parameter $n$ in the notation of
Equation \ref{ps:def}). 

\large
\begin{figure}[!h]
    \centering
\begin{prooftree}
      \AxiomC{ $PS(x) = \frac{1}{{\sqrt2}^{n+1}}
  \sum_{j=0}^{2}
  \sum_{k=0}^{2^n}
  e^{
    \frac{1}{2}\cdot\pi\cdot{i}\left(\frac{1}{2}  j(k_i + Q_k(x))+ P_k(x)\right)}\ket{\phi_{k}(x)}{}$}
      \RightLabel{$HH$}
      \UnaryInfC{$PS(x) =\frac{1}{{\sqrt2}^{n+1}}
  \sum_{k=0}^{2^n}
  e^{
    \frac{1}{2}\cdot\pi\cdot{i}\left( P[i:=Q_k(x)]_{k}(x)\right)}\ket{\phi[i:=Q_k(x)]_{k}(x)}{}$}
    \end{prooftree}
    \caption{The $HH$ path-sum transformation rule}
    \label{fig:hh_rule}
\end{figure}
\normalsize

We refer the desirous reader to ~\cite{amy2018towards,dblp:phd/basesearch/amy19} for an exhaustive exposition of the corresponding proof system. It was proved strongly normalizing, meaning
that every sequence of reduction rules application terminates with an
irreducible path-sum. Furthermore, finding and applying such a normalizing
sequence is feasible in time polynomial in the width of the
circuit at stake, which makes the overall reduction procedure tractable.

\subsubsection{Verified Properties}
\label{sec:ps_verified}
Hence, path-sums provide a human-readable formalism for the
interpretation of quantum circuits as ket data functions. Furthermore,
it is given a polynomial normalization procedure, based on a
restricted set of rewriting rules. The method was probed against both
circuit equivalence and functional specifications verification. More
precisely:

\begin{description}
\item[Translation validation] consists, for a given quantum algorithm,
  in (1) computing the path-sums for both a non-optimized and an
  optimized circuit realization and (2) using the normalization
  procedure for the automatic checking of their equivalence. It was
  performed on various quantum routine instances (Grover, modular
  adder, Galois field multiplication, \textit{etc}) of various size
  (up to several dozens of qubits). Interestingly, the methods proved
  as efficient for identifying non-equivalence (over erroneous
  instances) as for checking equivalence.

\item[Quantum algorithms verification] consists in verifying whether a
  given circuit instance respects its functional description.  It was
  performed for instances of similar case studies as for the
  translation validation (QFT, Hidden Shift~\cite{van2006quantum}),
  with up to a hundred qubits.
\end{description}

As conclusion, the path-sum formalism provides a fully automatized
procedure for verifying the equivalence between two circuits.
Hence, given two quantum circuits, the latter being a
supposed optimized version of the former, path-sums treatment enables
us to verify that they implement the same quantum function. Since
path-sums perform internal complexity reduction, an open direction is
a search for  efficient heuristics extracting an optimized quantum circuit from reduced
path-sums. As mentioned in Section~\ref{sec:zx}, this problem is closely linked to the reduction of ZX-Calculus diagrams. So, in its general form, it faces the same complexity limitations. The search for efficient reduction procedure applying to identified useful fragments could then both benefit from and feed advances in the ZX setting.

In its present state of development, using path-sums for quantum
algorithm verification is restricted to compilation time, when program
parameters are instantiated. Furthermore, a new run of the path-sum
reduction is required at each new call of a given quantum function. In
Section~\ref{sec:qbricks} we introduce the \qbricks language, whose
semantics is based on a parametrized extension of path-sums. It
enables the verification, once for all, of parametrized programs,
holding for any possible future parameter instances--yet at the
price of full automation as the manipulation of parametrized path-sums
requires first-order logic reasoning.

\subsection{Quantum abstract interpretation}

The techniques presented so far target an exhaustive functional description of a quantum circuit. Because of the intrinsic complexity of quantum computing, their use is limited, in particular by the size of  circuits. As illustrated in the benchmarks summed up in Section~\ref{sec:ps_verified}, path-sums reduction enabled, for instance, the verification of quantum circuits up to a hundred of qubits.

To push the boundary, a possible strategy comes from Abstract Interpretation (see Section~\ref{sec:fm_zoo}). There, one does  not target an exhaustive description 
of the functional behavior of a circuit, but an over-approximation of it. Such a framework relies on:
\begin{itemize}
    \item the identification of a conveniently structured set of properties of interest (the \emph{abstract domain});
    \item sound abstraction and concretization to/from this abstract domain;
    \item  reasoning tools for the abstract domain.
\end{itemize}

In~\cite{perdrix2008quantum}, the author introduces such an abstract interpretation of quantum states based on their entanglement structure. The technique enables to identify mutually separable sub-registers. It is useful, for example, for uncomputation, qubit discarding or  identifying a convenient decomposition for further analysis of a state. 

A more recent development on abstract interpretation applied to quantum process is~\cite{yu2021quantum}, in which the authors define abstract domains made of tuples of partial projections over quantum sub-registers. Intuitively, the idea is to overcome the exponential complexity of quantum states by decomposing them into sub-spaces. Interestingly, the method was implemented and evaluation results are provided. In particular, thus abstraction enabled to characterize the invariant in the main loop structuring the Grover search algorithm and to prove it for instances of width up to 300 qubits.

\subsection{Toward Integrated Verified Optimization: \voqc}
\label{sec:voqc}
A noticeable effort for an integrated verified quantum optimization
was recently led through the development of a \emph{Verified Optimizer
  for Quantum Circuits} (\voqc--pronounced `vox')
~\cite{hietala19:verif_optim_quant_circuit}. As main aspects, in
comparison to ZX- and path-sum calculus, VOQC:
\begin{itemize}
\item is integrated into a core programming environment and applies on
  circuits issued from parametrized programs;
\item not only validates the equivalence between an input quantum
  circuit and a candidate optimized version of it, but also provides
  the formally verified optimization procedure, directly generating
  this optimized version.
\end{itemize}

\begin{figure}[tbh]
    \centering
    \begin{tikzpicture}
\begin{scope}[yshift=2.5cm, xscale =1.1]
              \node(ss)[draw,rectangle ] at (0,0){\begin{tabular}{c}
               source  \\\sqir\\
           circuit
      \end{tabular}};
      
      \node(comp)[draw, ellipse] at (4,0){\begin{tabular}{c}
\voqc             \\ optimizers\\circuit mapper
            
      \end{tabular}};
      
      \node(ts)[draw,rectangle ] at (8,0){\begin{tabular}{c}
           target  \\\sqir\\
           circuit
      \end{tabular}};
      \end{scope}

\begin{scope}[xscale =1.1]
      
      \node(so)[draw,rectangle ] at (0,0){\begin{tabular}{c}
           source  \\ \oqasm\\
           circuit
      \end{tabular}};
      
      \node(to)[draw,rectangle ] at (8,0){\begin{tabular}{c}
           target  \\ \oqasm\\
           circuit
      \end{tabular}};
\end{scope}
\draw[very thick,-stealth](so)--(ss);      
\draw[ very thick,-stealth](ss)->(comp);      
\draw[ very thick,-stealth](comp)->(ts);      
\draw[ very thick,-stealth](ts)->(to);

    \end{tikzpicture}
    \caption{A simplified view of \voqc architecture}
    \label{fig:voqc_arch}
\end{figure}
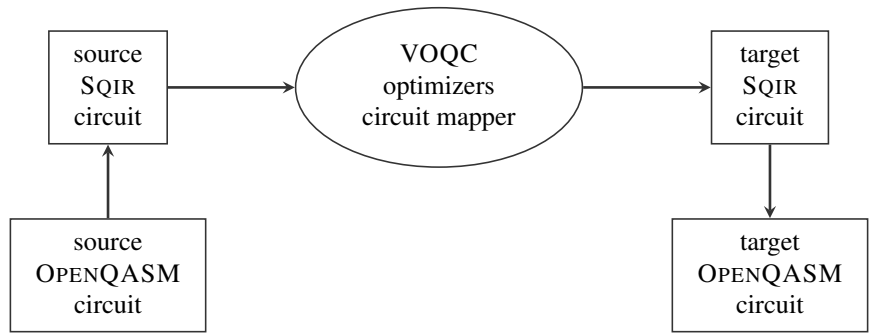

\subsubsection{Architecture}

In Figure~\ref{fig:voqc_arch} we give a simplified view of \voqc
architecture: it relies on an Intermediate Representation Language named
\sqir. Since it may also be used as a verified programming
environment, \sqir is introduced and developed \emph{per se} in a
dedicated section below (Section~\ref{sec:sqir}). In the present
context, we can think of it as a core language generating parametrized
quantum circuits. Then, for any instance of the parameters, \voqc
extracts the corresponding sequence of operations, applies an
optimization procedure upon this sequence of operations and extracts
back an optimized proved equivalent \sqir quantum circuit.
Furthermore,
\begin{itemize}
\item in addition to optimization, \voqc also contains some circuit
  mapping functionalities. They perform further circuit transformation so that the output circuits
  fit to specific quantum architecture qubit connectivity
  constraints. This functionality is a preliminary address to the problem raised as first bullet in
  Section~\ref{sec:compilation},
\item \voqc environment also provides both ways compilations between
  \sqir and the standard assembly language
  \oqasm~\cite{cross2017open}. Hence, it opens the way for a modular easy
  integration in any standard programming environment, in particular
  \qiskit~\cite{QiskitCommunity2017}, which uses \oqasm as
  assembly language.
\end{itemize}

\subsubsection{Optimization Procedure}
\voqc\ optimization process provides two functionalities, one is
deterministic (optimization by propagation and cancellation) and the
other one requires a replacing circuit input.

\paragraph*{Optimization by propagation and cancellation} is based on
local circuit rewriting schemes and self composition properties of
elementary gates, borrowed from \cite{nam2018automated} (eg., 
sequences of an even number of either \Had, \Cnot\ or X gate annihilate as the
identity, successive occurrences of R$_z$ gates melt by summing their
angle parameters, \textit{etc}). Hence, the procedure consists of two
successive steps:
\begin{itemize}
\item propagate: for any elementary gate, it 
\begin{itemize}
    \item considers several
  identified patterns enabling gate commutations,
\item finds all
  occurrences of these patterns,  
  \item and performs the related commutation, pushing any
  occurrence of this elementary gate to the end of the computation,
\end{itemize}\item cancellation then consists in deleting the resulting repetitive
  occurrences of the elementary gate at stake.
\end{itemize}

\paragraph*{Optimization by circuit replacement} consists in
substituting a part of a quantum circuit (a subcircuit) by another one
that is proven to be functionally equivalent. In this case, the
equivalence proof is led by  help of the path-sums semantics (see
Section~\ref{sec:path-sum-circuit}). Hence, in its most general form, the process requires an external equivalence proof oracle. Nevertheless, \voqc provides some instances of such proved equivalence patterns (eg, rotation merging), whose application is automatic.

\paragraph*{ Performance and Achievements.}  
\voqc performance has been evaluated against several standard
quantum computation routines, and compared with several existing
optimizers~\cite{amy2014polynomial,QiskitCommunity2017,%
nam2018automated,sivarajah2020t,kissinger2019pyzx}.
Note that in \voqc, since optimization is performed as a succession of
rewriting operations, the formal verification consists in assessing
the equivalence between the input and the output of the optimization,
it does not address the optimization performance.
Still, on reported experiments, \voqc performance competes with other
existing non-verified optimizers, both regarding computation time and
circuit complexity reduction--precise performance comparison tables
appear in~\cite{hietala19:verif_optim_quant_circuit}. Thus, in the
current state of the art, the benefits of formally verified circuit
optimization comes for free.

\subsection{Formally verified quantum compilation in an imperative setting: CertiQ}

CertiQ~\cite{shi2019certiq} is
another noticeable effort for  verified compilation. Interestingly, it applies to \qiskit language, which is certainly \emph{the most widely used quantum compiler}.  CertiQ proceeds similarly as \voqc, by applying successive optimization \emph{passes}, each consisting  in applying local circuit equivalence transformations over quantum circuits. CertiQ has been  evaluated against the very compilation environment provided by IBM \qiskit framework. And verified optimization went through 26 out of the 30 compilation passes the framework offered at writing time. Hence, the current limitations of formally verified compilation  is mainly inherited from the state of the art in  (non verified) quantum compilation.

The application to \qiskit also inevitably brings an additional drawback: while CertiQ can bring insurance that compilation respects the functional equivalence between an input and an output quantum circuits, the initial circuit building brick still lacks formal verification. Indeed,
the environment is not provided formal means to assess that the input circuit meets a given functional specification. An open work direction then is to develop formal circuit building verification methods similar as \sqir or \qbricks but applying to (mostly imperative) widely used development environments such as \qiskit, so as to complete the verified development chain. 

\section{Formal Quantum Programming Languages}
\label{sec:formal_qpl}

For reasoning  on a
concrete \emph{runs} of quantum algorithms  solving a specific problem
\emph{instances}, the notion of quantum circuit is  natural.
 Yet, a quantum circuit is only  a by-product of a specific run of a quantum algorithm, holding only for a specific instantiation of the algorithm parameters. 
 
 Hence, 
 a quantum algorithm is not reducible to a quantum
circuit.
To run  quantum algorithms and reason about them, one needs \emph{quantum
  programming languages} (QPL): this is the topic of the current section.

\subsection{Quantum Programming Languages Design}
\label{sec:qpl-design}

In Section~\ref{sec:low-level-verif}, circuits were merely seen as
sequences of elementary gates. However, in most quantum algorithms
circuits follow a more complex structure: they are built
compositionally from smaller sub-circuits and circuits combinators. 
Circuits are usually \textit{static} objects, buffered until completion
before being flushed to the quantum co-processor. Still, in some algorithms,
they are \emph{dynamically} generated: the tail of the circuit
depends on the result of former measurements.

In this section, we discuss the high-level structure of quantum
algorithms, the requirements for a quantum programming language, and
review some of the existing proposals.

\subsubsection{Structure of Quantum Algorithms}
\label{sec:struct-quantum-algo}
\begin{figure}[tbh]
  \begin{subfigure}{0.46\textwidth}
    \centering
    \fbox{\includegraphics[scale=.8]{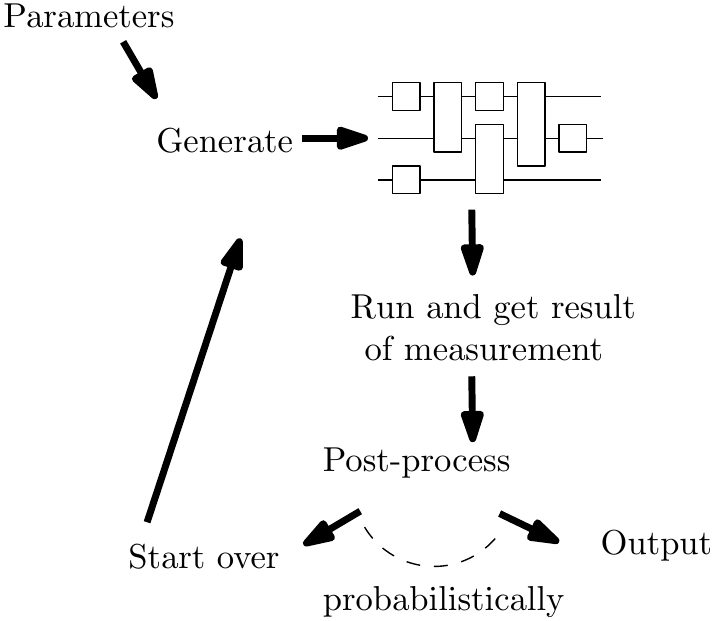}}
    \caption{Static scheme}
    \label{fig:wf-static}
  \end{subfigure}
  \hfill
  \begin{subfigure}{0.46\textwidth}
    \fbox{\includegraphics[scale=.8]{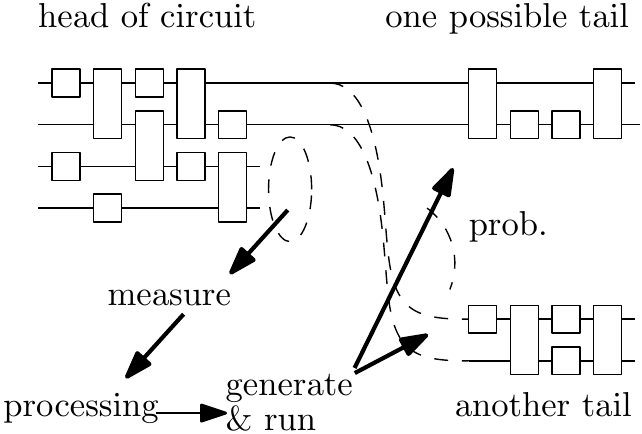}}
    \caption{Dynamic scheme}
    \label{fig:wf-dyn}
  \end{subfigure}
  \caption{Workflows for quantum algorithm}
  \label{fig:circ-workflow}
\end{figure}

The usual model for quantum computation was depicted in
Figure~\ref{qram}: a classical computer controls a quantum
co-processor, whose role is to hold a quantum memory. A programmatic
interface for interacting with the co-processor is provided to the
programmer sitting in front of the classical computer. The interface
gives methods to send instructions to the quantum memory to allocate
and initialize new quantum registers, apply unitary gates on qubits,
and eventually perform measurements. Even though the set of instructions is
commonly represented as a circuit, it is merely the result of a
\emph{trace of classical execution} of a classical program on the
classical computer.

Figure~\ref{fig:circ-workflow} presents two standard workflows with a
quantum co-processor. In Figure~\ref{fig:wf-static}, the classical
execution inputs some (classical) parameters, performs some
pre-processing, generates a circuit, sends the circuit to the
coprocessor, collects the result of the measurement, and finally
performs some post-processing to decide whether an output can be
produced or if one needs to start over. Shor's factoring
algorithm~\cite{shor1995scheme} or Grover's
algorithm~\cite{grover1996fast} fall into this scheme: the circuit is
used as a fancy probabilistic oracle. Most of the recent variational
algorithms~\cite{mcclean2016theory} also fall into this scheme, with
the subtlety that the circuit might be updated at each step.  The
other, less standard workflow is presented in
Figure~\ref{fig:wf-dyn}. In this scheme, the circuit is built ``on the
fly'', and measurements might be performed on a sub-part of the memory
along the course of execution of the circuit. The latter part of the
circuit might then depends on the result of classical processing in
the middle of the computation.

Understanding a quantum circuit as a by-product of the execution of
a classical program shines a fresh view on quantum algorithms: it cannot
be identified with a quantum circuit. Instead, in general, at the very
least a quantum algorithm describes a \emph{family} of quantum
circuits. Indeed, consider the setting of
Figure~\ref{fig:wf-static}. The algorithm is fed with some parameters
and then builds a circuit: the circuit will depend on the shape of the
parameters. If for instance, we were using Shor's factoring algorithm,
we would not build the same circuit for factoring 15 or
114,908,028,227. The bottom line is that a quantum programming language
should be able to describe \emph{parametrized} families of circuits.

\begin{figure}[tbh]
  \centering
  \includegraphics[scale=.8]{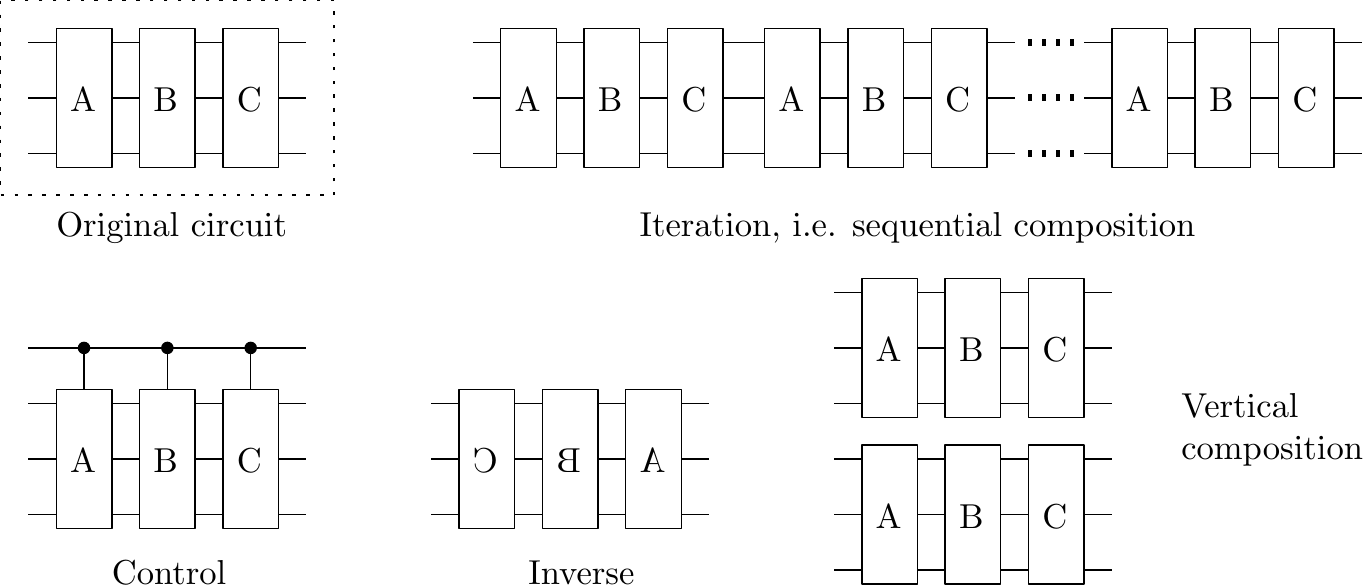}
  \caption{Standard Circuit Combinators}
  \label{fig:hl-circ-comb}
\end{figure}

\begin{figure}[tbh]
  \centering
  \includegraphics[scale=.8]{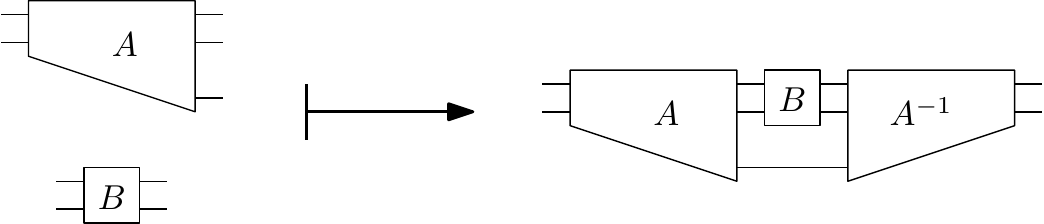}
  \caption{Example of a derived circuit combinator}
  \label{fig:hl-circ-comb-der}
\end{figure}

The circuits described by quantum algorithms are potentially very
large--a concrete instance of the HHL
algorithm~\cite{harrow2009quantum} for solving linear systems of
equations has been shown~\cite{scherer2017concrete} to count as much
as $\sim 10^{40}$ elementary gates, if not optimized. Unlike the
circuit-construction schemes hinted at in
Section~\ref{sec:low-level-verif}, this circuit is not uniquely given
as a list of elementary gates: it is built from sub-circuits
--possibly described as a list of elementary gates but not only--and
from high-level circuit combinators. These combinators build a circuit
by (classically) processing a possibly large sub-circuit. Some
standard such combinators are shown in Figure~\ref{fig:hl-circ-comb}
(where we represent inverse with reflected letters). There is a distinction
to be made between the combinator, applied on a sub-circuit, and its
semantics, which is an action on each elementary gate. Combinators are
abstractions that can be composed to build larger combinators, such as
the one presented in Figure~\ref{fig:hl-circ-comb-der}, built from
inversion, controlling and sequential composition.

\subsubsection{Requirements for Quantum Programming Languages}
\label{sec:qpl-req}

Any scalable quantum programming language should therefore allow the
following operations within a common framework.
\begin{itemize}
\item Manipulation of quantum registers and quantum circuits as
  first-class objects. The programmer should both be able to refer to
  ``wires'' in a natural manner and handle circuits as independent
  objects.
\item Description of \emph{parametric families} of quantum circuits,
  both in a procedural manner as a sequence of operations--gates or
  subcircuits--and in an applicative manner, using circuit
  combinators;
\item Classical processing. In our
  experience~\cite{green2013quipper}, quantum algorithms mostly
  consists of classical processing--processing the parameters,
  building the circuits, processing the result of the measurement.
\end{itemize}
This broad description of course calls for refinements. For instance
some of the classical processing might be performed on the quantum
co-processor--typically the simple classical controls involved in
quantum error correction. The level of classical processing performed
on the classical computer and performed on the quantum co-processor is
dependent on the physical implementation. Even though some recent proposals
such as Quingo~\cite{quingo} discuss the design of quantum programming
languages aware of the two levels of classical processing--in and
out of the co-processor, this is still a work in progress.

\subsubsection{Review of the Existing Approaches}
\label{sec:rev_of_ex_app}
Most of the current existing quantum programming languages 
follow the requirements discussed in Section~\ref{sec:qpl-req}.
In this section, we review some typical approaches followed both in
academic and industrial settings. This review is by no means meant to
be exhaustive: its only purpose is to discuss the possible strategies
for the design of QPLs.

When designing a realistic programming language from scratch, the main
problem is access to existing libraries and tools. In the context
of quantum computation, one would need for instance to access the
file system, make use of specific libraries such as Lapack or BLAS,
\textit{etc}. One can also rely on the well-maintained and optimized
compiler or interpreter of the host language. To quickly come
up with a scalable language, the easiest strategy consists in
\emph{embedding} the target language in a host language. Indeed, a
quantum programming language can be seen as a domain-specific language
(DSL), and it can be built over a regular language.

Even though the advantages of working inside a host language are clear, there
are two main drawbacks, The first one is the potential rigidity of the
host language: there might be constructs natural to the DSL that are
hardly realizable inside the host language. The second drawback has to
do with the compilation toolchain: the shallow embedding of the DSL
makes it impossible to access its abstract syntax tree, rendering
specific manipulation thereof impossible.

\paragraph*{Embedded QPLs.}
The first scalable embedded proposal is
\quipper~\cite{green2013quipper,green2013introduction}. Embedded in
Haskell, it capitalizes on \emph{monads} to model the interaction with
the quantum co-processor. \quipper's monadic semantics is meant to be
easily abstracted and reasoned over: it is the subject of
Section~\ref{sec:monsem}. Since \quipper, there has been a steady
stream of embedded quantum programming languages, often dedicated to a
specific quantum co-processor or attached to a specific vendor, and
mostly in Python: \qiskit~\cite{QiskitCommunity2017} and
\projectq~\cite{projectq} for IBMQ, CirQ~\cite{CirqDevelopers2018} for Google,
Strawberry Fields~\cite{strawberry-fields} for Xanadu,
AQASM for Atos, \textit{etc}. From a language-design point of
view, most of these approaches heavily rely on Python objects to
represent circuits and operations: the focus is on usability and
versatility rather than safety and well-foundedness.

\paragraph*{Standalone QPLs.}
On the other side of the spectrum, some quantum programming languages have
been designed as standalone languages, with their own parser and
abstract syntax tree. Maybe the first proposed scalable
language was Ömer's QCL~\cite{omer2003structured}. Ömer experimented
several features such as circuit-as-function, automatic inversion and
oracle generation. However, due to its non-modular approach, the
language did not have successors.

\liquid~\cite{wecker2014liqui} and its sequel Q\#
\cite{svore2018q}  developed by Microsoft are good examples of an
attempt at building a standalone language while keeping a tight link
with an existing programming environment, as  \liquid and Q\# are tightly linked with
the F\# framework (itself embedded in the  whole .NET framework), making it possible to easily ``reuse'' library
functions from within a Q\# piece of code. On the other hand, Q\# has
its  own syntax and type system, to capture run-time errors specific to
quantum computation.

ScaffCC~\cite{javadiabhari2015scaffcc} is another example of a
standalone QPL.  Even though the language is rather low-level its compiler has
been heavily optimized and experimented over, and it serves as a support
for a long stream of research on quantum compiler optimizations.

The last noteworthy language to cite in the series is \silq~\cite{silq},
as it serves as a good interface with the next paragraph: aimed at
capturing most of the best practices in terms of soundness and safety,
it is nonetheless targeted toward usability.

\paragraph*{Formal QPLs.}
The last line of works on QPLs we would like to mention here is
formal languages aimed at exploring and understanding the design
principles and the semantics of quantum algorithms. We shall be brief
as this is the topic of the remainder of this survey.  The initial
line of work was initiated by Selinger~\cite{selinger2004} with the
study of a small flow-chart language with primitive constructs to
interact with the quantum co-processor: qubit initialization,
elementary gate application and measurement. This language was later
extended to a simple Lambda-Calculus with similar primitive quantum
features~\cite{10.1007/11417170_26,Selinger2009}. Even though the language is not aimed at
full-scale quantum algorithms, it is nonetheless enough to serve as a
testbed for experimenting type systems and many
operational~\cite{10.1007/11417170_26,diaz-caro2017density,lago2017geometry}
and
denotational~\cite{10.1145/2535838.2535879,valiron2008categorical,Diaz-CaroM19,HASUO2017404}
semantics.

The study of formal QPLs took a turn toward circuit-description
languages \emph{à la} \quipper with the development of scalable quantum
languages. 
One of the first proposals of formalization is
\qwire~\cite{paykin2017qwire}, embedded in the proof assistant
\coq. \qwire uses Coq expressive system to encode the sophisticated
typing rules of \qwire. In a sense, \coq type system is expressive
enough to use \coq as a host language \emph{and} still be able to
manipulate the abstract syntax tree of a program.  The main design
choice for \qwire is to separate pure quantum computation with its
constraints such as no-cloning, from classical computation.

Albeit disconnected from \coq, the formalization of \quipper has
followed a similar root. This development is based on the formal
language \protoquipper~\cite{ross2015algebraic}, which extracts the
critical features out of \quipper: the creation and manipulation of
circuits using a minimal Lambda-Calculus. The language is equipped
with a linear type system and a simple operational semantics based on
circuit construction. The simple core proposed by \protoquipper
has stirred a line of research on the topic, including the
formalization of inductive datatypes, recursion and dependent types in this
context~\cite{Rios2017ACM,Rios2021,lindenhovius2018enriching,fu2020linear}.

The last class of formal programming language we want to mention
focuses on the specification and verification of high-level properties
of programs, and are solely based on circuit manipulation: unlike
\qwire or \quipper, qubits are not first-class objects and circuits
are simple ``bricks'' to be horizontally or vertically stacked. In
this class of languages, one can mention qPCF~\cite{paolini2017qpcf},
mainly a theoretical exploration of dependent type systems in this
setting, and \qbrick, presented in Section~\ref{sec:qbricks}.

\subsection{Formalizing the Operational Semantics}
\label{sec:op-sem}

In order to reason on quantum programming languages, one needs to have
a formal understanding of their operational semantics.

\subsubsection{Quantum Lambda-Calculi}
\label{sec:qlc}

\newcommand{\ifterm}[3]{{\tt if}\,{#1}\,{\tt then}\,{#2}\,{\tt else}\,{#3}}

The Lambda-Calculus~\cite{barendregt84lambda} is a formal language 
encapsulating the main property of higher-order functional languages:
functions are first-class citizens that can be passed as arguments to
other functions. Lambda-Calculus features many extensions to model
and reason about side-effects such as probabilistic or non-deterministic
behaviors, shared memory, read/write, \textit{etc}.

One of the first formal proposals of a quantum, functional language has
precisely been a \emph{quantum} extension of
Lambda-Calculus~\cite{10.1007/11417170_26}. On top of the regular
Lambda-Calculus constructs, the \emph{quantum Lambda-Calculus} features
constants to name the operations of qubit initialization, unitary maps
and measurements. A minimal system consists of the following terms: 
\[
  \begin{array}{l@{}l}
    M,N\quad{:}{:}{=}\quad{}
    &
      x\mid
      \lambda x.M
      \mid
      M\,N
      \mid\\
    &
      \ttrue
      \mid
      \ffalse
      \mid
      \ifterm{M}{N_1}{N_2}
      \mid\\
    &
      {\tt qinit}\mid{\tt U}\mid{\tt meas}.
  \end{array}
\]
Terms are represented with $M$ and $N$, while variables $x$ range over
an enumerable set of identifiers. The term $\lambda x.M$ is an
\emph{abstraction}: it stands for a function of argument $x$ and of
body $M$. The application of a function $M$ to an argument $N$ is
represented with $M\, N$. To this core Lambda-Calculus, we can add
constructs to deal with booleans: $\ttrue$ and $\ffalse$ are the
boolean constant values, while $\ifterm{M}{N_1}{N_2}$ is the usual
test. Finally, ${\tt qinit}$ stands for qubit initialization,
${\tt meas}$ for measurement and ${\tt U}$ ranges over a set of
 unitary operations. These three constants are functions: for
instance, ${\tt qinit}~\ttrue$ corresponds to $\ket{1}{}$ and
${\tt qinit}~\ffalse$ to $\ket{0}{}$, while ${\tt meas}$ applied to a
qubit stands for the measurement of this qubit. A fair coin can then
be represented by the term 
\begin{equation}\label{eq:faircoin}
  \texttt{meas}\,({\tt H}\,(\texttt{qinit}~\ttrue)),
\end{equation}
where ${\tt H}$ stands for the Hadamard gate.

The question is now: how do we formalize the evaluation of a piece of
code? In the regular Lambda-Calculus, evaluation is performed with
\emph{substitution} as follows:
\[
  (\lambda x.M)\,N \to M[x:=N]
\]
where $M[x:=N]$ stands for $M$ where all free occurrences of $x$
--i.e., those corresponding to the argument of the function--have
been replaced by $N$. Even though we can still require such a rule in the
context of the quantum Lambda-Calculus, this does not say how to deal
with the term $\texttt{qinit}~\ttrue$.

In order to give an operational semantics to the Lambda-Calculus, a
naive idea could be to add yet another construction: a set of
constants $c_{\ket{\phi}{}}$, once for every possible qubit state
$\ket{\phi}{}$. If--as shown by van Tonder~\cite{tonder04lambda}--
this can somehow be made to work, a more natural presentation consists
in mimicking the behavior of a quantum co-processor, in the style of
Knill's QRAM model~\cite{knill1996conventions}: we define an
\emph{abstract machine}
\(
  (\ket{\phi}{n}, L, M)
\)
consisting of a finite memory state $\ket{\phi}{n}$ of $n$ qubits, a
term $M$ with $n$ free variables $x_1,\ldots,x_n$, and a \emph{linking
  function} $L$, bijection between $\{x_1,\ldots,x_n\}$ and
the qubit indices $\{1,\ldots,n\}$. Variables of $M$ captured by $L$
are essentially pointers to qubits standing in the quantum memory.
The fair coin of Eq.~\eqref{eq:faircoin} then evaluates as follows.
\[
  \begin{array}{l}
    (\ket{}{0},\{\},\texttt{meas}\,({\tt
    H}\,(\texttt{qinit}~\ttrue)))
    \\
    \quad\to
    (\ket{1}{1},\{x\mapsto 1\},\texttt{meas}\,({\tt H}\,x))
    \\
    \quad\to
    (\frac1{\sqrt2}(\ket{0}{1}-\ket{1}{1}),\{x\mapsto
    1\},\texttt{meas}\,x)
    \\
    \quad\to
    \left\{
    \begin{array}{ll}
      (\ket{0}{1},\{\},\ffalse) & \text{with prob. $0.5$}
      \\
      (\ket{1}{1},\{\},\ttrue) & \text{with prob. $0.5$.}
    \end{array} 
                                 \right.
  \end{array}
\]
In this evaluation, most of quantum computation has been exemplified:
initialization of qubits, unitary operations and
measurements. Handling the latter in particular requires a
\emph{probabilistic evaluation}, and this requires some care--we
invite the interested reader to consult
for example Selinger \& Valiron~\cite{10.1007/11417170_26} for details.

\subsubsection{Monadic Semantics}
\label{sec:monsem}

The operational semantics of the quantum Lambda-Calculus is very
limited. Indeed, as discussed in Section~\ref{sec:qpl-design}, quantum
algorithms do not in general send operations one by one to the quantum
co-processor: instead, a quantum program must build circuits (or
pieces thereof) before sending them to the co-processor as batch
jobs. The quantum Lambda-Calculus does not allow to build circuits:
operations can only be sent one at a time. In particular, there is no
possibility to create, manipulate and process circuits: circuit
generation in the quantum Lambda-Calculus is a \emph{side-effect} that
is external to the language. One cannot interfere with it, and
embedding the quantum Lambda-Calculus as it stands inside a host
language, as suggested in Section~\ref{sec:rev_of_ex_app}, would not
help.

The solution devised by \quipper consists in relying on a special
language feature from Haskell called \emph{monad}. A monad is a type
operator encapsulating a side effect. Consider for instance a
probabilistic side effect. There are therefore two classes of terms:
terms without side-effect, with types e.g. {\tt Bool}, or {\tt Int},
and terms with side-effect, with types e.g. {\tt P(Bool)} or {\tt
  P(Int)} standing for ``term evaluating to a boolean/integer, possibly with a
probabilistic effect''. The
operator {\tt P(-)} captures the probabilistic side effect.

A monad comes with two standard maps. In the case of {\tt P} we would
have:
\begin{verbatim}
return :: A -> P(A)
bind :: P(A) -> (A -> P(B)) -> P(B)
\end{verbatim}
The {\tt return} operation says that an effect-free term can be
considered as having an effect--in the case of the probabilistic
effect, it just means ``with probability 1''. 
The  {\tt
  bind} operation\footnote{In Haskell, this map is denoted with {\tt >>=}. For
  the sake of legibility, here we denote it with {\tt bind}.}
says how to compose effectful operations: given a function
inputting {\tt A} and returning an object of type
{\tt B} with a probabilistic
side-effect, how to apply this function to a term of type {\tt A} also
having a probabilistic effect? We surely get something of type {\tt
  P(B)}, but the way to construct it is described by {\tt bind}.
A few equations have to be satisfied by {\tt return} and {\tt bind}
for them to describe a monad. For instance, {\tt bind~return} is the
identity on {\tt P(A)}. There can of course be more operations: for
instance, we can add to the signature of {\tt P} an operator {\tt
  coin} of type {\tt () -> P(Bool)}\footnote{In Haskell, the unit
  type is denoted with {\tt ()}.}.

A nice property of monads is that effectful operations can be written
with syntactic sugar in an imperative style:
\begin{verbatim}
do
  x <- coin ()
  if x then return 0 else return 1
\end{verbatim}
is a term of type {\tt P(Int)} equals to
\begin{center}
  \texttt{bind~(coin ())~ ($\lambda$\,x.if x then return 0 else return 1)}
\end{center}
once the syntactic sugar has been removed.

Following this approach, quantum computation can be understood as
side-effect: it combines both  
(1) Read/Write effect, since gates are
sent to the coprocessor, and results of measurements are received; 
(2)
Probabilistic effects, since measurement is a probabilistic
operation.   
The first attempt at formalizing this monad is Green's
quantum IO monad~\cite{altenkirch2010quantum}: it has then been
further developed in \quipper~\cite{green2013quipper}.

Internally, circuits in \quipper are represented using a simple inductive datatype akin to a list of gates\footnote{Technically a tree structure, as measurements entails branching.}. The interaction with the quantum-coprocessor is modeled using a specific I/O monad {\tt Circ}. This monad encapsulates the construction of circuits featuring wires holding qubits but also wires holding bits ---results of measurements. A bit is however uniquely useable ``inside'' the monad: to use it in Haskell ---in an if-then-else construct for instance--- we need to ``lift'' it into a regular Boolean. The signature of the monad in
particular includes
\begin{verbatim}
qinit :: Bool -> Circ(Qubit)
measure :: Qubit -> Circ(Bit)
hadamard :: Qubit -> Circ(Qubit)
\end{verbatim}
A fair coin can be implemented as a circuit returning a bit, of type $\texttt{() -> Circ Bit}$.
\begin{verbatim}
bitcoin () = do
  q <- qinit True
  q' <- hadamard q
  r <- measure q'
  return r
\end{verbatim}
The function {\tt bitcoin} will merely generate a computation ---producing a circuit--- waiting to be executed.
To implement the toss-coin of Section~\ref{sec:monsem}, we then need to ``run'' the circuit for lifting the bit into a Boolean. Provided that we have a function 
\begin{verbatim}
run :: Circ Bit -> P Bool
\end{verbatim}
one can then implement \texttt{coin ()}  as \texttt{run (bitcoin ())}.

Thanks to the monadic encapsulation, circuits can be manipulated
within Haskell. For instance, inversion and control can be coded in
Haskell as circuit combinators with the following types.
\begin{verbatim}
inverse :: (a -> Circ b) -> (b -> Circ a)
control :: (a -> Circ b) -> ((a,Qubit) -> Circ (b,Qubit))
\end{verbatim}
The compositionality of the monadic semantics also makes it possible to automatically construct oracles out of classical specification~\cite{green2013quipper,DBLP:conf/rc/Valiron16}.

Compared with the quantum $\lambda$-Calculus discussed in
Section~\ref{sec:qlc}, where the program can only send gates one by
one to the co-processor, {\quipper} gives to the programmer the
ability to manipulate circuits.  Although both the quantum
$\lambda$-Calculus and \quipper represent quantum computations
mathematically, \quipper provides a richer model, better-suited for
program specification and verification than the plain
$\lambda$-Calculus.

\subsection{Type Systems}
\label{sec:type-system}
In Section~\ref{sec:op-sem}, we discussed how to model the operational
behavior of a quantum program. We have however not mentioned yet the
run-time errors inherent to quantum computation. In the classical
world, type systems are a standard strategy to catch run-time errors at
compile-time. Several run-time errors specific to quantum computation
can also be caught with a type system, with a few specificities that
we discuss in this section.

\subsubsection{Quantum Data and Type Linearity}

The main problem with quantum information is that it is
\emph{non-duplicable} (a.k.a.~\textit{non clonable}, see Section~\ref{sec:build-quant-circ}). In terms of quantum programming language, this
means that a program cannot duplicate a quantum bit: if $U$ is a
unitary map acting on two qubits, the function
$\lambda x.U(x\otimes x)$ trying to feed $U$ with two copies of its
argument makes no sense. Similarly, it is not possible to control a
gate acting on a qubit with the same qubit. The \quipper code of
Figure~\ref{fig:example-erroenous-circuit} is, therefore, buggy.

\begin{figure}[tbh]
  \centering
  \begin{minipage}[b]{0.5\linewidth}
\begin{verbatim}
exp :: Circ Qubit
exp = do
  q1 <- qinit True
  q2 <- qinit True
  r <- qnot q1 `controlled` q1
  return r
\end{verbatim}
  \end{minipage}
\caption{An example of a buggy \quipper program}
  \label{fig:example-erroenous-circuit}
\end{figure}

Type systems, in a broad sense, provide a predicate that says that a
well-typed program does not have a certain class of bugs: in the case of
quantum programming languages, a large class of bugs comes from
duplicating non-duplicable objects. This calls for a \emph{linear}
type system, enforcing  at least the non-duplication of qubits. This
has been taken into account in recent scalable implementations such as
Silq~\cite{silq}.

On the theoretical side, type systems for quantum Lambda-Calculi and
 \protoquipper~\cite{ross2015algebraic}--the formalization of 
\quipper--are typically
based on linear logic. Originally  designed by
Girard~\cite{girard87linear} (as a continuation of~\cite{lambek1958mathematics}), linear logic assumes that formulas are
linear--i.e. non-duplicable and non-erasable--by default, and the
logic comes equipped with a logic constructor ``$!$'' to annotate
duplicable and erasable formulas. Linear logic also proposes a special
pairing constructor $\otimes$  replacing the usual product and
compatible with both the linearity constructs and the (linear)
implication $\multimap$.

A core type system for a quantum Lambda-Calculus with pairing
therefore consists of the following grammar:
\[
  A,B\quad{:}{:}{=}\quad
  {\tt qubit}
  \mid
  {\tt bool}
  \mid
  A\multimap B
  \mid
  A\otimes B
  \mid
  {!A}.
\]
Type $A\multimap B$ represents the type of (linear) functions, using
their argument only once. Type $A\otimes B$ represents the pair of
a term of type $A$ and a term of type $B$. Type $!A$ stands for a
duplicable term of type $A$. We give a few examples as follows.
\begin{itemize}
\item The identity function $\lambda x.x$ is of type $A\multimap A$,
  but also of type $!(A\multimap A)$ as it is duplicable (since it
  does not contain any non-duplicable object);  

\item If the pairing
  construct is represented with $\langle-,-\rangle$, the function
  $\lambda x.\langle x,x\rangle$ is of type
  $!A\multimap (!A\otimes !A)$: it asks for a duplicable argument;
\item The operator ${\tt qinit}$ is of type $!({\tt bool}\multimap{\tt
    qubit})$: it is duplicable but it does \emph{not} generate a duplicable qubit; 
    
\item The operator ${\tt meas}$ can however be typed with
  $!({\tt qubit}\multimap{!{\tt bool}})$ as a boolean should be
  duplicable; 

\item Provided that $U$ is a unitary acting on two qubits, one can
  type it in a functional manner with
  ${\tt qubit}\otimes{\tt qubit}\multimap{\tt qubit}\otimes{\tt
    qubit}$: it inputs two (non-duplicable) qubits and outputs the
  (still non-duplicable) modified qubits; 
  
\item In particular, provided that we assume implicit dereliction,
  casting duplicable elements of type $!A$ to $A$, the term
  $\lambda x.U\langle x,x\rangle$ can only be typed with
  $!{\tt qubit}\multimap {\tt qubit}\otimes{\tt qubit}$: its argument
  has to be duplicable. The fact that this program can never be
  actually used on a concrete qubit is a property of the type system
  (intuitively, {\tt qinit} only generates non-duplicable qubits).
\end{itemize}

\subsubsection{Example: Quantum Teleportation}

\begin{figure}[tbh]
  \centering
  \includegraphics{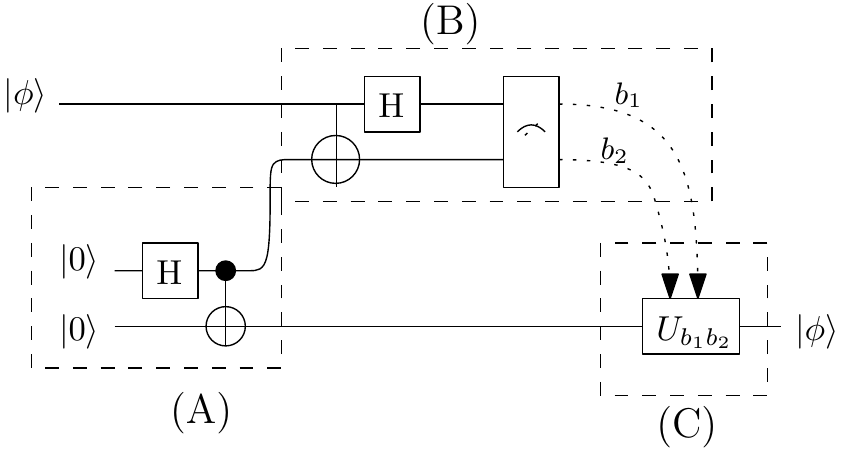}
  \caption{Teleportation algorithm}
  \label{fig:telep-type}
\end{figure}

Mixing quantum computation and higher-order objects can yield non-trivial
objects. The scheme of teleportation is given in
Figure~\ref{fig:telep-type}. It consists of three steps: (A) creation
of a Bell state, (B) measure in the Bell basis to retrieve two
booleans and (C) application of a gate $U_{b_1b_2}$ dependent on the
result of the measure. The state of the top wire is then
``teleported'' to the bottom wire.  It is possible to understand the
pieces of the quantum teleportation protocol as three (duplicable) functions:
\[
  \begin{array}{ll}
    \text{(A)} & !(() \multimap {\tt qubit}\otimes{\tt qubit})
    \\
    \text{(B)} & !({\tt qubit} \multimap ({\tt qubit} \multimap {\tt bool}\otimes{\tt bool}))
    \\
    \text{(C)} & !({\tt qubit} \multimap ({\tt bool}\otimes{\tt bool}
                 \multimap {\tt qubit}))
  \end{array}
\]
The parts $(B)$ and $(C)$ are duplicable functions producing two non-duplicable functions of type ${\tt qubit}\multimap {\tt bool}\otimes{\tt bool}$ and ${\tt bool}\otimes{\tt bool}\multimap{\tt qubit}$.
The teleportation algorithm then feeds the two qubits emitted by (A) to (B) and
(C); this gives a general type
\[
  !(() \multimap ({\tt qubit} \multimap {\tt bool}\otimes{\tt bool})
  \otimes ({\tt bool}\otimes{\tt bool} \multimap {\tt qubit}))
\]
for the protocol. It can be used several times (as it is duplicable). Each time it is run, it generates a pair of {\em non-duplicable} functions $\langle f,g\rangle$, and the
specification of the protocol states that these two functions are
inverse one of the other.

These two functions $f$ and $g$ of the pair are
\emph{non-duplicable}. Indeed, each of them holds a (non-duplicable)
qubit coming from $(A)$. Moreover, in a sense, these functions are entangled, since the Bell state from $(A)$ is entangled.

\subsubsection{Extending the Type System to Support Circuits}
\label{sec:Protoquipper}

Let us assume that the type system of the quantum Lambda-Calculus is extended
with lists: $[A]$ stands for lists of elements of type $A$ (see
e.g.~\cite{10.1007/11417170_26} to see how to do this).  A function
$[{\tt qubit}]\multimap[{\tt qubit}]$ inputs a list of qubits. It can
apply unitary gates to these qubit arguments: It can in fact describe
different circuits, depending on the size of the list. Such a function
, therefore, describes a \emph{family} of circuits.
The quantum Lambda-Calculus is not expressive enough to extract one
circuit out of this family of circuits and operate on it (e.g. by
inversing or controlling it).

The \protoquipper\ language ~\cite{ross2015algebraic} and its
successors~\cite{Rios2017ACM,Rios2021,fu2020linear,lindenhovius2018enriching}
are formalized fragments of the programming language \quipper. They
enforce structural properties of quantum programs using the linear
type system of the quantum Lambda-Calculus, yet extending it to
support circuit manipulation. \protoquipper comes with a new type
construct $Circ(A,B)$: the type of circuits from $A$ to $B$, and two
functions:
\begin{itemize} 
\item {\tt box} sends $A\multimap B$ to $Circ(A,B)$. It takes a
  function $A\multimap B$, partially evaluates it, and stores the emitted circuit in an object of type $Circ(A,B)$.
\item {\tt unbox} sends $Circ(A,B)$ to $A\multimap B$. It takes a
  circuit from $A$ to $B$ and reads it as a function of input $A$ and
  output $B$.
\end{itemize}
The behavior of {\tt box} and {\tt unbox} is specified by the operational
semantic  of the language~\cite{ross2015algebraic}.

One of the subtleties is the fact that {\tt box} turns a function--possibly representing a family of many circuits--to \emph{one}
circuit. In the case of a function of type
$[{\tt qubit}]\multimap[{\tt qubit}]$, this corresponds to choosing
one size of list and building the circuit for this input
size. Whenever the type system supports inductive types such as lists,
the operator {\tt box} then also takes a \emph{shape} as a second
argument, for deciding on the shape of the circuit to build.
In recent works~\cite{fu2020linear}, \protoquipper's type system has
been extended to very expressive \textit{dependent types}, in order to
characterize with a very fine-grain the shape structure of a family of
circuits.

For instance, suppose that the program $P$ sends $[{\tt qubit}]$ to
$[{\tt qubit}]$. It corresponds to a family of circuits, but if we
pick a choice of input size $n$, the type gives no information on the
output shape of the circuit--that is, the number of output
wires. Maybe $P$ duplicates each input wire? With dependent
types, we can for instance index list-types with size and type $P$
with
\[
\forall n . [{\tt qubit}]_n\multimap [{\tt qubit}]_{2n}.
\]
This type tells us that $P$ corresponds to a family of circuits of
even output wires. This makes it possible to catch errors when using circuit combinators: for instance, the inverse operator can be typed with
\[
  \forall n\ m. Circ([{\tt qubit}]_m,[{\tt qubit}]_n)\multimap
  Circ([{\tt qubit}]_n,[{\tt qubit}]_m)
\]
The inverse of $P$ then becomes a function of type
\[
\forall n . [{\tt qubit}]_{2n}\multimap [{\tt qubit}]_{n}.
\]
In particular, this function can only be applied to lists of even
sizes. This run-time error cannot be checked without shape information.

Although such a type system becomes very expressive, in general, it
fails to feature a type inference algorithm, as this would require to
be able to solve arbitrary arithmetic equations.

\subsubsection{Dependent types and Proofs of Programs}

To build a dependently typed language, an alternative approach
is to embed it inside an existing host language with this feature: QWIRE~\cite{paykin2017qwire,paykinPHD} follows this route and relies on the 
language and proof-assistant Coq~\cite{coq}.
While \quipper uses Haskell's monads to encode circuits, QWIRE capitalizes on
Coq's inductive types within the formalism of the Calculus of
Inductive Constructions (CiC)~\cite{cic}, the logical framework of Coq.

In Haskell, inductive types are limited in expressivity: without dependent types, it is not possible to impose constraints on the content of a datastructure.
As circuits in \quipper are internally made of elements of an inductive type, it makes it impossible to forbid ill-defined constructions such as re-using an erased wire, or using twice a wire on a controlled-not, as shown in Fig.~\ref{fig:example-erroenous-circuit}.

QWIRE can instead rely on dependent, inductive types to enforce such constraints on circuit constructions: instead of simply considering circuits as lists of gates, a circuit in QWIRE consists in a list of gates together with proofs that the gates are added in a sensible way ---in other words, the constructors of the inductive type of circuits in QWIRE corresponds to a set of \emph{typing rules} for writing valid circuits.

Interestingly enough, QWIRE does not have to rely on ``!'' type constructor to distinguish between duplicable and non-duplicable data. The idea is that instead of working in a linear-logic based type-system, QWIRE considers an equivalent linear-non-linear model~\cite{benton}. In this paradigm, there are two intertwined languages:
\begin{itemize}
\item A linear language aimed at qubit manipulation and gate application: a program in this language is a circuit. The linearity of the type system enforces the necessary constraints so that e.g. Fig.~\ref{fig:example-erroenous-circuit} is indeed invalid; 
\item A high-level, regular language--typically a lambda-calculus, with a regular type system. This language represents the ``usual'' programming paradigm where classical, conventional computation happens;
\item Then there are two operations, akin to {\tt box} and {\tt unbox}, to move from the linear language to the classical language.
\end{itemize}

In the context of QWIRE, the classical, regular language is Coq and the linear language is encoded using the inductive type of circuits. As Coq features dependent types, QWIRE can then be regarded as a dependently-typed quantum programming language. However, due to the sophisticated type system, QWIRE does not feature a type inference algorithm.

\subsubsection{Discussion}

Type systems for quantum programming languages provide very efficient
ways to encode and--whenever featuring a type inference algorithm--\textit{automatically} verify some important
properties of programs, and in particular to rule out at compile-time
large classes of run-time errors specific to quantum computation. In
particular, type systems have been used to characterize and enforce
\begin{itemize}
\item structure of parametric families of circuits;
\item linearity of non-duplicable elements;
\item control and inversion of only purely quantum circuits.
\end{itemize}
However, to be able to go further and characterize functional
correctness with respect to specification, or validate the number of
gates of a circuit, or catch subtle bugs involving concatenation of
inverted circuits, one needs to move away from the simple linear type systems of quantum lambda-calculi and shift towards sophisticated dependent
types, such as the extension ProtoQuipper-D~\cite{fu2020linear} of ProtoQuipper, or the approach of QWIRE. The gain in expressiveness is then at the expense of automation~\cite{singhal2020c,paykin2017qwire}.

The quest for a finer trade-off, permitting automation while capturing
some of what is currently only available with dependent type system is
an active research area in the community.

\section{High and Mid-Level Verification: Algorithms and Programs}
\label{sec:high-and-lid-level-verif}

Most {\it quantum programming languages}
(\qiskit~\cite{QiskitCommunity2017} \quipper
\cite{green2013quipper}, \liquid~\cite{wecker2014liqui}, Q\#
\cite{svore2018q}, \projectq~\cite{projectq}, \silq \cite{silq}, \textit{etc})
embed features for quantum circuit manipulations within a standard classical
programming language. 
Such {\it circuit-building quantum languages} is the current
consensus for high-level executable quantum programming languages.
A current major challenge is to link this language design paradigm with formally verified programming. 
In the present section, we introduce the main existing propositions in that direction.

\subsection{Quantum Hoare Logic}
\label{qhl}
\label{sec:high-level-algorithm}

Quantum Hoare logic (QHL)
\cite{chadha2006reasoning,feng2007proof,kakutani2009logic,dblp:journals/toplas/ying11,DBLP:journals/pacmpl/Unruh19,unruh2019quantumghost,BartheHYYZ20}
is a general framework for reasoning about the classical control instructions over unitary operations in quantum algorithms. Referring again to Figure~\ref{qram}, the focus is on the interaction between the classical computer and the quantum co-processor, instead of the circuit analysis  as in Section~\ref{sec:low-level-verif}, or the gate to gate circuit building functions as in Sections~\ref{sec:qbricks} and ~\ref{sec:sqir}. Therefore, we consider it as a high-level  description of
quantum algorithms. 

   It is based on the  assertion method of Floyd and Hoare
\cite{Flo67,Hoa69}--attach each program
point with an assertion and whenever the data flow reaches a program
point the attached assertion should be satisfied--which was
originated with Alan Turing \cite{AptOlderog19}.  Hoare's approach
enables (interactive) theorem proving for high-level algorithmic
description verification that proceeds at the same abstraction level
as the language itself.  This makes verification more human-friendly
than lower-level (machine-friendly) verification.

\subsubsection{Quantum Programming Language: Quantum \WHILE-Programs}

The guideline for the hybrid model introduced in
Section~\ref{sec:hybrid-model} is summed up by the slogan
\textit{``quantum data and classical control"} \cite{selinger2004}: quantum
data can be superposed and entangled, they are manipulated by basic
quantum operations--unitary evolution and measurement, but the
high-level control is still classical (e.g. branch, loops, \textit{etc}).

In light of this slogan, QHL introduces a minimal programming language
for describing quantum algorithms \cite{ying2016foundations}.  We follow ~\cite{dblp:journals/toplas/ying11} for the introduction of QHL technical environment. Let $q$ (resp. $\lst{q}$) be
a quantum variable (resp. a list of quantum variables); let $U$ be a
unitary operator acting on the qubits $\lst{q}$ and let
$M \triangleq \{M_m\}_m$ with $\sum_m M_m^\dag M_m = I$ be a
measurement on the qubits $\bar{q}$, each $M_m$ corresponding to a
measurement result $m$ (see Section~\ref{sec:quant-circ-semant}).  As a special case, let
$M' \triangleq \{M_0, M_1\}$ with $M_0^\dag M_0 + M_1^\dag M_1 = I$.
Then quantum \WHILE-programs are generated by the following syntax.
\begin{displaymath}
\begin{array}{rcll}
  S & \triangleq & \SKIP & \mbox{No operation} \\
    & \mid & q := \ket{0}{} & \mbox{Initialization} \\
    & \mid & \lst{q} \starequal U & \mbox{Unitary operation} \\
    & \mid & S_1;S_2 & \mbox{Sequential composition} \\
    & \mid & \ifStat{\Box m\cdot M[\lst{q}] = m \rightarrow S_m} & \mbox{Probabilistic branching} \\
    & \mid & \whileStat{M'[\lst{q}] = 1}{S_0} & \mbox{Probabilistic while loop}
\end{array}
\end{displaymath}
The intended semantics of language constructs above is similar to that
of their classical counterparts.  To
illustrate the quantum features contained in these constructs, we make
the following comments:
\begin{description}
\item[(i)] in the initialization, the choice of a fixed state
  $\ket{0}{}$ is without loss of generality, since any known quantum state can be
  prepared by applying a unitary operator to $\ket{0}{}$;
\item[(ii)] according to Born rule, measurement results follow a probabilistic law. Since they lie on measurement result observations, branching (resp. while loops) is therefore
probabilistic. It creates
  different branches $\{S_m\}_m$ \big(resp. $\{\SKIP, S_0\}$\big),
  chosen according to  outcomes of the
  measurement $M$ (resp. $M'$) on the qubits $\bar{q}$.
\end{description}
We  refer the reader to
\cite{dblp:journals/toplas/ying11} for a detailed exposition of the syntax above.

\begin{example}[Preparation of the Bell state $\bellzz$]
  \label{exam_prep}
  Let $p$ and $q$ be quantum variables, each denoting one qubit. Then
  the following program initiates them to $\ket{0}{}$ and implements
  the circuit from Figure~\ref{fig:circuit_bell_states}, preparing
  state \bellzz from Example~\ref{ex:bell-states}.:
  \[
    \beta_{00}  \quad \triangleq \quad p := \ket{0}{};\
    q := \ket{0}{};\ p \starequal H;\ (p,q) \starequal \Cnot
  \]
\end{example}

Note that the quantum programming language defined above is in the
spirit of the hybrid circuit model presented in
Section~\ref{sec:gen-back-qc}.  Indeed, the basic sequence of quantum
operations (initialization, unitary operation, and measurement) are
meant to be interpreted as a generalized quantum circuit to be
executed on a quantum co-processor; and post-measurement branchings
(in, e.g., probabilistic branching and while loop) are meant to be
controlled by a classical computer.

\subsubsection{Quantum States, Operations and Predicates}
\label{sec:qhl-quantum-states}
 
Measuring a quantum state transforms it, following the Born rule (see
Section~\ref{sec:qd}). The resulting probability distribution over
quantum states is formalized as a \emph{mixed state} (as opposed to
\emph{pure states}, see Section~\ref{sec:quant-circ-semant}). For example, a measurement on
any pure  quantum state
$\ket{+}{} \triangleq \frac{1}{\sqrt{2}}(\ket{0}{} + \ket{1}{})$ or
$\ket{-}{} \triangleq \frac{1}{\sqrt{2}}(\ket{0}{} - \ket{1}{})$ will
result in the mixed state
$\Ee = \big\{ \big(\frac{1}{2}, \ket{0}{}\big), \big(\frac{1}{2},
\ket{1}{}\big) \big\}$, with states $\ket{0}{}$ and $\ket{1}{}$
occurring with an equal probability of $\frac{1}{2}$ (notice that this
observation makes both states $\ket{+}{}$ $\ket{-}{}$ impossible to
distinguish by simple measurement). 

In this way, the representation of the final state of applying a series of
measurements to a quantum state could expand exponentially.  To
address this issue, a square-matrix representation of quantum states,
i.e. partial density operator, is adopted instead.  For example, pure
quantum state $\ket{+}{}$ is represented as $\outprod{+}{+}$, and
mixed quantum state $\Ee$ as
$\frac{1}{2} \outprod{0}{0} + \frac{1}{2} \outprod{1}{1}$.  See Section~\ref{sec:quant-circ-semant} for a brief introduction to  partial density operators, and
\cite{selinger2004, ying2016foundations,nielsen2002quantum} for further details.

If we see the matrix representation of a quantum state (partial
density operator) as a linear operator, then a quantum operation--
initialization, unitary evolution and measurement--can be thought of
as a super operator, \ie a function from linear operators to linear
operators.  What's interesting is that every quantum \WHILE-program
defined above can be interpreted as a quantum operation, and partial
density operators are closed under quantum operations.  This justifies
the success of representing quantum states as partial density
operators and defining the denotational semantics of quantum programs
as quantum operations \cite{selinger2004, ying2016foundations}.

Following \cite{panangaden2006}, a quantum predicate on vector space
$\Hh$ is defined as a Hermitian operator $M$ between the zero operator
$0_{\Hh}$ (representing the contradiction) and the identity operator
$I_{\Hh}$ (representing the tautology).  Instead of the usual binary
satisfaction judgment, QHL evaluates the satisfaction of a predicate
by a state as a real value between $0$ (false) and $1$ (true). It is
defined as the trace $\tr(M \rho)$  of
the product $M \rho$.  Intuitively, it represents the expectation for
the truth value of $M$ in the mixed state $\rho$ (which is, again, a
probability distribution over pure states).

Then, the intuition of implication between predicates is also
probabilistic. It is filled by the \emph{L\"{o}wner order}
$M \sqsubseteq N$, relating operators $M$ and $N$, if and only if, for
any state $\rho$, the expectation truth value of $N$ in $\rho$ is more
or equal to that of $M$ in $\rho$. This condition is formalized as
$\tr(M \rho) \leq \tr(N \rho)$ for all states $\rho$ (See, e.g.,
\cite[Lemma 2.1]{dblp:journals/toplas/ying11}).

Adopting such quantum predicates as assertions, among many others
(e.g., interpreted as physical observables), provides simple
expression means for many properties of quantum effects.  For example,
quantum predicate $\outprod{+}{+}$ expresses that a state $\rho$ is in
the equal superposition $\ket{+}{}$ with probability
$\tr(\outprod{+}{+} \rho)$; quantum predicate
$\outprod{\beta_{00}}{\beta_{00}}$ expresses that a state $\rho$ is in
the maximal entanglement $\bellzz$ with probability
$\tr(\outprod{\beta_{00}}{\beta_{00}} \rho)$, \textit{etc}.

\subsubsection{Quantum Program Verification}

For now, a quantum (partial) correctness formula can be the Hoare's
triple $\pcor{P}{S}{Q}$, where $S$ is a quantum \WHILE-program, and
$P,Q$ are quantum predicates.  To define the partial-correctness
semantics of quantum Hoare's triples, in the sequel, let
$\llbracket S \rrbracket$ denote the semantic function of $S$ (Note
that $\llbracket S \rrbracket$ is a quantum operation defined by
induction on $S$, cf. \cite{ying2016foundations}), and
$\llbracket S \rrbracket (\rho)$ the output of $S$ on the input
$\rho$.

\begin{definition}[Semantics of partial correctness,
  cf. \cite{dblp:journals/toplas/ying11}]
  Let $P$, $Q$ be quantum predicates and $S$ a quantum
  \WHILE-program.  We say that $S$ is (partially) correct
  w.r.t. precondition $P$ and postcondition $Q$, written
  $\models\pcor{P}{S}{Q}$, if
  \begin{eqnarray}\label{eqt_par}
                       \forall \rho, \quad    \tr(P\rho) &\leq& \tr\big(Q \llbracket S \rrbracket {(\rho)}\big) +
                      \big[ \tr(\rho) - \tr\big( \llbracket S \rrbracket (\rho)\big) \big].
  \end{eqnarray}
\end{definition}

Note that Inequality ($\ref{eqt_par}$) can be seen as a probabilistic
version of the following statement: if state $\rho$ satisfies
predicate $P$, then, executing program $S$ on input $\rho$, either $S$
fails to terminate or the resulting state
$\llbracket S \rrbracket (\rho)$ satisfies predicate $Q$.

\begin{table}[tbh]
  \caption{Proof system for partial correctness.}
  \label{qPW}
  \centering
  \begin{tabular}{rl}
    (Skip Axiom) & $\pcor{P}{\SKIP}{P}$ \\
    \\
    (Init Axiom) & $\pcor{ {\sum}_{i} \ket{i}{q} \bra{0}{} P \ket{0}{q} \bra{i}{} }{q := \ket{0}{}}{ P }$ \\
    \\
    (Unit Axiom) & $\pcor{U^\dag P U}{\lst{q} \starequal U }{ P }$ \\
    \\
    (Comp Rule) & $\frac{\pcor{P}{S_1}{Q} \quad \pcor{Q}{S_2}{R}}{\pcor{P}{ S_1;S_2 }{R}}$ \\
    \\
    (If Rule) & $\frac{\pcor{P_m}{S_m}{Q} \mbox{ \small for all } m}{\pcor{\sum_m M_m^\dag P_m M_m}{\ifStat{\Box m\cdot M[\lst{q}] = m \rightarrow S_m}}{Q}}$ \\
    \\
    (Par Loop Rule) & $\frac{\pcor{P}{S_0}{M_0^{\dag} Q M_0 + M_1^{\dag} P M_1}}
    {\pcor{M_0^{\dag} Q M_0 + M_1^{\dag} P M_1}{\whileStat{M'[\lst{q}] = 1}{S_0}}{Q}}$ \\
    \\
    (Order Rule) & $\frac{P\sqsubseteq P' \quad \pcor{P'}{S}{Q'} \quad Q'\sqsubseteq Q}{\pcor{P}{S}{Q}}$
  \end{tabular}

\end{table}

The axiom system for proving partial correctness of quantum
\WHILE-programs is composed of axioms and inference rules
manipulating quantum Hoare's triples
\cite{dblp:journals/toplas/ying11}. It is shown in Table \ref{qPW}
\big(where $\{\ket{i}{}\}_q$ is the computational basis for quantum
variable $q$\big).  Remark that each of these rules and axioms follows
the assertion method.  Here we only show how to derive
the most complex rule (Par Loop Rule). The derivation of other proof
rules can be done similarly.

\paragraph*{Intuition of (Par Loop Rule).}
To derive (Par Loop Rule), by   assertion method, we attach each program point, say $l_1,l_2,l_3$, of a \WHILE-statement with an assertion, say $R,P,Q$, respectively:
\[
  \{l_1\colon R\}\ \textbf{while}\ M'[\lst{q}] = 1\ \textbf{do}\ \{l_2\colon
  P\}\ S\ \textbf{od}\ \{l_3\colon Q\}
\]

Fix the input $\rho$ at the program point $l_1$ satisfying the
assertion $R$.  By semantics of a \WHILE\ loop, after the measurement
$M'$, one part $M_1 \rho M_1^\dag$ of the input will go to the loop
body through the program point $l_2$ where the assertion $P$ will be
satisfied; the other part $M_0 \rho M_0^\dag$ will leave the while
loop through the program point $l_3$ in which the assertion $Q$ will
be satisfied. Hence:
\begin{eqnarray*}
  \tr(R \rho) &\leq& \tr\big( Q (M_0 \rho M_0^\dag) \big) + \tr\big( P (M_1 \rho M_1^\dag) \big)
\end{eqnarray*}
Due to the arbitrariness of $\rho$, by properties of the trace function
and L\"{o}wner order, we have that
$R\sqsubseteq M_0^\dag Q M_0 + M_1^\dag P M_1$.  Then, by weakening
$R$ to $M_0^\dag Q M_0 + M_1^\dag P M_1$ and lifting the above
reasoning process into an inference rule, (Par Loop Rule) follows.

The following example illustrates how to derive a partially correct
quantum Hoare triple using the axioms and inference rules presented above.

\begin{example}[Specification and correctness proof for the  \bellzz\
  state construction program]

  Recall from Example \ref{exam_prep} the definition of quantum program $\beta_{00}$:
  \[
    \beta_{00}  \quad \triangleq \quad p := \ket{0}{};\
    q := \ket{0}{};\ p \starequal H;\ (p,q) \starequal \Cnot
  \]
  To show (partial) correctness of this program,
  it suffices to prove
  \begin{equation}\label{assert_1}
    \pcor{I_p \otimes I_q}{\beta_{00}}{\ket{\beta_{00}}{p,q}\bra{\beta_{00}}{}}
  \end{equation}
  This can be done as follows. By sake of space we need to decompose the derivation tree. We first derive specifications for the initialization instructions:
  \begin{prooftree}
\AxiomC{\hspace{3.5cm}\textit{\small Init Axiom}}
\UnaryInfC{$\pcor{I_p \otimes I_q}{p := \ket{0}{}}{ \ket{0}{p}\bra{0}{}\otimes
      I_q}$}
\AxiomC{}
      \RightLabel{\small Init Axiom}
  \UnaryInfC{$\pcor{\ket{0}{p}\bra{0}{} \otimes I_q}{q := \ket{0}{}}{
      \ket{0}{p}\bra{0}{}\otimes \ket{0}{q}\bra{0}{}}$}
      \RightLabel{\small Comp Rule}
\BinaryInfC{$\pcor{I_p \otimes I_q}{p := \ket{0}{};\ q := \ket{0}{}}{\ket{0}{p}\bra{0}{} \otimes \ket{0}{q}\bra{0}{}}\qquad(i)$}
\end{prooftree}

Then come the unitary application instructions. We set the following abbreviations for Hoare triples:

$
\begin{array}{lll}
(ii)    &:= &   \pcor{\ket{0}{p}\bra{0}{} \otimes \ket{0}{q}\bra{0}{}}{p \starequal H}{ \ket{+}{p}\bra{+}{} \otimes \ket{0}{q}\bra{0}{} }\\
(iii)     &:=& \pcor{ \ket{+}{p}\bra{+}{} \otimes \ket{0}{q}\bra{0}{} }{ (p,q) \starequal \Cnot }{ \ket{\beta_{00}}{p,q}\bra{\beta_{00}}{} }
\end{array}$

They are instances of axioms and we can combine them:
\begin{prooftree}
  
\AxiomC{}
      \RightLabel{Init Axiom}
  \UnaryInfC{$ (ii)$}
\AxiomC{}
      \RightLabel{Init Axiom}
  \UnaryInfC{$(iii)$}
      \RightLabel{Comp Rule}
\BinaryInfC{$    \pcor{\ket{0}{p}\bra{0}{} \otimes \ket{0}{q}\bra{0}{}}{p \starequal H;\ (p,q) \starequal \Cnot }{\ket{\beta_{00}}{p,q}\bra{\beta_{00}}{}}\qquad (iv)$}
\end{prooftree}

Finally, the two preceding proof tree branch together via the sequential composition rule, which achieves the derivation:

\begin{prooftree}
  
\AxiomC{(i)}
\AxiomC{(iv)}
      \RightLabel{Comp Rule}
\BinaryInfC{$    \pcor{I_p \otimes I_q}{\beta_{00}}{\ket{\beta_{00}}{p,q}\bra{\beta_{00}}{}}
$}
\end{prooftree}
\end{example}

\subsubsection{Implementations and Extensions}

Several works have taken advantage of extended Quantum Hoare Logic, e.g. algorithmic
analysis of termination problem \cite{li2017algorithmic} or 
characterization and generation of loop invariants \big(i.e.
$M_0^\dag Q M_0 + M_1^\dag P M_1$ in (Par Loop Rule)\big)
\cite{DBLP:conf/popl/YingYW17}. 

The practical illustration of QHL
can be found in Liu et al. paper ~\cite{dblp:journals/corr/liulwyz16}, containing an implementation
in \isabelle together with a formalization of Grover~\cite{grover1996fast} and Quantum Phase Estimation (QPE~\cite{kitaev1995quantum})
algorithms. Nevertheless in these examples, the central verification
part is assumed through Python libraries uses.

More recent work \cite{10.1007/978-3-030-25543-5_12} includes full
proof for a parametrized version of Grover's search algorithm. It constitutes an
illustration of QHL use on a non-trivial example. 

\subsubsection{Other Quantum Hoare Logics}

In addition to the framework introduced above,  applying Hoare Logic methods to quantum processes brought several additional developments, focusing on different aspects of quantum computations. We introduce a few of these QHL related framework in the following paragraphs. 
\paragraph{Quantum Hoare Logic with ghost variables.}
One of the principal shortcuts of  \cite{dblp:journals/toplas/ying11} comes from the limitations of the specification language. In~\cite{unruh2019quantumghost}, the author in particular targets the possibility to characterize probabilistic distributions of values. His proposition introduces \emph{ghost} variables in the specifications language: a  \emph{ghost}  variable does not occur in the program but only in its specifications.
In~\cite{unruh2019quantumghost}, a ghost variable is interpreted under an implicit existential quantification. Ghost variables enable, in particular, to explicitly refer to discarded, measured or overwritten qubits. 
In addition to probability distribution definitions, it brings several features to the expressive power of the specifications language, such as  separability--unentanglement--of variables or the fact, for a given variable, to hold for a classical data value.



\paragraph{Quantum Hoare Logic with classical variables.}
With similar concern,~\cite{DBLP:journals/corr/abs-2008-06812} extends QHL specifications with classical variables. In the specification and verification of algorithms of practical use, holding classical information is indeed crucial. It enables, for example,  
describing and specifying an algorithm holding classical parameters or a hybrid program intertwining classical and quantum instructions (see~Section\ref{sec:hybrid-model}).

Hence, the semantics relies on so-called \emph{cq-states} (classical/quantum states) made of both:
\begin{itemize}
    \item a classical variable assignment for the interpretation of classical variables;
    \item a density operator interpreting quantum variables.
\end{itemize}

This extension of the semantics induces an extension of the proof system, which is proved sound and complete--with respect to partial correctness. Interestingly, the paper presents detailed specified case studies, including Grover's and Shor's algorithms.

\paragraph{Robustness analysis}
is another important line of work, initiated in~\cite{hung2019quantitative} as a continuation of QHL. 
The different solutions presented so far rely on the implicit assumption that quantum gates are applied deterministically as indicated by their matrix semantics.
Still, as stated in Section~\ref{sec:quantum-noise}, this noise-free modeling is not 
perfectly accurate in the NISQ era.

A more realistic description of the behavior of a gate would consider several different possible behaviors, weighed with their respective probabilities of occurrence: the \emph{intended} one, expected  by the semantics, and one or several additional erroneous behaviors\footnote{Strictly speaking, the language in ~\ref{sec:quantum-noise} is limited to considering up to one possible error per gate application. But this is without lost of generality since  more sophisticated scenarios can be encoded through, eg., the replacement of gate application intervals by potentially erroneous identity transformations.}. 

Interestingly, in this setting, the application of a quantum gate is formally represented as a probability distribution over unitaries. Hence, a non-deterministic gate application naturally formalizes in the formalism of super operators acting over density operators.

Then, a metrics is defined for measuring the difference between the behavior of a quantum system
 under a given error scheme (for each gate application, the mention of a possible erroneous behavior together with its probability of occurrence) and its intended behavior. It is called the \emph{trace distance} and  serves as an evaluation for the robustness of the implementation.
 
 Among other case studies, the method is illustrated by evaluating a minimal quantum error correcting scheme.

\paragraph{Quantum separation logic}~\cite{zhou2021quantum} aims at simplifying quantum programs specifications. The leading observation is that, while quantum programs often manipulate big quantum circuits, with matrix semantics growing exponentially over their width, many quantum algorithms proceed via sequences of local manipulations over quantum sub-registers. The authors exhibit the examples of Quantum Machine Learning~\cite{biamonte2017quantum} and Variational Quantum Algorithms (VQA~\cite{mcclean2016theory}), which are among the most promising classes of algorithms in the NISQ era~(see Section~\ref{sec:quantum-noise}).

To efficiently reason about state evolution in such implementations, a Quantum Separation Logic is proposed, together with a dedicated proof system. 
This logic allows the expression of local manipulation on separated quantum registers while maintaining the state of the rest of the register.
 The approach is probed with case studies from both quantum programming analysis (VQA) and communication protocols security checks (one-time pad and secret sharing).

\paragraph{Quantum relational Hoare logic} \cite{DBLP:journals/pacmpl/Unruh19,BartheHYYZ20}
allows to reason about how the outputs of two quantum programs relate to each other  
given a relation between their inputs, which can be used to analyze
security of post-quantum cryptography and quantum protocols. 

\paragraph{Quantum Hoare type theory} \cite{singhal2020c} is inspired by classical
Hoare type theory and extends the Quantum IO
Monad~\cite{altenkirch2010quantum} by indexing it with pre- and
postconditions that serve as program specifications, which has the
potential to be a unified system for programming, specifying, and
reasoning about quantum programs.

\paragraph{Quantum dynamic logic.}
We end up this section by mentioning a line of work that is not a strict extension or application of Hoare Logic~\cite{Hoa69}, but shares similar concerns and related solutions.
Just as QHL and its extensions, dynamic logic formalizes the evolution of a state along the execution of a process acting over it.

Dynamic logic inherits from the modal logic apparatus where,  in addition to propositional logic connectives,  modalities $\Box$ and $\Diamond$ are intriduced. Intuitively, given a formula $\phi, \Box\phi$ means that $\varphi$ is \emph{necessarily} true and $\Diamond\varphi$ means  that $\varphi$ is \emph{possibly} true.
The standard semantics, based on Kripke models~\cite{Kripke1963-KRISCO}, helps catching this intuition. A Kripke model $\mathcal{K}$ is made of a set of states $S$ holding, each, a valuation $P : S \to \{\mathbf{true},\mathbf{false}\}$ for a set of propositional variables $P$. And the modality is interpreted through an accessibility relation $R\subseteq S\times S$ For example, given  a propositional variable $p$,  $\Box p$ (resp. $\Diamond p$) is true in state s, written $\mathcal{K},s\models \Box p$ iff $p$ (resp. $\mathcal{K},s\models\Diamond p$) is true if $p$ is true in every (resp. at least one) state $s'$ such that $(s,s')\in R$.

In dynamic logic, several modalities coexist, formalizing a set of \emph{actions}. Hence, given an action $a, formula [a]\varphi$ is true iff $\varphi$ is always true after performing action $a$. Actions can combine together, for example $[a;a']\varphi$ means that $\varphi$ is always true after the successive performance of actions $a$ and $a'$.

  In their \emph{Logic of Quantum Programs} (LQP~\cite{baltag2006lqp,DBLP:journals/corr/abs-2109-06792}), Baltag
and Smets designed a quantum version of dynamic logic. Here, states correspond to one-dimensional subspaces of a Hilbert space for a quantum register. Actions are of two kinds: a \emph{test} encodes a measurement through the set of projection transitions corresponding to the different possible outcomes (for example, given formulas $\varphi$ and $\psi$, $\varphi?\psi$ is satisfied in the state of evaluation if and only if any state satisfying $\varphi$ after measurement--any successful test for $\varphi$--also satisfies $\psi$), and proper \emph{actions} deterministically encode unitary state transformations). In~ \cite{baltag2006lqp}, the framework is illustrated by a  correction proof for the teleportation and quantum secret sharing protocols.
Baltag and Smets link their work to similar  quantum logics as in~\cite{brunet2004dynamic}.

Note that to treat measurement, LQP formalizes informations about the set of states that are possibly reachable, but the language does not hold any notion of probability. This shortcut is overcome in ~\cite{baltag2014plqp}, where the authors introduce a probabilistic test modality characterizing, for any formula $\varphi$ and rational $x\in[0,1[$, that a test for $\varphi$ has probability at least $x$ to succeed from the evaluation state.
In~\cite{baltag2014plqp},  the authors also provide a decidability proof for  their setting and
 illustrate its expressive power by formally specifying  the   
Grover’s search algorithm and  a leader election protocol. Further work in the community has used
this probabilistic logic setting for the verification of the BB84 Quantum Key Distribution algorithm~\cite{DBLP:journals/soco/BergfeldS17}.
 
\subsection{\qbrick}
\label{sec:qbricks}

\qbrick~\cite{chareton2021automated} is a recently proposed circuit
description language together with a deductive verification framework. It enables
automated proof support for program specifications, reducing the
required human effort for the development of verified programs.

\qbrick object language (\qbrickDSL) consists in a minimal functional
language with features for the design of circuit families.
Similarly to the formal contract style of algorithm descriptions (see
Section~\ref{sec:quantum-algorithm-design}), \qbrick\ functions are
written with explicit pre-and postconditions, specifying their
complexity and the parametrized input-output quantum data registers
function they implement.  These specifications are written in a
dedicated formal language, called \qbrickSPEC.

To support proofs, \qbrick is given a Hoare style derivation rules system
including rules for each parametrized circuit constructor.
These  rules are enriched with equational theories
enabling, in particular, reasoning about measurement and
probabilities.

\qbricks is a domain-specific language, embedded in the
Why3~\cite{filliatre2013why3} deductive verification framework :
programs are written in ML language and annotated with specifications
in \qbrickDSL (pre- and postconditions, loop invariants, calls for
lemmas, \textit{etc}). Compiling a \qbrick program interprets these
specifications as proof obligations.  Then, a dedicated interface
enables to directly access these proof obligations and either send
them to a set of automatic SMT-solvers (CVC4, Alt-Ergo, Z3, \textit{etc}.), or
enter some interactive proof transformation commands
(additional calls for lemmas or hypotheses, term substitutions, \textit{etc}.)
or even to proof assistants (\coq, \isabelle).

\subsubsection{Writing Quantum Circuits Functions in \qbrick: \qbrickDSL}

\qbrickDSL makes use of a regular inductive datatype for circuits,
where the data constructors are elementary gates, sequential and
parallel composition, and ancilla creation.  In particular, unlike in
e.g. \quipper or \qwire, a quantum circuit in \qbrick is not a
function acting on qubits: it is a simple, static object. Nonetheless,
for the sake of implementing quantum circuits from the literature, this
does not restrict expressiveness as they are usually precisely
represented as sequences of blocks. 

The core of \qbrickDSL is presented in Figure~\ref{tab:qbrickdsl}.  It
is a small first-order functional, call-by-value language. To the elementary gates presented in
Section~\ref{sec:quant-circ-semant}, \qbricks adds the qubit swapping
gate \texttt{SWAP} and the identity \texttt{ID}.  The constructors for
high-level circuit operations are sequential composition \texttt{SEQ},
parallel composition \texttt{PAR} and ancilla creation/termination
\texttt{ANC}.

\begin{figure}[tbh]
\scalebox{.9}{
  $
  \begin{array}{r@{\quad}lll}
    \text{Expression}
    & e
    & {:}{:}{=}
    & x \bor c \bor
      f(e_1,\ldots,e_n)
      \bor \letinterm{\tuple{x_1,\ldots,x_n}}{e}{e'}\bor \\
    &&& \iftermx{e_1}{e_2}{e_3}\bor
        \texttt{iter}~f~{e_1}~{e_2}
    \\[1ex]
    \text{Data Constructor}
    & c
    & {:}{:}{=}
    & \underline{n}\bor\ttrue\bor\ffalse\bor \tuple{e_1,\ldots, e_n}
      \bor \texttt{CNOT} \bor\texttt{SWAP} \bor \texttt{ID}\bor \texttt{H}\bor\texttt{Ph}(e)\bor\texttt{R}_z(e)
        \bor \\
    &&&
    \texttt{ANC}(e) \bor \texttt{SEQ}(e_1,e_2) \bor
        \texttt{PAR}(e_1,e_2)
    \\[1ex]
    \text{Function}
    & f
    & {:}{:}{=} 
    & f_{\text{d}} \bor f_{\text{c}}
    \\[1ex]
    \text{Declaration}
    & d
    & {:}{:}{=} 
    & \decl{f_{\text{d}}}{x_1,\ldots,x_n}{e}
    \\[1ex]

  \end{array}  
$}
\caption{The syntax for \qbrickDSL}
\label{tab:qbrickdsl}
\end{figure}

The term constructs are limited to function calls, \texttt{let}-style
composition, test with the ternary construct \texttt{if-then-else} and simple iteration:
$\texttt{iter}~f~n~a$ stands for $f(f(\cdots f(a)\cdots))$, a succession of 
$n$ calls to $f$.

Even though the language does not feature measurement, it is nonetheless
possible to \emph{reason} on probabilistic outputs of circuits, if we
were to measure its output. This is expressed
in a regular theory of real and complex numbers in the specification
language (see Section~\ref{sec:probabilistic-reasoning} below for
details).

\subsubsection{Parametrized Path-Sums}
\label{pps}

To interpret circuit description functions, \qbricks uses parametrized
path-sums (\pps), that is an extension of
 path-sums~\cite{amy2018towards}  (Section~\ref{sec:path-sum-circuit} ).

In \qbricks setting, a path-sum $P$ is an object from an opaque type
with four parameters, presented in Table~\ref{pps_par} with their
types and identifier shortcuts: two integer constants $\pwidth(P)$ 
(the width of the target circuit)  and $\prange(P)$ (the
\emph{range}, meaning that the output sum of kets has a term for each
bit vector $\vec{y}$ of length $\prange(P)$--written
$\vec{y}\in \Bv_{\prangeab(P)}$) and two functions
$\pangle(P) $ and
$\pket(P) $ : for any
input bit vector $\vec{x}$ of length $\pwidth(p)$ (standing for a
basis ket input to the target circuit) and for any index bit vector
$\vec{y}$  of length $\prange(p)$, functions $\pangle(P)$ and $\pket(P)$ respectively define a
real scalar and a bit vector of length $\pwidth(p)$ (standing for a
basis ket output to the target circuit)%
\footnote{In the rest of this section, type $\bitvector$ corresponds
  to bit vectors $\vec{x}$. They are encoded in an abstract type that provides a positive integer
  $\texttt{length}(\vec{x})$ (the length of the vector) and a value function
  $\texttt{get\_bv}(\vec{x})$ : $\inttype \to \inttype$.  For any
  integer $i$, we commonly abbreviate $\texttt{get\_bv}(\vec{x}) i $
  as $\vec{x}_i$; if $i\in \tofset{\texttt{length}(\vec{x})}$ (that is, if $i$ is actually in the range of the bit vector) then it
  is such that $0\leq \vec{x}_i <2$.}.

\begin{table}[tbh]
  \centering
  \caption{\Pps\ accessors and types}
  \begin{tabular}{|lcc|}\hline
    Identifier &Type&Id-abbreviation\\
    \hline        \pwidth  &   \inttype & \pwidthab\\
    \prange  &   \inttype & \prangeab\\
    \pangle  &   $\bitvector \to\bitvector \to\realtype$  & \pangleab\\
    \pket  &   $\bitvector \to\bitvector \to\bitvector$ & \pketab\\
    \hline
  \end{tabular}
  \label{pps_par}
\end{table}
 
Then for any bit vector $\bitvec{x}$ of size $\pwidthab(P)$, the
expression \begin{equation}Ps(h,\ket{\bitvec{x}}{}) =
  \frac{1}{\sqrt{2^{\prangeab(P)}}} \sum_{\bitvec{y}\in
    \Bv_{\prangeab(P)}} e^{2\cdot \pi i. \pangleab(P)(\bitvec{x},
    \bitvec{y})}\ket{\pketab (P)(\bitvec{x},
    \bitvec{y})}{\pwidthab(P)}\label{pps_expr}\end{equation} combines
these different element together to define a linear application for
quantum state vectors. Function $Ps$ is extended by linearity to any
ket $\ket{u}{}$ of length $\pwidthab(p)$%
\footnote{That is, to any linear combination of basis kets
  $\ket{\vec{x}}{}$ of length $\pwidthab(p)$.}.
 
A path-sum $P$ is said to \emph{correctly interpret a given circuit
  $C$} (written $\rcorrect{P}{C}$) if and only if $C$ has width
$n= \pwidth(P)$ and for any bit vector $\vec{x}$ of length $n$,
$\Mat{C}\cdot \ket{\vec{x}}{}=Ps(P,\ket{\vec{x}}{})$.  The relation
$\rcorrect{\_}{\_}$ enjoys nice composition properties along
\qbrickDSL circuit constructors (see~\cite{chareton2021automated} ).
 
\qbrickSPEC generalizes path-sums by introducing Parametrized
path-sums (\pps). A \pps\  is a function that inputs a set
of parameters  and outputs a path-sum. Then, it can be seen as a
family of path-sums (one for each possible value of its parameters)
describing the  effects of the different  members in a family of quantum
circuits. Hence, it is well-fitted for the specification of
parametrized algorithms such as Shor order finding (\shorof see  Figure~\ref{ncpe}).

The main strength of \pps\ semantics, with regards to formal
verification, is that  each of the path-sum
parametrized accessors (see Table~\ref{pps_par})  combines compositionally along with circuit constructors.

Hence,  it enables  reasoning  about parametrized  quantum circuits and their semantics without
manipulating sum terms or other higher-order objects. 
Thanks to this tool, the automatic generation of proof obligations for \qbricks
specifications results in only first-order formulas, enabling a high
level of automation when sent to SMT-solvers.

\subsubsection{From Quantum Circuits to  Path-Sums}
\label{syntax-qbdsl}

The specification language for \qbrick is a first-order predicate
language, equipped with various equational theories.  For any quantum
circuit family $C$, \qbrickDSL enables to identify a \pps\
$\circtopps(C)$. Each of its parametrized accessors is defined
inductively upon the structure of $C$ and it is proved that
$\rcorrect{\circtopps(C)}{C}$, for any instance of circuit. These
accessors are listed and given abbreviations in
Table~\ref{tab:my_label}.

\begin{table}[tbh]
    \caption{Function \circtopps and accessors}
    \label{tab:my_label}
    \centering
    \begin{tabular}{|c|c|}
    \hline
    Accessor& Abbreviation
    \\\hline
    $\pwidth(\circtopps(C))$     & $\Cw(C)$ \\
$\prange(\circtopps(C))$         & $\Cr(C)$\\
    $\pangle(\circtopps(C))$     & $\Ca(C,-,-)$ \\
$\pket(\circtopps(C))$         & $\Ck(C,-,-)$\\\hline
    \end{tabular}

\end{table}

\subsubsection{Probabilistic Reasoning}
\label{sec:probabilistic-reasoning}

\qbrickDSL does not contain any constructor for the measurement of quantum
registers. Nevertheless, \qbrickSPEC provides reasoning tools about
it. In particular, function
\[\probmeaspart: \circtype\times \kettype \times \inttype \to \realtype\]
inputs a circuit $C$, a quantum data register $\ket{v}{n}$ and an
index $j\in \tofset{2^n}$. It outputs the probability, for one
measuring the quantum register resulting from applying circuit $C$ to
$\ket{v}{}$, to get the bit vector representing integer $j$ as a result. Function \probmeaspart is
defined, for any input ket $\ket{u}{}$ of length $\pwidth(C)$, by
application of the Born rule (see Section~\ref{sec:qd}), as
\[
  \probmeaspart(C,\ket{i}{n},j) = \vert
  \big(Ps(\circtopps,\ket{i}{n})\big (j)\vert^2.
\]

\qbrickSPEC also provides similar reasoning ghost functions for discussing the effect of partial measurement over quantum sub-registers.
\subsubsection{Verified Properties}
\label{sec:qbricks_spec}

The \qbrick framework aims at providing tools for writing and
verifying the standard format of quantum algorithm specifications as
they appear in algorithms  (see, e.g., Figure~\ref {ncpe}): to
perform a given computation task with a given amount of resources.
Hence, as illustrated above, \qbrickSPEC is designed for the
formalization of both:
\begin{itemize}
\item parametrized input/output relations for families of
  circuits. For a family of circuits, they typically consist in
  characterizing their parametrized output ket vector. Thanks to function \probmeaspart, \qbricks
  enables to identify the probability to get a given result after
  measurement and derivated probability reasoning (such as bounding
  the parametrized probability of success of a computation, \wrt a
  pre-defined success condition). Proof support for these
  specifications is processed through the \pps\ formalism,
\item complexity specifications: \qbrick\ enables to specify the
  parametrized width, number of required ancilla qubits and number of
  elementary gates of a circuit family.
\end{itemize}

\begin{example}[{\Pps} specifications for the Bell generating circuit]
  For example, a specification for the Bell generating circuit can be
  written using accessors of \pps\ \circtopps(\textit{Bell-circuit}),
  as below
  \[\begin{array}{c}
      \multicolumn{1}{l}{\Gamma, \bitvec{x},\bitvec{y}:
      \texttt{bit\_vector},j : \inttype \vdash}\\
      \{\begin{array}{lrlrl}
          \bvlength(\vec{x}) = 2
          &\wedge
          &\bvlength(\vec{y}) = 1 &\wedge& j\in\tofset{2}
        \end{array}\}
      \\
      \textit{Bell-circuit}
      \\
      \left\{
      \begin{array}{c}
        \begin{array}{lrlr}
          \Cw(\result) = 2
          &\wedge
          & \Ck(\result,\bitvec{x},\bitvec{y}) = \bitvec{x(0) \cdot
            (1-x(1))}
          & \wedge
          \\
          \Cr(\result) = 1
          &\wedge
          &\Ca(\result \bitvec{x},  \bitvec{y}) = x(0)*y(0)\
          &\wedge \\
        \end{array}
        \\
        \begin{array}{lrlrlr}
          \width(\result) = 2
          & \wedge
          & \noeg(\result) = 2\
          & \wedge
          &\texttt{ancillas}(\result) = 0 
        \end{array}
      \end{array}\right\}
    \end{array}
  \]
\end{example}
This Hoare style notation, $\{\textit{Pre}\}p\{\textit{Post}\}$ states
that whenever \textit{Pre} is satisfied, then running $p$ ensures that
\textit{Post} is satisfied. Formula \textit{Post} uses $\result$ as a
variable standing for program $p$.
The specification uses free--\emph{ghost}--bit vectors variables $\vec{x}$ and
$\vec{y}$ and integer variable $j$.  It requires $\vec{x}$ and
$\vec{y}$ to have respective lengths $2$ and $1$ and $j$ to be in
$\tofset{2}$. Given these preconditions, the specifications ensures that
the angle $\Ca$ outputs $\bitvec{x}_0*\bitvec{y}_0$ for inputs
$\vec{x}$ and $\vec{y}$ and the ket function $\Ck$ outputs
$\bitvec{x}_0\cdot(1-\bitvec{x}_1)$ for inputs $\vec{x},\vec{y}$ and
$j$. Then, one easily verifies that, applying equation~\ref{pps_expr}
on each bit-vector $\overrightarrow{a\cdot b}$ of length $2$ results in the
corresponding output $\ket{\beta_{ab}}{}$,
formally:
\[
  Ps\big(\circtopps(\textit{Bell-circuit}), \ket{ab}{}\big) =
  \Mat{\textit{Bell-circuit}}\cdot\ket{ab}{} )
\]

So the postcondition, by use of $\Cw,\Cr,\Ca$ and $\Ck$, enables a
complete characterization of the input/output function performed by
the Bell circuit. In addition, the specification brings some
complexity related postcondition: the circuit has width $2$ and size
$2$ (\ie, length of the required quantum register and number of performed elementary operations), and does not use any additional ancilla qubit.

\subsubsection{Deduction and Proof Support}

Proof support in \qbrick strongly relies on the compositional structure
of quantum circuits, enabling compositional reasoning on both \pps\
and complexity features. As an example, in
Figure~\ref{tab:qbrickdsl-nog-size} we give some of the rules used for
the characterization of circuits size.  Gates \texttt{ID} and
\texttt{SWAP} are considered as free, so they count for null. The
other elementary gates have size $1$, both sequence and parallel
compositions sum the size of their components and ancilla
creation/termination does not affect circuit size.

Similar deduction rules are defined for circuit width, number of
ancilla qubits and \pps\ accessors. We do not introduce them here for
sake of concision but we refer the desirous reader
to~\cite{chareton2021automated}.

\begin{figure}[tbh]
  \begin{center}
    \begin{prooftree}
    \hfill
      \AxiomC{$C\in \{\texttt{ID,SWAP}\}$}
      \RightLabel{$\texttt{(Id-SWAP-size)}$}
      \UnaryInfC{$\size{C} = 0$}
      \DisplayProof
      \hfill
      \AxiomC{$C\in \{\texttt{H}, \texttt{Ph}(n), \texttt{R}_z(n)\}$}
      \RightLabel{$\texttt{(H-Ph-R}_z\texttt{-size)}$}
      \UnaryInfC{$\noeg(C) = 1$}
      \hfill
    \end{prooftree}
    \hfill
    
    \begin{prooftree}
      \AxiomC{$\Gamma \vdash\noeg(C_1) = n_1$}
      \AxiomC{$\Gamma \vdash\noeg(C_2) = n_2$}
      \AxiomC{$\Gamma \vdash\size{C_1} = \size{C_2} $}
      \TrinaryInfC{$\Gamma \vdash\noeg(\texttt{SEQ}(C_1,C_2)) = n_1 +  n_2$\hspace*{8ex}{\texttt{(seq-size)}}\hspace*{-21ex}}
    \end{prooftree}
    \begin{prooftree}
      \AxiomC{$\Gamma \vdash\noeg(C_1) = n_1$}
      \AxiomC{$\Gamma \vdash\noeg(C_2) = n_2$}
      \RightLabel{$\texttt{(par-size)}$}
      \BinaryInfC{$\Gamma \vdash\noeg(\texttt{PAR}(C_1,C_2)) = n_1 +  n_2$}
    \end{prooftree}
    
    \begin{prooftree}
      \AxiomC{$\Gamma \vdash\noeg(C) = n$}
      \RightLabel{$\texttt{(anc-size)}$}
      \UnaryInfC{$\Gamma \vdash\noeg(\texttt{ANC}(C)) = n$}
    \end{prooftree}
    
    \caption{Deduction rules for \qbrick: size (number of  gates)}
    \label{tab:qbrickdsl-nog-size}
  \end{center}
\end{figure}

\subsubsection{Implementation and Case Studies}

\qbricks enabled to implement,
specify and formalize parametrized versions for Quantum Fourier
Transform, Grover search algorithm, QPE, Shor order finding, \textit{etc}.
The main specificity of these \qbricks implementations is to
hold a high level of proof support automation. This is largely due
to the reduction of proof
obligations to first-order logic predicates by use of \pps\
characterizations. 

 \subsection{\sqir}

\label{sec:sqir}
The \sqir
language~\cite{hietala19:verif_optim_quant_circuit,hietala2020proving} is the representation language used by the \voqc optimizer (see Section~\ref{sec:voqc}).
On its own, it also constitutes a solution for formally proved correct quantum
programs. It is developed concurrently with \qbrick and holds similar
concerns: basically, to  reduce the expressivity of programming
languages such as \quipper and \qwire so as to (1) still enable the
whole implementation of  emblematic algorithms  (2) enable formal proof of specifications.

The development of \sqir followed that of \qwire (see Section~\ref{sec:rev_of_ex_app}) when the authors
observed the difficulty to hold formal verification, mainly linked to
the management of memory wires. 
Hence \sqir and \qwire
have overlapping  author and developer ship, they are both deeply embedded in
the \coq proof assistant and they share the same mathematical libraries
for, e.g., matrices and complex numbers. Schematically, compared to \qwire, \sqir has a reduced expressivity (disabling, eg., the identification of qubit held values), making tractable the formal verification of functional program properties.

\subsubsection{Programming Language}

Just as \qbricks, the programming part of \sqir is reduced to the
minimum enabling implementation of main quantum programming
features. It is two-layered: the \emph{unitary part} corresponds to
the design of unitary circuits and a generalized
circuit layer adds branching measurement and ket initialization.

In \sqir, quantum circuits have type $\mathbb{N} \to \texttt{Set}$,
with a positive integer parameter corresponding to their width.
Quantum memory wires are identified by integer indexes, bounded by a
global register size parameter \texttt{d}. Then, the type for
unitary operators in \sqir (\texttt{ucom}) features the application of
a circuit to a quantum register with \texttt{d} wires.  It is defined
inductively as follows.
\begin{center}
$\begin{array}{l}
\texttt{Inductive ucom (U: }\mathbb{N} \to \texttt{Set)  (d: }\mathbb{N}\texttt{) : Set :=}\\
| \texttt{ useq : ucom U d $\to$ ucom U d $\to$ ucom U d}\\
| \texttt{ uapp1 : U 1 $\to$ $\mathbb{N}$ $\to$ ucom U d}\\
| \texttt{ uapp2 : U 2 $\to$ $\mathbb{N}$ $\to$ $\mathbb{N}$ $\to$ ucom U d}\\
| \texttt{ uapp3 : U 3 $\to$ $\mathbb{N}$ $\to$ $\mathbb{N}$ $\to$ $\mathbb{N}$ $\to$ ucom U d}\\
\end{array}$
\end{center}

This definition holds two kinds of operations:
\begin{itemize}
\item the sequential composition, \texttt{useq}, which
  inputs two circuits and outputs their sequential composition,
\item the application of an elementary gates to (a) given wire(s),
  depending on the width of this gate. There are three different
  versions of this operations, for gates of width $1,2$ or $3$, named
  respectively \texttt{uapp1}, \texttt{uapp2} and \texttt{uapp3}. As an example, 
  $\texttt{uapp1}$ inputs a gate $\texttt{U}$ of width $1$ and a
  parameter \texttt{i}. It outputs the result of applying
  $\texttt{U}$ on wire $\texttt{i}$ in a register of size
  \texttt{d}\footnote{Note that, for this construction to make sense,
    the parameter $\texttt{i}$ should not be greater than
    $\texttt{d}$. This condition is encoded by the semantics of
    \sqir.}.
\end{itemize}

Then, \sqir provides a generalized circuit building layer, enabling
the sequential composition of unitary commands, their initialization
and branching measurement. Again, a circuit is an object \texttt{com}
of type $\mathbb{N} \to$ \texttt{Set} applied on a register of
specified size \texttt{d}. It is defined, inductively, as follows.
\begin{center}
$\begin{array}{l}
\texttt{Inductive com (U: }\mathbb{N} \to \texttt{Set)  (d: }\mathbb{N}\texttt{) : Set :=}\\
| \texttt{ uc : ucom U d $\to$ com U d $\to$ ucom U d}\\
| \texttt{ skip : com U d}\\
| \texttt{ meas : $\mathbb{N}$ $\to$ com U d  $\to$ com U d  $\to$ com U d}\\
| \texttt{ seq : com U d $\to$ com U d $\to$ com U d}\\
\end{array}$
\end{center}

A generalized circuit is built as either the lifting \texttt{uc} of a
circuit into a generalized circuit, the empty \texttt{skip} operation,
the branching measurement \texttt{meas} (with inputs a wire identifier
for a qubit to measure and two generalized circuits to execute,
depending on the measurement result) and the sequence \texttt{seq} of
two different generalized circuits.

\subsubsection{Matrix Semantics and Specifications}

The semantics for \sqir programs is based on the standard matrix
apparatus. For the unitary fragment, it uses the matrix semantics presented in
Section~\ref{sec:quant-circ-semant}. It is extended for
generalized circuits by density operators semantics, similarly as in
Section~\ref{qhl}.

In its present state of development, \sqir enables to specify  functional properties, describing the input-output relation,
similarly to the one for \qbricks introduced in
Section~\ref{sec:qbricks_spec}

\subsubsection{Implementation and Case Studies}

\sqir is implemented as an embedded DSL into the \coq proof assistant. It
was illustrated with specified and proved parametrized implementations
of Simon's algorithm, QPE and Grover's Search
Algorithm.

\subsubsection{Comparison Between \qbricks and \sqir}

\qbricks and \sqir are being developed concurrently, with very similar
objectives. In particular they both trade-off between offering user-friendly
programming features and
reducing the language expressivity to the minimal, to enable
functional formal verification.  The solutions they 
provide share many common points. We discuss their main 
design differences.

\begin{itemize}
\item \sqir elementary operations consist in applying quantum gates on
  given wires of a quantum register, whereas \qbricks proceeds by
  assembling quantum gates together into a quantum circuit, just as
  bricks of a wall.  Both views are 
  inter-simulable: \qbricks provides a macro \texttt{place} with
  integer parameters specifying the wire identifier a given
  sub-circuit should be applied to and the size of the overall
  circuit (corresponding to the size of the available quantum register). 
  This macro is built by the parallel combination
  of its sub-circuit arguments with the appropriate number of
  occurrences of \texttt{ID} gate. It is of similar use as  \sqir
  function \texttt{uapp}. On the other hand, \qbricks gates assemblage
  is trivially simulable through \sqir \texttt{uapp};


\item On the other hand, \sqir provides a generalized circuit building
  layer, including measurement and classical control. Nevertheless,
  this upper layer is formalized through density operators, which are
  cumbersome objects for formal reasoning. So far, this part of the
  language only received illustrations with toy examples, such as
  superdense coding or quantum teleportation. In more involved
  implementations (such as QPE or Grover), \sqir authors followed a
  specification and proof strategy similar to that of \qbricks: by
  reasoning on the quantum data outputs of circuits, specifying over
  the probability distribution of result if a measurement were
  performed. Hence, designing a generalized circuit
  building language semantics probed against actual implementations of real usage algorithms  is still an open
  challenge; 
  
\item As introduced in Section~\ref{sec:complexity-req}, complexity
  properties of circuits constitute a fundamental aspect of quantum
  certification, decisive with regards to both the physical
  reliability of a computation and the quantum advantage it may
  provide. In the present state of development, \sqir does not offer a
  solution for this type of specifications. Still, this could be
  merely implemented in \sqir as additional functionality.
\end{itemize}

\subsection{Conclusion about Formal Verification of Quantum Programs}
\label{sec:xps-comp}

In Table~\ref{mat-ps-comp} we sum up the main concrete case study
realizations of formally verified quantum algorithms: instances of
Grover algorithm from \qbricks, \sqir and QHL, Deutsch-Jozsa and QPE
instances from \qbricks and \sqir, and \shorof implementation from
\qbricks. For each of these implementations, we give the length of the
code (column LoC) and a measure of the human proof effort required for
the specification proofs. It was obtained by adding the length of the
program specifications (Spec stands for the number of lines of
specifications and intermediary lemmas) and the number of proof
commands that were required to prove these specifications (column Cmd).

\begin{table}[tbh]
  \caption{Compared implementations of formally verified quantum algorithms}
  \label{mat-ps-comp}
  \begin{center}
    {\footnotesize
      \begin{tabular}{|l|cc|cc|cc|} \hline
        \rowcolor{gray!50}
\multicolumn{1}{|c}{}     &
          \multicolumn{2}{c}{\qbrickCORE{}\cite{chareton2021automated}  }
        &
          \multicolumn{2}{c}{\sqir\cite{hietala2020proving,hietala19:verif_optim_quant_circuit}}
        & \multicolumn{2}{c|}{QHL\cite{10.1007/978-3-030-25543-5_12}}\\
        \rowcolor{gray!50} &LoC & Spec+Cmd
        &LoC
                           & Spec+Cmd &LoC &
                                             Spec+Cmd
        \\
        \hline
        DJ&11&85&10&261&&\\
        \rowcolor{gray!25}
        Grover & 39&279&15&926&90&2975\\
        QPE &23&246&40&812&&\\
        \rowcolor{gray!25}
        \shorof &132&1212&&&&\\
        \hline
      \end{tabular}
    }
  \end{center} 

\end{table}

To the best of our knowledge Table~\ref{mat-ps-comp} is comprehensive
regarding parametrized formally proved quantum algorithm~\footnote{An additional formalization of Deutsch Jozsa algorithm is presented by Bordg et al.~\cite{bordg2021certified}. We do not include it in Table~\ref{mat-ps-comp} since it is not generated by a programming language but directly led as an algebraic proof. The total length of the proof is over 1 700 lines. Additionally, the online \sqir repository contains a Shor algorithm folder, but  to the best of our knowledge, it was never explicitly presented nor described. It seems, by the way, hardly comparable with the \qbricks implementation since it focuses on different aspects by bringing additional classical post-processing functions but lacking the oracle implementation.}. Let us stress out how young the field is (in complement to
the reduced number of concrete realizations, note that none of them is
dated earlier than 2019). 
Nevertheless, it has already brought promising results.

One of the main challenges for formal verification is to reduce the
human proof effort that is required for the certification of
programs. As Table~\ref{mat-ps-comp} shows, comparing this effort to
the length of effective programs, \qbricks offers a quite stable ratio
$\simeq 10$. In
a quite regular way, \sqir adds a $\simeq 3.5$ factor to this ratio
and QHL, for the case of Grover algorithm, requires $\simeq 10$ times
more human effort \footnote{Note that the QHL implementation of Grover
  algorithm concerns a restricted case, with regards to the two others figuring in
  Table~\ref{mat-ps-comp}. Furthermore, it does not contain the
  gate-to-gate circuit building but uses large circuit portions as
  primitives instead. Therefore, factor $\simeq 10$ is actually an
  underestimation.}

\section{Discussion and Bibliographical Notes}
\label{sec:discussion}
We end up this survey by providing some additional references for
usage of formal methods in quantum information and quantum computing.

\subsection{Deductive Verification}

Deductive verification appears to be the most promising direction for
the development of formal methods in quantum computing.  In
particular, it is particularly adequate for the formal verification of
functional specifications, which is crucial for quantum programming
and prone to play, there, a role similar to that of testing and
debugging in classical computing (see
Section~\ref{sec:need-formal-methods}).  It is worth noting that, currently, 
all existing  formally verified quantum
algorithms descriptions~\cite{10.1007/978-3-030-25543-5_12} or 
implementations
\cite{chareton2021automated,hietala19:verif_optim_quant_circuit}
are
 based either on deductive verification or interactive proofs.

\subsection{Model Checking}

Attempts for functional verification of quantum algorithm with
model checking techniques were also led before these developments
\cite{10.1007/978-3-540-70545-1_51,ying:2014:mlp:2648783.2629680,ying2021model}.
They enabled to verify toy examples of quantum processes in a completely
automated way. Nevertheless, this direction is limited by its high
scale sensitivity, which is specifically problematic for quantum
programs, since they are designed to tackle large problems instances.

\subsection{Type Checking}

Apart from functional specification, specialized type systems for
quantum programming languages also facilitate programming and
debugging.  In Section~\ref{sec:need-formal-methods} we introduced the
verification of structural constraints and the non-duplicability of
quantum information. Type checking may also have further use in
quantum computing. 

Recently, the  \silq\ language was proposed. It is based on a linear
type system which, upon other features, enables to
verify, for any quantum circuit, whether it can be \emph{uncomputed}, a
computation feature required at many stages of quantum
implementations.  Based on this type system, \silq enables to
automatically generate the uncomputation steps of circuits.  This
partial automation of the development lowers the expertise requirements
for developers and the length of programs with regards to languages
such as Q\# or \quipper.

\label{sec:further}

\subsection{Runtime assertion Checking}
\label{sec:assertion_checking}
First, recall from Section~\ref{sec:intro} that proving quantum
programs is mainly meant to replace the standard classical debugging
method of testing and assertion checking. Apart from the development
of formal proofs as an alternative, efforts are led to adapt this
classical strategy to the quantum case. There, we still decorate
programs with formal specifications (called \emph{assertions} in this
context) describing the evolution of the system state through an
execution.  But instead of mathematical proofs, these assertions are
probed by statistical testing over program fragments.  The challenges
faced by such methods are mainly twofold:

\begin{description}
\item[Destructive measurement :] memory reading destructs the
  superposition of a quantum state, therefore one cannot continue the
  execution after checking. Hence, assertion checking can be applied
  only to fragments of an execution.
\item[Non-determinism :] what we aim to check is a superposition of
  states, which induces a probability distribution of outcomes when
  a measurement is performed. Then, checking an assertion requires a
  number of testing runs large enough to build up a representative
  statistical distribution.
\end{description}
To overcome these difficulties, a first strategy is to reduce the
specifications so as to express only properties that may be handled by
assertion checking.  Huang and Martonosi propose a “runtime-monitoring
like” verification method for quantum
circuits~\cite{huang:2019:sav:3307650.3322213}.  The annotation
language is reduced so as to specify, for a quantum register, to be
either in a classical state, in a superposition or entangled, without
any concern about further description of the state.

More recently, Li et al.~\cite{li2020projection} developed an
assertion-checking based method for the verification of fine quantum
registers states properties, including functional descriptions of
circuit behaviors.
This method is based on:
\begin{itemize}
\item an assertion language, based on QHL projections (see
  Section~\ref{qhl}), enabling functional specifications about
  computations at stake;
    
\item the use and formalization of \textit{gentle measurements}:
  quantum registers are not measured in the usual computation basis,
  but in a basis containing an output that is very close to the
  expected state;
    
\item an appropriate formalization of the notion of distance between
  quantum states, enabling verification in terms of confidence
  interval between the current state of the system and its expected
  value.
\end{itemize}
Hence, gentle measurement enables to test executions over full
functional state specifications. In addition to bringing expressivity to
the specifications, it lowers the undesired effects of destructive
measurements: since a gentle measurement operator contains an
eigenstate that is an approximation of the expected quantum state, in
most cases the measurement effect on the system state can be
considered negligible and the execution can be pursued. Furthermore,
the test result probability distribution is centered on a specific
value. Therefore, a much-reduced set of runs bring valuable statistical
conclusions.

However, verification following this strategy only holds for
a particular instance of a circuit, instead of a family of quantum
circuits with unassigned parameters as in propositions from
Section~\ref{sec:high-and-lid-level-verif}. Furthermore, in the
general case, gentle measurements are implemented by applying (1) a
unitary $U$ uncomputing the system state into a ket of the
computational basis (2) a measurement in the computational basis (3)
unitary $U^{\dagger}$, to recover the initial state. Then,
assertion checking inputs unitaries $U$ and $U^{\dagger}$ that are
themselves prone to error. More precisely, to test against the exact
expected value of the state, operator $U^{\dagger}$ should be
equivalent to the computation under test. Practically, gentle
measurements approximate the state under test. This enables a
simplification of the measurement operator, at  the cost  of the
robustness of the procedure.

\subsection{Verification of Quantum Communication Protocols}
Another challenge for formal methods is the verification of quantum
information processes concerns quantum protocols.

Quantum key distribution
protocols~~\cite{scarani2009security,bennett2020quantum,liao2017satellite,lo2005decoy}
enable secured information exchanges between two parties. These
protocols exploit the fact that, due to the destructive
measurement, physics laws prevent almost any \footnote{These protocols are based on the fact that, in the general case, a potential eavesdropper destructs a quantum message he attempts to intercept, so that the parties can detect the attempt. Nevertheless, it is based on probabilities and there is always a chance for an eavesdropper to perform only conservative measurements. The corresponding probability is bounded by $r^n$, with $r\in ]0,1[$ and $n$ the length of the sent message.} possibility for a potential
eavesdropping in the exchange of quantum information.
In particular, several formally verified implementations of the BB84
quantum key distribution protocol~\cite{bennett2020quantum} have been
proposed in the literature, based either on process
calculus~\cite{nagarajan2002formal,kubota2016semi}, formalization in
\coq~\cite{dblp:journals/corr/boenderkn15} or
model checking~\cite{elboukhari2010verification,fernandez2011formal}.

Bordg et al.~\cite{bordg2021certified} propose  a
formalization of quantum information in the \isabelle proof
assistant. They illustrate their methods through the cases of quantum
teleportation and the quantum prisoner dilemma. This work also
contains a formally verified implementation of the Deutsch-Jozsa
algorithm in \isabelle.

In a more fundamental prospect, Echenim et
al.~\cite{echenim2021quantum} provide an \isabelle proof for the
CHSH inequalities~\cite{clauser1969proposed}. These are probability
distributions about crossed measurement results of quantum
observables. They provide a proof for the Bell
theorem~\cite{bell1964einstein}, inducing that no classical theory
could account for the entanglement phenomenon (hence that quantum
physics cannot be reduced to local classical theories).

\section{Conclusion}

\label{sec:conclusion}

\subsection{Summary}
Throughout this survey, we introduced the context, the main
challenges and the most promising results in formally verified quantum
programmatic. The current state of affairs in this emerging domain can
be summed up as follows:

\begin{itemize}
\item Quantum computing is an emerging domain, with huge potential
  application fields and promises. Progresses in the development of
  concrete machines are reaching the practical relevance landmark:
  prototypes are getting powerful enough to overpass classical
  computers. Consequently, quantum software development is becoming a
  crucial industrial short-term need;
  
\item Quantum software deals with an entirely new programming
  paradigm. Upon its main particularities are the dual nature of
  information (either classical or quantum), the destructive
  measurement and irreducibly probabilistic computations;

\item These specificities make programming particularly non-intuitive
  and prone to error. Furthermore, they make it very hard to directly
  import usual debugging methods from the classical practice (based on
  test and assertion checking). The technique presented in
  Section~\ref{sec:assertion_checking} might bring some hope, but this
  is still so far at a very preliminary stage;

\item Formal methods appear as the privileged alternative for
  debugging strategies. Apart from providing solutions to the
  destructive measurement challenge, they have additional decisive
  advantages: mainly, they provide absolute guarantee of the
  correction of programs, and they hold for any instance of programs
  they verify;

\item During the last ten years (the \emph{genesis} of formal quantum
  programming), this new field has shown promising results in the
  different stages of software development: high-level program
  designs, circuit building languages, verification, compilation, optimization,
  \textit{etc}.

\end{itemize}

\subsection{Main Current Challenges}

Although encouraging, these early successes  draw the road map for rising the field from academic proof of concepts to practically usable programming solutions. We present  the main coming challenges for this development in three categories: providing relevant integrated development solutions for the NISQ era, offering practical wide spreadable user experience and developing a full-fledged formally verified quantum compilation toolchain. 

\paragraph*{Provide relevant integrated development solutions for the NISQ era.}

So far, quantum formal verification mainly proved its relevance by
offering solutions for the unitary core of quantum computations. As a
matter of fact, illustrations and concrete implementations using these
techniques primarily treated historical emblematic algorithms such as
Shor~\cite{shor1994algorithms}, QPE~\cite{kitaev1995quantum} or Grover search~\cite{grover1996fast}. The classical treatment in
these algorithms can be completely decoupled from their quantum core.

However, the first generation of quantum computing machine (the
\emph{NISQ era},~see Section~\ref{sec:quantum-noise}) hints towards a radically distinct mode of
operation. NISQ machines will have limited, noisy resources.  A major
consequence is that these quantum processors are too small to support
the error correction mechanisms required for Shor's and Grover's
algorithms. Such NISQ processors aim instead at different kinds of
algorithms: hybrid algorithms such as variational algorithms
\cite{mcclean2016theory}. Hybrid algorithms tightly mix quantum and
classical data treatments: one cannot decouple the quantum part of the
algorithm from its classical part.

Adapting the quantum formal methods to the NISQ setting to
support hybrid quantum/classical computation is a challenge and an
active current research avenue.

\paragraph*{Offer practical wide spreadable user experience:
  formalism, language design, automation.}

In the present state of development, programming languages enabling
formal verification usually sacrifice their expressivity to formal
reasoning. In particular, all the solutions that have been probed
against actual parametric quantum algorithms fail to satisfy essential
elements of the requirements listed in Section~\ref{sec:qpl-req} for
scalable quantum programming languages: in addition to classical
processing mentioned above, they cannot manipulate 
quantum registers and wire references. Verified programming should
address these limitations.

Another issue, while concerning any method of quantum programming, is
especially critical in the case of formally verified programming. It
concerns the level of qualifications required from developers. The
interpretation of quantum computations indeed requires unusual and
non-intuitive mathematical formalism (including Kronecker products,
complex phase amplitudes, probabilistic reasoning,
\textit{etc}). While the need for qualified programmers is prone to
grow rapidly in the coming years, integrating formal verification
should come with the development of user-friendly specification
languages and highly automated mathematical reasoning engines.

\paragraph*{Formally verified quantum compilation toolchain.}
In addition to the preceding considerations, in its early academic
ages, quantum formal verification focused on idealized representations
of quantum circuits, directly extracted from algorithm descriptions.
As introduced in Section~\ref{sec:quantum-noise}, these logical qubits
are merely abstract models for actual computations to be run.
A major addition that is left for future works in quantum formal
verification is error correction, with formal verification that the
state of the system preserves the functional correctness of
computations (assuming a given error model, and possibly with
probabilistic specifications). Presented in Section~\ref{sec:voqc},
\voqc is a first step towards this goal.

Another future direction concerns integration in a widespread
classical development environment. Recall from
Section~\ref{sec:rev_of_ex_app} that many widespread quantum
programming languages benefit from embeddings in usual programming
languages--such as Python. Today most formal verification solutions
are embedded in a more academic functional development environment (such
as Haskell, Ocaml, proof assistants or deductive verification
environments). Interfaces should be developed to integrate
formally verified quantum computations into comprehensive projects.

At the other extremity of the development stack, formal verification
should accompany the implementation of compiled programs on concrete
machines. This induces verified solutions for the qubit mapping problem and
gate simulation (see Section~\ref{sec:compilation}), depending on
 the particular target material and its proper
constraints.

\section*{Acknowledgments}

This work was supported in part by the French National
Research Agency (ANR) under the research project SoftQPRO
ANR-17-CE25-0009-02, by the DGE of the French Ministry of
Industry under the research project PIA-GDN/QuantEx
P163746-484124, by the STIC-AmSud project Qapla' 21-SITC-10 and by the Carnot project Qstack.
We thank the anonymous reviewers for helpful comments on earlier drafts of the manuscript.


\bibliographystyle{plain}
\bibliography{main}

\setcounter{page}{16}

\end{document}